\documentclass[a4paper,11pt]{article}

\usepackage{jheppub}
\usepackage{hyperref}
\usepackage[T1]{fontenc}
\usepackage{caption}
\usepackage{subcaption}
\usepackage{dsfont}
\usepackage{extarrows}
\usepackage[aligntableaux=center,boxsize=0.8em]{ytableau}
\usepackage[section]{placeins}
\usepackage{graphicx}
\usepackage{tikz}

\usepackage{xcolor}

\allowdisplaybreaks[1]

\def\({\left(}
\def\){\right)}
\def\[{\left[}
\def\]{\right]}
\def\<{\left\langle}
\def\>{\right\rangle}

\def\be{\begin{eqnarray}}
\def\ee{\end{eqnarray}}

\def\a{\alpha}

\def\D{\Delta}
\def\e{\epsilon}

\def\k{\kappa}

\title{Bootstrapping mixed correlators in the 3d $\mathrm{O}(N)$ normal boundary universality class}

\author{Wenliang Li, Yuheng Yao}
\affiliation{School of Physics, Sun Yat-Sen University, Guangzhou 510275, China}

\emailAdd{liwliang3@mail.sysu.edu.cn}

\abstract{We study the normal boundary universality class of the three-dimensional $\mathrm{O}(N)$ model using the truncated bootstrap with $\eta$-minimization.
This universality class is realized in spin systems by applying a symmetry-breaking field on the boundary and is closely related to the extraordinary-log transition in three dimensions.
We separately consider bulk two-point functions involving the relevant $\mathrm{O}(N)$ vector $\phi_a$ and singlet $S$: $\langle\phi_a\phi_b\rangle$ for $N=6$, $\langle\phi_a S\rangle$ for $N=2,3,4,5,6,8,10,12$, and $\langle S S\rangle$ for $N=1,2,3$.
The analyses yield estimates of low-lying bulk and boundary scaling dimensions, as well as the coefficients in the conformal-block decompositions of these correlators.
The results from the mixed correlator $\langle\phi_a S\rangle$, together with those from $\langle\phi_a\phi_b\rangle$, allow us to further disentangle several low-lying bulk and boundary operator expansion coefficients.
The results from $\langle SS\rangle$ and $\langle\phi_a\phi_b\rangle$ provide estimates for the ratio of the bulk OPE coefficients involving the subleading singlet.
Combining our $N=6$ result from $\langle\phi_a\phi_b\rangle$ with previous results for $N=2,3,4,5$, we estimate the extraordinary-log stability threshold to be $N_c\approx 5$, consistent with previous studies.
We also compare the dependence of selected conformal data on $N$ with analytic large-$N$ predictions.
These results further illustrate the utility of $\eta$-minimization for boundary bootstrap analyses without assuming positivity.}

\begin{document} 
\maketitle
\flushbottom
\section{Introduction}
Boundaries have attracted considerable attention owing to their close connections with a wide range of physical phenomena \cite{Andrei:2018die}, including the Casimir effect \cite{Casimir:1948dh,Bordag:2001qi}, the Kondo problem \cite{Affleck:1995ge}, D-branes in string theory \cite{Recknagel:2013uja}, and topological phases \cite{Cho:2016kcc,Cho:2016xjw,Han:2017hdv,Fliss:2017wop,Wong:2017pdm,Wen:2016snr,Qi:2012ngv}. 
In this work, we focus on boundary criticality \cite{Diehl:1996kd,Binder:1983,Diehl:1986,Bray:1977fvl,Burkhardt:1987,Diehl:1994a,Burkhardt:1994,Diehl:1994b,Metlitski:2020cqy}.
At a fixed bulk critical point, different boundary conditions can give rise to distinct boundary universality classes.
The $\mathrm{O}(N)$ model with a surface provides a prototypical example of boundary criticality.
In its classical lattice realization, the $\mathrm{O}(N)$ model consists of $N$-component unit-vector spins, with $N=1,2,3$ corresponding to the Ising, XY, and Heisenberg models, respectively.
At the bulk transition, the system can exhibit various boundary universality classes, including the normal, extraordinary, special and ordinary classes.

In the presence of an explicit boundary field that breaks $\mathrm{O}(N)$ to $\mathrm{O}(N-1)$, the boundary transition at bulk criticality is known as the normal transition.
The corresponding boundary fixed point is stable, with no relevant boundary perturbations that preserve the residual $\mathrm{O}(N-1)$ symmetry \cite{Metlitski:2020cqy}.
Since the boundary enhancement need not be tuned, the normal transition is readily accessible in both experiments and numerical simulations.
This transition is realized in spin systems when a symmetry-breaking magnetic field is applied to the surface.
For $N=1$, the normal universality class is also relevant to critical adsorption in binary fluid systems \cite{Diehl:1994b}.\footnote{This connection follows from the equivalence between the Ising and lattice-gas models, under which the magnetization corresponds to the fluid density. The nonzero surface field can be interpreted as a preferential affinity for one of the two components.}

Without an explicit symmetry-breaking field at the boundary, the system can exhibit the extraordinary, special, or ordinary boundary universality class for spatial dimensions $d>3$.
The boundary coupling determines which of these classes is realized.
At sufficiently strong coupling, spontaneous boundary order gives rise to the extraordinary transition, which shares the same boundary critical behavior as the normal transition \cite{Burkhardt:1994,Diehl:1994b,Metlitski:2020cqy}.
For weaker boundary coupling, the system exhibits the ordinary transition, while the special transition occurs at the multicritical point separating the ordinary and extraordinary regimes.
Among the four corresponding fixed points, the special fixed point is the least stable, since it requires fine-tuning the boundary enhancement to its critical value.
Moreover, a small symmetry-breaking boundary magnetic field can drive the system from the ordinary or special fixed point toward the normal fixed point.

For $d=3$, boundary critical behavior associated with spontaneous boundary ordering may differ from that in $d>3$.
If the two-dimensional boundary were an isolated system with short-range interactions, the Mermin-Wagner theorem would forbid spontaneous breaking of a continuous symmetry.
At the bulk critical point, however, critical bulk fluctuations mediate effective long-range interactions along the boundary, making the situation more subtle.
The resulting boundary behavior can be described by an effective theory that couples the boundary Goldstone modes to the critical theory at the normal boundary fixed point.
In Ref.~\cite{Metlitski:2020cqy}, Metlitski showed that this effective theory leads to extraordinary-log boundary criticality for $N\ge2$ below a critical value $N_c$.
Unlike conventional extraordinary boundary criticality, this universality class does not exhibit true boundary long-range order.
Instead, the boundary correlations decay to zero logarithmically.
The stability threshold $N_c$ for the extraordinary-log phase is determined by universal data at the normal boundary fixed point.
This connection provides further motivation for studying the normal boundary universality class of the three-dimensional $\mathrm{O}(N)$ model.
Refs.~\cite{Toldin:2020wbn,Hu:2021xdy,Padayasi:2021sik,Toldin:2021kun,Zhang:2022hpz,Zou:2022mhr,Krishnan:2023cff,Sun:2023vwy,Cuomo:2023qvp,Toldin:2024pqi} contain subsequent studies of the logarithmic critical behavior.

At criticality, the long-distance universal behavior near a surface can be described by a boundary conformal field theory (BCFT).
In this framework, the universal information is encoded in the scaling dimensions and the operator expansion coefficients.
Various approaches have been developed to study boundary criticality. 
Beyond the mean-field approximation \cite{Bray:1977fvl}, fluctuation effects can be studied using perturbative methods, including the $\epsilon=d-2$ \cite{Diehl:1986mu,Padayasi:2021sik}, $\epsilon=4-d$ \cite{Dey:2020lwp,Padayasi:2021sik,Giombi:2025pxx,Sun:2026mib}, and large-$N$ expansions \cite{Metlitski:2020cqy,Ohno:1984,Padayasi:2021sik,Giombi:2020rmc,Ohno:2025qfv,Csipes:2026nyo}.
These methods provide analytic approximations to the associated BCFT data.
However, the expansion parameters $\epsilon$ and $1/N$ may not be sufficiently small in three dimensions and at finite $N$. 
Consequently, the perturbative estimates may deviate substantially from the exact values.

Common nonperturbative approaches to BCFTs include Monte Carlo simulations \cite{Przetakiewicz:2025gzi,Toldin:2021kun,ParisenToldin:2010,Hasenbusch:2010,Deng:2005dh,Toldin:2023fny} and the conformal bootstrap \cite{Liendo:2012hy,Padayasi:2021sik,Behan:2020nsf,Behan:2021tcn,Collier:2021ngi,Meineri:2025qwz,Gliozzi:2015qsa,Gliozzi:2016cmg,Hu:2025yrs,Toldin:2023fny}.\footnote{Recently, the fuzzy-sphere method has also been used to study BCFTs \cite{Dedushenko:2024nwi,Zhou:2024dbt,Feng:2026iii}.}
Monte Carlo analyses of boundary criticality involve boundary-sensitive observables and additional boundary corrections,
which can make the extraction of some boundary data less straightforward.
The conformal bootstrap provides a complementary nonperturbative approach, which can be broadly divided into the positive bootstrap and the truncated bootstrap.

The conformal bootstrap based on positivity was proposed in Ref.~\cite{Rattazzi:2008pe}. 
Instead of solving the crossing equations directly, this approach uses them to derive rigorous bounds on CFT data \cite{Poland:2018epd,Rychkov:2023wsd}.
Subsequent works have extended this program in various directions, including applications to three-dimensional $\mathrm{O}(N)$ bulk CFTs \cite{El-Showk:2012cjh,Kos:2013tga,El-Showk:2014dwa,Kos:2014bka,Simmons-Duffin:2015qma,Kos:2015mba,Kos:2016ysd,Simmons-Duffin:2016wlq,Chester:2019ifh,Chester:2020iyt,Reehorst:2021hmp,Chang:2024whx}.
The positive bootstrap was first applied to BCFTs in Ref.~\cite{Liendo:2012hy}, and further applications can be found in Refs.~\cite{Padayasi:2021sik,Behan:2020nsf,Behan:2021tcn,Collier:2021ngi,Meineri:2025qwz}.
This boundary bootstrap method uses crossing symmetry and positivity assumptions to set up a linear or semidefinite programming problem.
Refs.~\cite{Liendo:2012hy,Padayasi:2021sik} applied the positive bootstrap to bulk two-point functions in 3d BCFTs.
These two studies made the additional assumption that the bulk-channel conformal block coefficients are nonnegative.
Ref.~\cite{Liendo:2012hy} found evidence for this assumption at the extraordinary and special transitions, although it has not been rigorously established.\footnote{Our results for the normal transition are also consistent with positivity, as we find no negative coefficients within the truncations considered in this work.}
The same study also noted that this assumption can fail at the ordinary transition, indicating the limited applicability of this method.

In Ref.~\cite{Gliozzi:2013ysa}, Gliozzi introduced the truncated bootstrap, which does not rely on positivity. 
In that work, the resulting truncated crossing equations were solved using the determinant method.
After truncating the conformal-block expansion, one evaluates derivatives of the truncated crossing equation.
The derivative constraints form a homogeneous linear system for the expansion coefficients and can be assembled into a constraint matrix whose elements depend on the unknown scaling dimensions.
A nontrivial solution exists only when the matrix is rank deficient, or equivalently, when the relevant minors vanish.
To determine the truncated spectrum, one selects and solves independent determinant conditions equal in number to the unknown scaling dimensions.
Applications of the determinant method to BCFTs can be found in Refs.~\cite{Gliozzi:2015qsa,Gliozzi:2016cmg}.
At higher truncation orders, selecting suitable independent minors and solving the resulting determinant conditions becomes increasingly difficult.
Ref.~\cite{Esterlis:2016psv} proposed a method based on minimizing the smallest singular value of the constraint matrix, whose vanishing is equivalent to the determinant conditions.
Refs.~\cite{Padayasi:2021sik,Toldin:2023fny} applied this method to the 3d O($N$) BCFT.
The singular value method is more robust than finding simultaneous roots of minors and can reduce ambiguities arising from different subsets of minors \cite{Esterlis:2016psv}.
These early studies were restricted to relatively low truncation orders, and some estimates for the normal boundary universality class showed noticeable deviations from Monte Carlo results \cite{Toldin:2021kun,Przetakiewicz:2025gzi}.

Ref.~\cite{Li:2017ukc} generalized the determinant method by introducing $\eta$-minimization, in which an objective function $\eta$ measures the residual violation of the truncated crossing equations.
When the retained constraints and free parameters are equal in number, we search for the minima with vanishing $\eta$, which is equivalent to solving the determinant conditions.
The $\eta$-minimization can also be applied when the number of constraints exceeds that of the free parameters \cite{Li:2017ukc,Li:2023tic,Poland:2023vpn}, in which case the minimum residual is generally above zero.
The computational cost of $\eta$-minimization also grows with the truncation order.
But at higher orders, minimizing $\eta$ is more manageable than solving the determinant equations.
In Ref.~\cite{Hu:2025yrs}, it was applied to the three-dimensional $\mathrm{O}(N)$ model at the normal boundary transition. 
The sparse low-lying boundary spectrum with no relevant $\mathrm{O}(N-1)$ singlet and vector operators makes the normal universality class particularly amenable to a truncated bootstrap analysis.
By substantially increasing the truncation order, that work obtained highly accurate Ising results, especially from its mixed-correlator analysis.
It also studied the identical correlator $\langle\phi_a\phi_b\rangle$ at $N=2,3,4,5$, but did not consider the corresponding mixed correlator of the bulk vector $\phi_a$ and singlet $S$.
Motivated by these encouraging results and other available bulk input data, we extend the boundary bootstrap study to additional correlators and values of $N$, especially the mixed correlator $\langle\phi_a S\rangle$ for $N\ge2$.

In this work, we study the two-point correlators of the relevant bulk vector $\phi_a$ and singlet $S$ in the normal boundary universality class. 
We apply the truncated bootstrap with $\eta$-minimization to each correlator separately.
We focus primarily on the mixed correlator $\langle\phi_a S\rangle$ for $N=2,3,4,5,6,8,10,12$, whose bulk channel similarly contains relatively few low-lying operators.
We also study the correlators $\langle\phi_a \phi_b\rangle$ for $N=6$ and $\langle SS\rangle$ for $N=1,2,3$.
Using our $N=6$ data from $\langle\phi_a\phi_b\rangle$ and the $N=2,3,4,5$ results of Ref.~\cite{Hu:2025yrs}, we estimate the extraordinary-log stability threshold $N_c$ and compare it with previous estimates.
For $N=2,3,4,5,6$, we combine the $\langle\phi_a\phi_b\rangle$ and $\langle\phi_a S\rangle$ bootstrap results to disentangle important bulk and boundary operator expansion coefficients.
For $N=1,2,3$, combining our $\langle SS\rangle$ results with the $\langle\phi_a\phi_b\rangle$ results of Ref.~\cite{Hu:2025yrs} provides information on bulk coefficients associated with the subleading singlet operator.
We compare results obtained from different correlators as a consistency check on our bootstrap analysis.
We also compare selected BCFT data with previous numerical estimates, as well as with analytic predictions from the $\e=4-d$, $\e=d-2$, and large-$N$ expansions.
Using our results for several values of $N$, we extrapolate selected quantities to $N\to\infty$ and compare the resulting limits with the analytic large-$N$ predictions.

This paper is organized as follows. 
Sec.~\ref{sec:2} introduces the theoretical framework and numerical approach used in this work.
We present the main numerical results in Sec.~\ref{sec:3} and discuss their $N$-dependence in Sec.~\ref{sec:largen}.
We conclude with a summary of the results and outline possible directions for future work in Sec.~\ref{sec:4}.
Supplementary technical details and additional numerical results are collected in the appendices.

\section{Preliminaries}
\label{sec:2}
Before presenting our numerical results, we summarize the physical setting and methodological framework underlying our analysis.
We review the boundary universality classes of the $\mathrm{O}(N)$ model and formulate the crossing equations for the three correlators.
Finally, we describe our truncated-bootstrap procedure based on $\eta$-minimization and discuss some subtle aspects of truncation choices.
\subsection{Boundary universality classes}
\label{sec:2:1}
\begin{figure}[tbp]
\centering 
\begin{subfigure}[t]{0.49\textwidth}
\centering
\includegraphics[width=\textwidth]{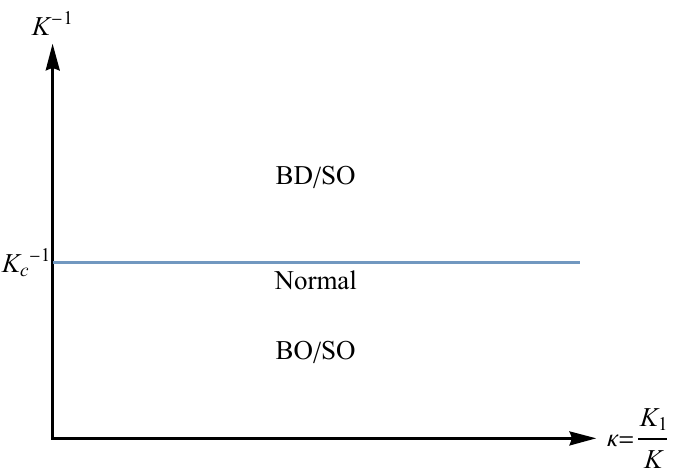}
\caption{$h_1\neq0$}
\label{fig:a}
\end{subfigure}
\hfill
\begin{subfigure}[t]{0.49\textwidth}
\centering
\includegraphics[width=\textwidth]{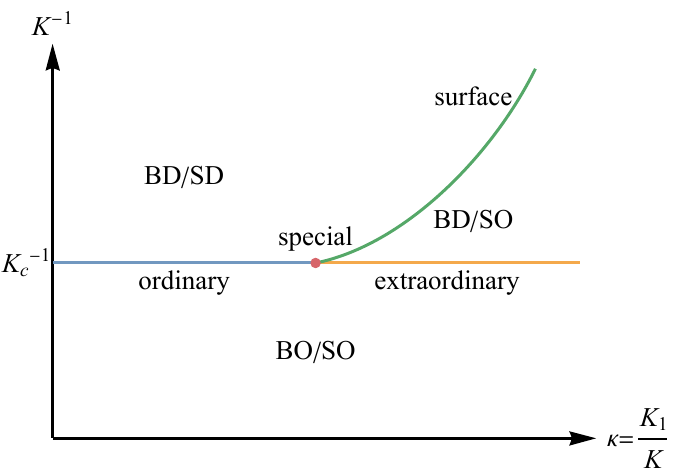}
\caption{$h_1=0$}
\label{fig:b}
\end{subfigure}
\caption{\label{fig2:1:1}Conventionally accepted phase diagram for the classical $\mathrm{O}(N)$ model with a boundary. 
``BO'', ``BD'', ``SO'', and ``SD'' denote bulk-ordered, bulk-disordered, surface-ordered, and surface-disordered phases, respectively.
(a) The $h_1\neq0$ phase diagram.
(b) The conventional $h_1=0$ phase diagram for $d>3$.
For $d=3$, the $N=1$ phase diagram remains unchanged. 
The $N=2$ case has the same topology, but its surface-ordered phase exhibits only quasi-long-range order.}
\end{figure}

\FloatBarrier
We consider the classical $\mathrm{O}(N)$ lattice model with a flat boundary, described by the reduced Hamiltonian
\begin{equation}
\label{eq2:1}
\beta H=-\sum_{\langle i,j \rangle}K_{ij}\vec{S}_{i}\cdot \vec{S}_{j}-\sum_{i\in \mathrm{boundary}}\vec{h}_{1}\cdot\vec{S}_{i},
\end{equation}
where $\beta=1/(k_{\mathrm B}T)$ is the inverse temperature, $\vec{S}_i$ is an $N$-component unit vector, and $K_{ij}$ is the nearest-neighbor coupling.
Here $\vec{h}_1$ denotes a spatially uniform boundary field.
We define $h_1\equiv|\vec{h}_1|$ as its magnitude.
A nonzero $h_1$ explicitly breaks the $\mathrm{O}(N)$ symmetry.
The coupling $K_{ij}$ is set to 
\begin{equation}
K_{ij}=
\begin{cases}
K_{1} & i,j\in\mathrm{boundary},\\
K & \text{otherwise}.
\end{cases}
\end{equation}
The conventionally accepted phase diagram for this model is shown schematically in Fig.~\ref{fig2:1:1}, where $K_c$ marks the bulk critical point
and $\k=K_{1}/K$ is the surface-to-bulk coupling ratio.
This ratio measures the enhancement of the surface interactions relative to those in the bulk.

A nonzero boundary symmetry-breaking field $h_1$ realizes the normal boundary universality class, independently of the surface coupling \cite{Metlitski:2020cqy}.
The field breaks the boundary $\mathrm{O}(N)$ symmetry to $\mathrm{O}(N-1)$.
As a relevant perturbation of the $\mathrm{O}(N)$-symmetric fixed points, this boundary field drives the RG flow toward the normal fixed point.
This fixed point is stable, since it has no relevant boundary perturbations preserving the residual $\mathrm{O}(N-1)$ symmetry.\footnote{For $N=1$, the normal fixed point has no residual internal symmetry, and Ref.~\cite{Padayasi:2021sik} assumes that the lowest nonidentity boundary operator is the displacement operator with scaling dimension $3$ in $d=3$.
For $N\ge2$, the lowest nonidentity boundary operator is assumed to be the tilt operator, with scaling dimension $2$ in $d=3$.
We adopt this spectral assumption.
In this case, no boundary perturbation at the normal fixed point is relevant, even if it breaks the residual O($N-1$) symmetry.
Coupling to the tilt operator gives a marginal boundary perturbation.}
The normal universality class can be realized whenever the bulk $\mathrm{O}(N)$ model has a finite-temperature critical point. 
In $d=3$, such a critical point exists for all $N$.

For $d>3$ and $h_1=0$, there are three boundary universality classes: ordinary, special, and extraordinary.
At the ordinary transition, the bulk and the boundary order simultaneously. 
As the surface coupling increases, the system passes through the special transition and reaches the extraordinary universality class.
When the surface coupling exceeds the critical value at the special transition, the system undergoes a separate surface transition above the bulk critical temperature.
The bulk transition therefore takes place in the presence of a symmetry-broken boundary, corresponding to the so-called extraordinary transition.
For $d=3$, the phase diagram for $N=1$ remains the same. 
For $N=2$, although the phase diagram still has the same topology as that shown in Fig.~\hyperref[fig2:1:1]{\ref*{fig2:1:1}(b)}, the boundary exhibits only quasi-long-range order in the ``BD/SO'' region \cite{Parga:1981,Landau:1989,Peczak:1991}. 
For $2\le N < N_c$, Ref.~\cite{Metlitski:2020cqy} suggested the existence of the ``extraordinary-log'' universality class, where the correlation function decays logarithmically with distance.
For sufficiently large $N$, the large-$N$ analysis in Ref.~\cite{Metlitski:2020cqy} suggests that only the ordinary boundary universality class remains at bulk criticality.

For $d>3$, the extraordinary universality class coincides with the normal boundary universality class \cite{Burkhardt:1994,Diehl:1994b,Metlitski:2020cqy}.
At the extraordinary transition, the spontaneous surface order acts as an effective symmetry-breaking boundary field.
The associated boundary Goldstone modes decouple from the bulk fluctuations in the infrared for $N\ge 2$ \cite{Metlitski:2020cqy}.
However, the situation is more subtle in $d=3$.
Fluctuations on the two-dimensional boundary can destroy long-range order for continuous symmetries.
Near the normal fixed point, the boundary Goldstone modes are described by a nonlinear sigma model coupled to the normal BCFT.
In the normalization of Refs.~\cite{Metlitski:2020cqy,Padayasi:2021sik}, its coupling $g$ obeys
\begin{equation}
\beta(g)\equiv\mu\frac{dg}{d\mu}
=\a g^2+O(g^3),
\end{equation}
where $\mu$ is the inverse length scale.
The coefficient $\alpha$ is given by
\begin{equation}
\label{eq:2:4}
\a=\frac{a_{\phi}^2}{32\pi b_{\phi t}^2}
-\frac{N-2}{2\pi},
\end{equation}
which is determined by the universal amplitudes $(a_{\phi},b_{\phi t})$ in the normal universality class. 
When $\a>0$, a small $g$ flows logarithmically to zero in the infrared, and the extraordinary-log universality class is realized. 
By contrast, when $\a<0$, a small $g$ flows toward the strong-coupling regime, indicating the $g=0$ fixed point is unstable. 
Thus, the stability of the extraordinary-log fixed point is controlled by universal BCFT data at the normal fixed point.
The critical value $N_c$ is determined by $\alpha(N_c)=0$.
Numerical estimates of the normal BCFT data can be obtained using various approaches, including Monte Carlo simulations \cite{ParisenToldin:2010,Toldin:2021kun,Hasenbusch:2010,Przetakiewicz:2025gzi} and the conformal bootstrap \cite{Liendo:2012hy,Padayasi:2021sik,Gliozzi:2015qsa,Hu:2025yrs}.

The Monte Carlo method is a well-established nonperturbative tool for studying critical phenomena.
Applied to a specific lattice model, it allows one to extract universal properties through finite-size scaling near criticality.
In this approach, lattice correlators are measured and fitted to the corresponding boundary scaling forms.
Once the normalization convention is fixed, the fitted amplitudes can be translated into universal BCFT data.
Monte Carlo studies of boundary criticality often rely on observables that are sensitive to boundary effects.
Their analysis must also account for additional finite-size corrections associated with boundary operators.

The conformal bootstrap provides a complementary approach that directly constrains BCFT data through conformal symmetry and consistency conditions.
It can provide estimates of quantities that may be difficult to extract from Monte Carlo simulations.
For quantities estimated by both methods, comparisons of bootstrap and Monte Carlo results provide useful consistency checks.
In this work, we use the conformal bootstrap to estimate BCFT data for the normal boundary universality class.

\subsection{Boundary conformal bootstrap}
\label{sec:2:2}
The boundary conformal bootstrap exploits the symmetry principles and consistency conditions intrinsic to BCFTs.
In the following, we briefly review the aspects of boundary conformal field theory relevant to this work. We then present the crossing equations used in our bootstrap analysis, following Refs.~\cite{Liendo:2012hy,McAvity:1995zd,McAvity:1993ue,Billo:2016cpy,Lauria:2017wav,Burkhardt:1987}.

For a $d$-dimensional CFT with a flat boundary at $x_{\perp}=0$, we write $x=(\mathbf{x},x_{\perp})$, where $\mathbf{x}$ denotes the coordinates parallel to the boundary.
The conformal group SO($d+1,1$) is then reduced to the subgroup SO($d,1$) that preserves the boundary. 
This SO($d,1$) subgroup is generated by dilatations, translations and special conformal transformations parallel to the boundary, as well as rotations within the boundary plane.
We focus on the bulk two-point function of scalar primary operators,
\begin{equation}
\label{eq2:2:6}
\langle \mathcal{O}_{1}(x)\mathcal{O}_{2}(y)\rangle=\frac{1}{(2x_{\perp})^{\Delta_1}(2y_{\perp})^{\Delta_2}}\xi^{-\frac{\Delta_{1}+\Delta_{2}}{2}}G_{12}(\xi).
\end{equation}
The function $G_{12}(\xi)$ depends only on the conformal cross-ratio $\xi=\frac{\left(x-y\right)^2}{4x_{\perp}y_{\perp}}$, which is invariant under boundary conformal transformations. 

In a BCFT, the same bulk two-point function can be decomposed in two ways. 
Following Ref.~\cite{Padayasi:2021sik}, we adopt unit two-point-function normalization for all bulk and boundary primary operators entering these conformal-block decompositions.
The first decomposition, known as the bulk channel, is given by the standard bulk operator product expansion (OPE):
\begin{equation}
\mathcal{O}_{1}(x)\mathcal{O}_{2}(y)=\sum_{k}\lambda_{12k}C_{12k}(x-y,\partial_y)\mathcal{O}_{k}(y).
\end{equation}
Here $k$ labels bulk conformal primary operators, $\lambda_{12k}$ are the bulk OPE coefficients, and the differential operators $C_{12k}(x-y,\partial_y)$ are fixed by conformal symmetry \cite{Ferrara:1973yt}. 
Taking the expectation value of the bulk OPE introduces the one-point functions of the bulk primaries. 
The bulk one-point function takes the form
\begin{equation}
\langle \mathcal{O}_{k}(x) \rangle=\frac{a_{k}}{\left(2 x_{\perp} \right)^{\Delta_k}}.
\end{equation}
Here $a_k$ denotes the one-point-function coefficient of the bulk primary $\mathcal O_k$.
Boundary conformal symmetry allows nonzero one-point functions only for scalar bulk primaries.
Another channel, known as the boundary channel, arises from the boundary operator expansion (BOE). 
In a BCFT, a bulk operator approaching the boundary can be expanded in terms of boundary primary operators $\hat{\mathcal{O}}_{n}(\mathbf{x})$:
\begin{equation}
\mathcal{O}_{k}(x)=\sum_{n}b_{kn}\hat{C}_{kn}(x_{\perp},\partial_{\mathbf{x}})\hat{\mathcal{O}}_{n}(\mathbf{x}).
\end{equation}
Here $b_{kn}$ is the coefficient of the $\hat{\mathcal O}_n$ term in the BOE for $\mathcal O_k$.
The differential operators $\hat C_{kn}(x_\perp,\partial_{\mathbf x})$ are fixed by boundary conformal symmetry.
Throughout this work, we generally use hats to denote boundary operators and related quantities, such as their scaling dimensions and conformal blocks.
The BOE coefficient associated with the boundary identity operator $\hat{I}$ is equal to the bulk one-point coefficient, namely $b_{k\hat{I}}=a_{k}$. 
Applying the BOE to both bulk operators expresses the correlator in Eq.~\eqref{eq2:2:6} in terms of boundary two-point functions.

\begin{figure}[tbp]
\centering 
\begin{tikzpicture}[scale=1.0]
\node at (-3,2) {Bulk Channel};
\node at (4,2) {Boundary Channel};
\draw [thick] (-4.5,-1) -- (-1.5,-1);
\node at (-5.5,0) {\Large $\displaystyle\sum_k$};
\draw [thick] (-3,0) -- node [anchor=east] {$\mathcal{O}_k$} (-3,-1) ;
\draw [thick] (-3,0) -- (-4,1) node [anchor=east] {$\mathcal{O}_1$};
\draw [thick] (-3,0) -- (-2,1) node [anchor=west] {$\mathcal{O}_2$};
\node at (1.5,0) {\Large $\displaystyle\sum_n$};
\draw [thick] (2.5,-1) --  (5.5,-1);
\draw [thick] (3,-1) -- (3,1) node [anchor=east] {$\mathcal{O}_1$};
\draw [thick] (5,-1) -- (5,1) node [anchor=west] {$\mathcal{O}_2$};
\node at (4,-0.6) {$\hat{\mathcal{O}}_n$};
\node at (0,0) {\Large $=$};
\end{tikzpicture}
\caption{\label{fig:crossing}Crossing symmetry for a two-point function of bulk scalar primary operators in a BCFT.}
\end{figure}
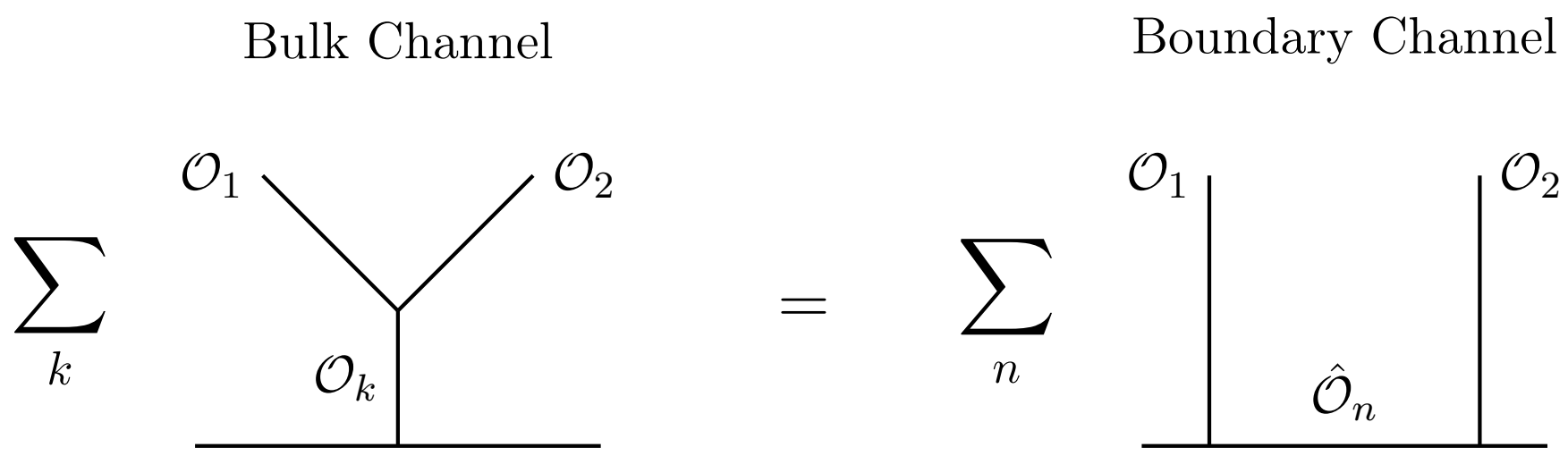

As illustrated in Fig.~\ref{fig:crossing}, the consistency between the two decompositions leads to the crossing equation \cite{Liendo:2012hy},
\begin{equation}
\label{eq2:2:10}
\sum_{k}\lambda_{12k}a_{k}f_{\Delta_k}^{\Delta_{12}}(\xi)=\xi^{\frac{\Delta_{1}+\Delta_{2}}{2}}\sum_{n}b_{1n}b_{2n}\hat{f}_{\hat{\Delta}_n}(\xi),
\end{equation}
where $\Delta_{12}=\Delta_1-\Delta_2$.
We denote the conformal blocks in the bulk and boundary channels by $f_{\Delta}^{\Delta_{12}}(\xi)$ and $\hat f_{\hat{\Delta}}(\xi)$, respectively.
Each conformal block collects contributions from a primary operator and its descendants.
The descendant terms are generated by the corresponding differential operator.
With the conventions used here, the two conformal blocks take the following forms \cite{McAvity:1995zd}:
\begin{align}
f_{\Delta}^{\Delta_{12}}(\xi)&=\xi^{\Delta/2}\ {}_2F_{1}\(\frac{\Delta+\Delta_{12}}{2},\frac{\Delta-\Delta_{12}}{2};\Delta-\frac{d-2}{2};-\xi\),  \label{eq2:2:11}\\
\hat{f}_{\hat{\Delta}}(\xi)&=\xi^{-\hat{\Delta}}\ {}_2F_{1}\(\hat{\Delta},\hat{\Delta}-\frac{d}{2}+1;2\hat{\Delta}-d+2;-\frac{1}{\xi}\). 
\label{eq2:2:12}
\end{align}
For $d=3$, 
the boundary conformal blocks in Eq.~\eqref{eq2:2:12} with $\hat{\Delta}>0$
can be written in elementary form \cite{Gliozzi:2015qsa},\footnote{At $\hat{\Delta}=1/2$, the equivalence between the hypergeometric
and elementary functions depends on how the limit is taken. However, no such operator appears in the boundary spectrum considered in this work.}
\begin{equation}
\label{eq2:2:13}
\hat{f}_{\hat{\Delta}}(\xi)\bigr|_{d=3}=
\xi^{-\hat{\Delta}}\ \frac{4^{\hat{\Delta}-1}\(1+\sqrt{1+1/\xi}\)^{-2(\hat{\Delta}-1)}}{\sqrt{1+1/\xi}},
\end{equation}
while
$\hat f_{\hat{\Delta}=0}(\xi)=1$ for the boundary identity operator.
The crossing equation holds in the region where both channel expansions converge.
Previous bootstrap studies commonly carry out the analysis around $\xi=1$.
We adopt the same expansion point.

In this work, we focus on the two-point functions involving the relevant vector and singlet primaries, denoted by $\phi_a$ and $S$.\footnote{In addition to $\phi_a$ and $S$, bulk primaries with $\Delta<3$ also occur in the antisymmetric and traceless symmetric sectors.
OPEs involving these primaries can contain additional $\mathrm{O}(N)$ irreducible representations.
The current operator in the antisymmetric sector has spin one, and its two-point functions involve structures different from those of scalar correlators.
In principle, we can also consider $\<S T\>$ as only operators $\mathcal{O}\in\mathcal{T}$ contribute in the bulk channel.
}
Here $\phi_a$ is the fundamental vector operator associated with the order parameter, while $S$ is the leading scalar singlet associated with the local energy-density operator.
Specifically, we study $\langle\phi_a\phi_b\rangle$, $\langle\phi_a S\rangle$ and $\langle S S\rangle$ at the normal fixed point. 
We first consider the bulk channel. The bulk fusion rules follow from the tensor product decomposition of $\mathrm{O}(N)$ representations \cite{Grans-Samuelsson:2021uor},
\begin{equation}
\label{eq:fusion:bulk}
\begin{split}
\phi_a\times\phi_b
&\sim
\delta_{ab}\sum_{\mathcal S}\mathcal O
+\sum_{\mathcal T}\mathcal O_{(ab)}
+\sum_{\mathcal A}\mathcal O_{[ab]}
\\
&=\delta_{ab}(I+S+S'+\dots)
 +(T_{(ab)}+T'_{(ab)}+\dots)
 +(A_{[ab]}+A'_{[ab]}+\dots),
\\
\phi_a\times S
&\sim
\sum_{\mathcal V}\mathcal O_{a}
=\phi_a+\phi'_a+\phi''_a+\dots,
\\
S\times S
&\sim
\sum_{\mathcal S}\mathcal O
=I+S+S'+\dots,
\end{split}
\end{equation}
where $\mathcal S$, $\mathcal V$, $\mathcal T$, and $\mathcal A$ label the sets of primary operators transforming in the singlet, vector, traceless-symmetric, and antisymmetric irreducible representations of $\mathrm{O}(N)$.
Only spin-zero bulk operators can have nonzero one-point functions in the presence of a boundary.
In the $\phi_a\times\phi_b$ OPE, Bose symmetry restricts operators in the antisymmetric sector $\mathcal A$ to odd spacetime spin, so this sector does not contribute.
Throughout this work, we denote the nonidentity primaries in the $\mathcal S$, $\mathcal T$, and $\mathcal V$ sectors by $S,S',\ldots$, $T_{(ab)},T'_{(ab)},\ldots$, and $\phi_a,\phi'_a,\ldots$, respectively.
We omit the internal-symmetry indices when no confusion can arise.
The identity operator is denoted by $I$.
The mixed correlator is particularly well suited to the truncated bootstrap because its bulk-channel spectrum has a relatively large gap.\footnote{This is because $\phi^3$ becomes a descendant of $\phi$ in the interacting theory \cite{Rychkov:2015naa}.}

We next turn to the boundary channel, where the normal boundary condition breaks $\mathrm{O}(N)$ to $\mathrm{O}(N-1)$.
We split the index $a$ as $a=(i,N)$, where $i=1,2,\ldots,N-1$ labels the directions transforming under the unbroken subgroup. 
Accordingly, the field $\phi_a$ decomposes into a singlet $\phi_N$ and a vector $\phi_i$ under $\mathrm{O}(N-1)$. 
In the boundary channel, the selection rules are
\begin{equation}
\label{eq:fusion:bdy}
\begin{split}
\phi_{N},S&\sim \hat{I}+\sum_{\hat{\mathcal{S}}}\hat{\mathcal{O}}=\hat{I}+D+\hat{S}+\hat{S}'+\dots,
\\
\phi_{i}&\sim \sum_{\hat{\mathcal{V}}}\hat{\mathcal{O}}_i=t_i+\hat{V}_i+\hat{V}'_i+\dots.
\end{split}
\end{equation}
We use hats for boundary operators, except for $D$ and $t_i$.
Here $\hat{\mathcal{S}}$ denotes $\mathrm{O}(N-1)$ nonidentity singlet primaries and $\hat{\mathcal{V}}$ denotes $\mathrm{O}(N-1)$ vector primaries.
We label them as $D,\hat{S},\hat{S}',\ldots$ and $t_i,\hat{V}_i,\hat{V}'_i,\ldots$, respectively.
The boundary identity operator is denoted separately by $\hat I$.
Following Ref.~\cite{Padayasi:2021sik}, we assume that $D$ and $t_i$ are the lowest nonidentity primary operators in the boundary singlet and vector sectors.
The leading operator $D$ in the boundary singlet sector is the displacement operator, which characterizes the response to local normal displacements of the boundary \cite{Gliozzi:2015qsa,Billo:2016cpy}.
It has the protected scaling dimension $\hat{\Delta}_D=3$.
We denote the Zamolodchikov norm of the displacement operator by $C_D$.
In the unit-normalization convention of Refs.~\cite{Padayasi:2021sik,Toldin:2021kun}, the stress-tensor Ward identity \cite{Cardy:1990xm} gives
\begin{equation}
    \label{eq:2:23}
C_{D}=\(\frac{\Delta_{\mathcal{O}} a_{\mathcal{O}}}{4\pi b_{\mathcal{O}D}}\)^2.
\end{equation}
Here, $\mathcal O$ is a nonidentity bulk scalar with a nonzero one-point coefficient $a_{\mathcal O}$, and $b_{\mathcal O D}$ is its BOE coefficient of the displacement operator.
The value of $C_D$ does not depend on the choice of $\mathcal O$.
We will later evaluate $C_D$ from this relation for $\mathcal O=\phi_N$ and for $\mathcal O=S$.
Agreement between the resulting estimates provides a self-consistency check of the extracted BCFT data.
The leading operator $t_i$ in the boundary vector sector is known as the tilt operator.
It couples to an infinitesimal rotation of the boundary magnetic field. 
Its scaling dimension is also protected, $\hat{\Delta}_t=2$ \cite{Bray:1977fvl,Padayasi:2021sik}. 
For a bulk $\mathrm{O}(N)$ vector operator $\mathcal{O}_a$, the tilt Ward identity gives \cite{Padayasi:2021sik}
\begin{equation}
\label{eq:2:24}
C_{t}=\(\frac{a_{\mathcal{O}}}{4\pi b_{\mathcal{O} t}}\)^2,
\end{equation}
where $a_{\mathcal O}$ is the longitudinal one-point coefficient, and $b_{\mathcal O t}$ is the transverse BOE coefficient of the tilt operator.
This quantity is directly related to the RG parameter $\alpha$ \cite{Metlitski:2020cqy,Padayasi:2021sik},
\begin{equation}
\alpha=\frac{\pi}{2}C_t-\frac{N-2}{2\pi}.
\end{equation}

With the fusion rules in Eqs.~\eqref{eq:fusion:bulk}--\eqref{eq:fusion:bdy}, the boundary bootstrap equations take the following explicit forms. Below, we write the contributions from the bulk and boundary identity operators separately.
For $\langle\phi_a\phi_b\rangle$, there are two independent crossing equations when $N\ge 2$,
\begin{align}
\label{eq2:19}
1+\sum_{\substack{k\in \mathcal{S}\\ k\neq I}}\lambda_{\phi\phi k}a_k f_{\Delta_k}^{\Delta_{12}}(\xi)+\sum_{l\in \mathcal{T}}\lambda_{\phi\phi l}a_l f_{\Delta_l}^{\Delta_{12}}(\xi)=\xi^{\Delta_{\phi}}\(a_{\phi}^2+\sum_{n\in\hat{\mathcal{S}}}b_{\phi n}^2  \hat{f}_{\hat{\Delta}_n}(\xi)\),
\\
1+\sum_{\substack{k\in \mathcal{S}\\ k\neq I}}\lambda_{\phi\phi k}a_k f_{\Delta_k}^{\Delta_{12}}(\xi)-\frac{1}{N-1}\sum_{l\in \mathcal{T}}\lambda_{\phi\phi l}a_l f_{\Delta_l}^{\Delta_{12}}(\xi)=\xi^{\Delta_{\phi}}\(\sum_{m\in\hat{\mathcal{V}}}b_{\phi m}^2  \hat{f}_{\hat{\Delta}_m}(\xi)\),
\label{eq2:20}
\end{align}
which correspond to the $\mathrm{O}(N-1)$ singlets $\langle\phi_{N}\phi_{N}\rangle$ and $\frac{1}{N-1}\sum_i \langle\phi_i\phi_i\rangle$, respectively. 
The rank-two symmetric traceless primaries have components labeled by two $\mathrm{O}(N)$ indices.
For each $l\in\mathcal{T}$, we define $a_l$ as the one-point amplitude of its $NN$ component.
Residual $\mathrm{O}(N-1)$ symmetry requires all transverse diagonal one-point amplitudes to be equal.
Tracelessness then fixes each of them to $-a_l/(N-1)$, giving the corresponding factor in Eq.~\eqref{eq2:20}.
For the mixed correlator $\langle\phi_a S\rangle$, residual $\mathrm{O}(N-1)$ symmetry implies $\langle\phi_i S\rangle=0$.
The singlet component $\langle\phi_N S\rangle$ gives the single nontrivial crossing equation,
\begin{equation}
\label{eq2:21}
\sum_{k\in \mathcal{V}}\lambda_{\phi S k}a_k f_{\Delta_k}^{\Delta_{12}}(\xi)=\xi^{\frac{\Delta_{\phi}+\Delta_{S}}{2}}\(a_{\phi}a_{S}+\sum_{n\in\hat{\mathcal{S}}}b_{\phi n} b_{S n}  \hat{f}_{\hat{\Delta}_n}(\xi)\).
\end{equation}
For $k\in\mathcal V$, the coefficient $a_k$ denotes the one-point amplitude of its $N$-th component.
This equation is homogeneous in the expansion coefficients and therefore determines only their ratios.
In the numerical analysis, we divide Eq.~\eqref{eq2:21} by $\lambda_{\phi S\phi}a_\phi$, thereby setting the coefficient of the $\phi$ block in the bulk channel to unity.
Accordingly, the coefficient of the boundary-channel identity contribution becomes $a_S/\lambda_{\phi S\phi}$.
For $\langle S S\rangle$, the crossing equation is
\begin{equation}
\label{eq2:22}
1+\sum_{\substack{k\in \mathcal{S}\\ k\neq I}}\lambda_{SSk}a_{k} f_{\Delta_k}^{\Delta_{12}}(\xi)=\xi^{\Delta_S}\(a_S^2+\sum_{n\in\hat{\mathcal{S}}}b_{Sn}^2 \hat{f}_{\hat{\Delta}_n}(\xi)\).
\end{equation}
These crossing equations provide the starting point for our bootstrap analysis. 
The input values $(\Delta_{\phi},\Delta_{S},\Delta_{S'},\Delta_{T})$ are taken from recent studies and summarized in Table~\ref{ta:1} of Appendix~\ref{app:tech}.
With these inputs fixed, we analyze each correlator separately within the truncated-bootstrap framework and numerically solve the corresponding truncated crossing equations.

\subsection{Truncated boundary bootstrap with $\eta$-minimization}
\label{sec:2:3}
We follow the truncated boundary bootstrap approach with $\eta$-minimization introduced in Refs.~\cite{Li:2017ukc,Hu:2025yrs}.
Operators with higher scaling dimensions are expected to make subleading contributions to the conformal block expansion.
Crossing symmetry can therefore be approximately satisfied using a finite set of low-lying operators.
The objective function $\eta$ quantifies the residual violation of the resulting truncated crossing equations.
The main idea is to obtain approximate values of the BCFT data by minimizing $\eta$.

We first define the objective function $\eta$ using the boundary crossing equation. 
In the truncated version of Eq.~\eqref{eq2:2:10}, 
we retain $n_{\mathrm{bulk}}$ operators in each $\mathrm{O}(N)$ representation sector of the bulk channel and $n_{\mathrm{bdy}}$ operators in each $\mathrm{O}(N-1)$ representation sector of the boundary channel.\footnote{Our counting differs slightly from that of Ref.~\cite{Hu:2025yrs}. We exclude both the bulk and boundary identity operators from the truncation counts.}
Taylor expanding the truncated crossing equation around $\xi=1$ yields an infinite set of derivative constraints.
For a truncation with $\Lambda$ unknown parameters, we select $\Lambda$ constraints starting with those at the lowest derivative orders.
We also refer to $\Lambda$ as the truncation order.
Based on the spectral truncation, we introduce the $\eta$ function as a sum of squares:
\begin{equation}
\eta(\{\lambda_{12k}a_{k},\Delta_{k}\},\{b_{1n}b_{2n},\hat{\Delta}_{n}\})=\sum_{j} \sum_{m=0}^{M_j} L_{m,j}^{2},
\end{equation}
where $\{L_{m,j}\}$ is a set of residuals
\begin{equation}
\label{eq:residual}
L_{m,j}=\left| \partial_{\xi}^{m}\(
\mathrm{truncated\ crossing\ equation}
\)_{j}\right|_{\xi=1}.
\end{equation}
Here $j$ labels the crossing equations, and $M_j$ is the highest derivative order used for the corresponding equation. 
The crossing equations are given in Eqs.~\eqref{eq2:19}--\eqref{eq2:20} for $\<\phi_a \phi_b\>$, Eq.~\eqref{eq2:21} for $\<\phi_a S\>$, and Eq.~\eqref{eq2:22} for $\<SS\>$.
For $\langle\phi_a\phi_b\rangle$ at $N\ge 2$, we have $j=1,2$ and $M_1=M_2=\Lambda/2-1$.
For each of $\langle\phi_a S\rangle$ and $\langle SS\rangle$, we have $j=1$ and $M_1=\Lambda-1$.

We then explain how to solve the truncated crossing constraints $\{L_{m,j}=0\}$ by minimizing the $\eta$ function.
In general, the $\eta$ minimization can lead to different local minima. 
We focus on those satisfying $\eta=0$. 
As $\eta$ is defined as the squared sum of $L_{m,j}$, 
the condition $\eta=0$ is equivalent to requiring all residuals to vanish, i.e., $\{L_{m,j}=0\}$.
Furthermore, a local minimum with $\eta=0$ is also a global minimum. 
For numerical implementation, we carry out the $\eta$-minimization using \texttt{FindMinimum} in \texttt{Mathematica}. 
In practice, we treat $\eta$ as zero when $\eta<10^{-500}$. 
The results of \texttt{FindMinimum} depend strongly on the starting point, 
so we need to provide a suitable initial guess to reach a zero of $\eta$, which will be discussed later.

We further explain why we use $\eta$-minimization instead of solving the system $\{L_{m,j}=0\}$ directly.
Solving $\{L_{m,j}=0\}$ directly is difficult because the conformal blocks contain transcendental functions.
One approach is to approximate the bulk conformal blocks by rational functions \cite{Lauria:2017wav,Penedones:2015aga}. 
These approximations are constructed using recursion relations derived from the singularity structure of the blocks.
Clearing the denominators then converts the approximate truncated crossing equations into a polynomial system.
For more accurate approximations, we need to perform more recursion steps and increase the degrees of polynomials. 
As the truncation and polynomial orders increase, the nonlinear algebraic system typically admits more solutions and becomes increasingly difficult to solve. 
In particular, few of them are physically acceptable, as explained later. 
Following Refs.~\cite{Li:2017ukc,Hu:2025yrs}, we transform the task of solving $\{L_{m,j}=0\}$ into a nonlinear least-squares problem in terms of $\eta$. 
This local minimization formulation reduces the computational cost by focusing the search on physical parameter regions. 
In this way, we avoid the computationally demanding part of finding all truncated crossing solutions, whose number increases rapidly at higher orders.  

Our bootstrap procedure includes four main steps: 
finding solutions at low truncation orders and selecting the physically acceptable ones, 
gradually extending the calculation to higher truncation orders, 
extrapolating the high-order solutions to the $\Lambda\to\infty$ limit, 
and estimating the uncertainties in the $\Lambda\rightarrow \infty$ results. 
The procedure is broadly similar to that described in Ref.~\cite{Hu:2025yrs}.
Below is an explanation of the first three steps.
\begin{itemize}
\item
At low truncation orders, we reduce $\{L_{m,j}=0\}$ to a polynomial system.
Solving the polynomial system may yield multiple solutions, but not every solution is physically acceptable.
We select those with positive real scaling dimensions and the assumed spectral ordering.
Each newly introduced operator must lie above the fixed input operators in its sector.
In the bulk, this gives $\Delta_{\mathcal{O}\in\mathcal{S}}>\Delta_{S'}$, $\Delta_{\mathcal{O}\in\mathcal{V}}>\Delta_{\phi}$ and $\Delta_{\mathcal{O}\in\mathcal{T}}>\Delta_T$.
We further require $\Delta_{\mathcal{O}\in\mathcal{V}}>3$, since $\phi_a$ is assumed to be the only relevant bulk vector scalar.\footnote{Although we do not impose the condition $\Delta_{\mathcal{O}\in\mathcal{T}}>3$ on newly introduced operators $\mathcal{O}\in\mathcal{T}$, we find that their scaling dimensions exceed 3 in all numerical solutions that satisfy the other selection criteria.}
For boundary operators, the corresponding bounds are $\hat{\Delta}_{\hat{\mathcal{O}}\in\hat{\mathcal{S}}}>3$ and $\hat{\Delta}_{\hat{\mathcal{O}}\in\hat{\mathcal{V}}}>2$.
\item
When increasing the truncation order, we use the solution from the preceding truncation order to set most of the starting values in \texttt{FindMinimum}. 
For newly introduced operators, initial guesses for the scaling dimensions are inferred from features of the lower-lying spectrum, such as the level spacings. 
For most expansion coefficients, we can choose $1$ as the initial value in \texttt{FindMinimum}.\footnote{For some higher-order truncations in the $\langle SS\rangle$ case, the first few coefficients should be initialized with the corresponding values from the solution at the preceding truncation.}
For each truncation choice, our search finds at most one solution satisfying the stated spectral assumptions.
\item
Repeating the preceding step yields solutions at successively higher truncation orders.
We then extrapolate them to the infinite-truncation limit.
We do so by fitting their dependence on $1/\Lambda$ to the power-law form $p+q(1/\Lambda)^c$, where $p$, $q$, and $c$ are fit parameters.
For some quantities, fixing $c=1$ gives a more stable extrapolation.
This choice corresponds to a linear fit in $1/\Lambda$. 

\end{itemize}
In the last step, we include two sources of uncertainty: input-induced variations and sensitivity to the extrapolation procedure.
\begin{itemize}
\item
To estimate the input-induced uncertainty, we repeat the calculations at the endpoints of the input ranges. This procedure provides a reasonable estimate when the input ranges are sufficiently small and the results vary monotonically within these ranges. In such cases, the endpoint calculations capture the largest deviations from the central estimates. In practice, we find that many low-lying quantities show this behavior.
\item
The extrapolation uncertainty is estimated from fits to subsets of the data. 
Each subset is formed by removing one or two points from either end and retaining at least three points.
The extrapolation uncertainty generally decreases as the truncation order $\Lambda$ increases.
\end{itemize}
We expect our results to be only weakly affected by the extrapolation.
For all results presented in Sec.~\ref{sec:3}, the extrapolation uncertainty is smaller than the corresponding input-induced uncertainty.
Since the input uncertainty is the main source of error, it is unlikely to improve the precision substantially by further increasing the truncation order. 

\begin{figure}[tbp]
\includegraphics[width=0.48\textwidth]{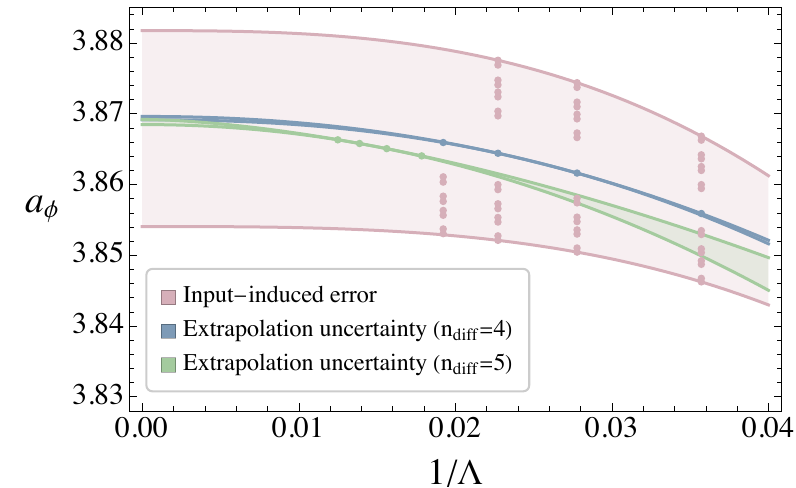}
\hfill
\includegraphics[width=0.48\textwidth]{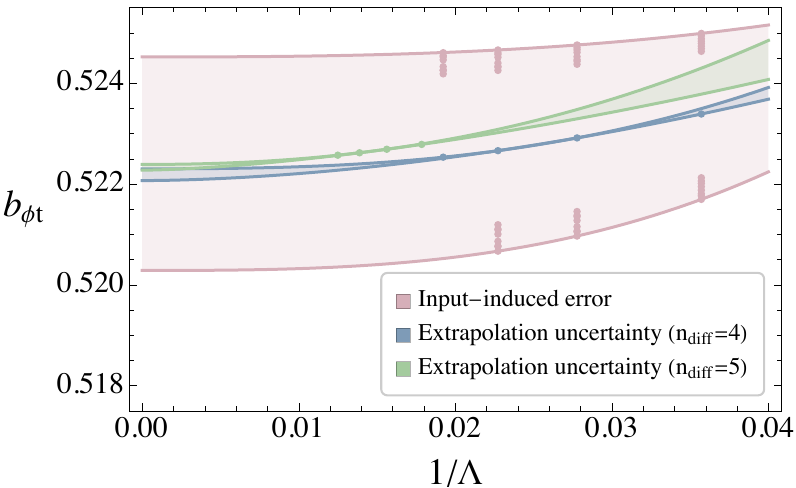}
\caption{\label{fig:2}Power-law fits of $a_{\phi}$ and $b_{\phi t}$ from the $N=6$ $\langle\phi_a\phi_b\rangle$ bootstrap. 
The fits use the form $p+q(1/\Lambda)^c$, where $p$, $q$, and $c$ are fit parameters.
Here a truncation family is a sequence of truncation schemes with fixed $n_{\mathrm{diff}}\equiv n_{\mathrm{bulk}}-n_{\mathrm{bdy}}$. 
Blue points use the $n_{\mathrm{diff}}=4$ family with $(n_{\mathrm{bulk}},n_{\mathrm{bdy}})=\{(6,2),(7,3),(8,4),(9,5)\}$, while green points use the $n_{\mathrm{diff}}=5$ family with $(n_{\mathrm{bulk}},n_{\mathrm{bdy}})=\{(10,5),(11,6),(12,7),(13,8)\}$. 
The blue and green bands show the extrapolation uncertainty estimated by varying the range of points included in the fits.
The purple region represents the uncertainty obtained by varying the input values, using the same truncation choices as the blue points whenever possible.
When the input $\Delta_T$ is set to its lower bound, no $(9,5)$ result is obtained and the fit uses only the first three truncations.
}
\end{figure}

The extrapolation results, together with their uncertainty estimates, are presented in Sec.~\ref{sec:3}. 
We discuss the truncation choices used for these extrapolations later. 
In Fig.~\ref{fig:2}, we use the concrete examples of $a_{\phi}$ and $b_{\phi t}$ obtained from the $N=6$ $\langle\phi_a\phi_b\rangle$ bootstrap to illustrate the $\Lambda\to\infty$ extrapolation and uncertainty estimation.
These examples also show that some truncation schemes may fail to yield reliable solutions for certain input values, motivating our discussion of truncation choices.

In general, the choice of truncation can affect whether we find a physical solution.
Our search often finds physical solutions at several successive truncation orders when we increase both $n_{\mathrm{bulk}}$ and $n_{\mathrm{bdy}}$ by one.
We refer to such a sequence of truncation choices as a truncation family, characterized by a fixed value of
\begin{equation}
n_{\mathrm{diff}}
\equiv n_{\mathrm{bulk}}-n_{\mathrm{bdy}}.
\end{equation}
When moving to a higher truncation order, we first search within the same family.
However, it is usually impossible to use a single truncation family all the way to high truncation orders.
The resulting change in truncation families can be seen in Fig.~\ref{fig:2}.
In practice, we begin with a truncation family of a relatively small $n_{\mathrm{diff}}>0$.
If the search fails within the current family as the truncation order increases, we generally increase $n_{\mathrm{diff}}$ by one and continue the search in the new family.\footnote{For a given small $n_{\mathrm{bulk}}$, our search can find physically acceptable solutions for several values of $n_{\mathrm{bdy}}$, corresponding to different truncation families.
For a given larger $n_{\mathrm{bulk}}$, it may find solutions in only one family.
Similar behavior is observed when $n_{\mathrm{bdy}}$ is fixed and $n_{\mathrm{bulk}}$ is varied.}

The following discussion of truncation choices uses the case shown in Fig.~\ref{fig:2} as an example.
This example concerns $a_\phi$ and $b_{\phi t}$ in the $N=6$ $\langle\phi_a\phi_b\rangle$ bootstrap.
Similar considerations apply to other quantities and correlators.
For the central input values, we obtain solutions in the $n_{\mathrm{diff}}=4$ truncation family up to the truncation choice $(n_{\mathrm{bulk}},n_{\mathrm{bdy}})=(9,5)$. 
With an initial guess based on the $(9,5)$ solution, we find no solution with $\eta=0$ at $(10,6)$.
However, the computation can be continued by switching to the $n_{\mathrm{diff}}=5$ truncation family, $\{(10,5),(11,6),\ldots\}$.
Although the results from the two truncation families do not fall on a single smooth curve, the difference between their extrapolated values is small.
This indicates that the choice of truncation family has little influence on the extrapolation results.\footnote{Among the cases considered, the scaling dimensions of boundary operators extracted from the $N=6$ $\langle\phi_a \phi_b\rangle$ correlator show a somewhat stronger dependence on the truncation family.
The $n_{\mathrm{diff}}=4$ and $n_{\mathrm{diff}}=5$ families give the central estimates $(\hat{\Delta}_{\hat S},\hat{\Delta}_{\hat S'})\approx (5.3,7)$ and $(5.6,8)$, respectively.
This may be related to the relatively small values of $n_{\mathrm{bdy}}$ in the truncations studied so far.}

\begin{figure}[tbp]
\centering 
\includegraphics[width=0.7\textwidth]{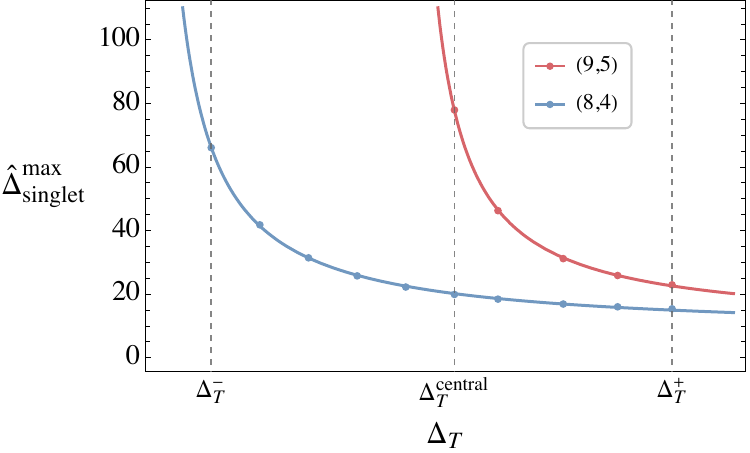}
\caption{\label{fig:23}The dependence of the largest scaling dimension in the boundary singlet sector, denoted by $\hat{\Delta}_{\mathrm{singlet}}^{\mathrm{max}}$, on $\Delta_{T}$ from the $N=6$ $\langle\phi_a\phi_b\rangle$ bootstrap. 
The blue and red points correspond to the truncation choices $(n_{\mathrm{bulk}},n_{\mathrm{bdy}})=(8,4)$ and $(9,5)$, respectively.
The gray vertical dashed lines mark the lower ($\Delta_{T}^{-}=1.1306$), central ($\Delta_{T}^{\text{central}}=1.1401$) and upper ($\Delta_{T}^{+}=1.1486$) input values of $\Delta_T$.}
\end{figure}

In addition, the truncation choice at which the analysis must switch to an alternative family may depend on the input values.
In the $N=6$ $\langle\phi_a\phi_b\rangle$ analysis, we may already need to switch from the $n_{\mathrm{diff}}=4$ family to the $n_{\mathrm{diff}}=5$ family after the $(n_{\mathrm{bulk}},n_{\mathrm{bdy}})=(8,4)$ truncation, as shown in Fig.~\ref{fig:2}.
Whether our search finds a physical solution depends primarily on $\Delta_{T}$.
Starting from the solution obtained with the central input values, we track the evolution of the solutions as we vary $\Delta_T$ toward either its lower bound $\Delta_T^{-}$ or its upper bound $\Delta_T^{+}$.
During this deformation, all quantities change slightly except for the largest retained scaling dimension in the boundary singlet sector.
We denote this quantity by $\hat{\Delta}_{\mathrm{singlet}}^{\mathrm{max}}$. 
Fig.~\ref{fig:23} shows the dependence of $\hat{\Delta}_{\mathrm{singlet}}^{\mathrm{max}}$ on $\Delta_{T}$ under the truncation choices $(8,4)$ and $(9,5)$. 
For the $(9,5)$ truncation, the scaling dimension $\hat{\Delta}_{\mathrm{singlet}}^{\mathrm{max}}$ grows rapidly as $\Delta_T$ is varied toward $\Delta_T^{-}$.
This indicates that the truncation choice $(9,5)$ is no longer applicable near $\Delta_T^{-}$, and explains why we cannot find a solution there.
In comparison, the results obtained with the truncation choice $(8,4)$ remain relatively stable over the same range of input values. 
This comparison suggests that the sensitivity to inputs becomes more pronounced at higher truncation orders.

Nevertheless, when numerical solutions can still be found, the unusually large values of $\hat{\Delta}_{\mathrm{singlet}}^{\mathrm{max}}$ are not accompanied by appreciable changes in the low-lying data.
Thus, such anomalous high-lying behavior may not undermine the reliability of the low-lying estimates.
Within a truncation family, most low-lying quantities show good convergence as $\Lambda$ increases.
We therefore perform the $\Lambda\to\infty$ extrapolations using results from one truncation family at a time. 
Table~\ref{ta:truncation} in Appendix~\ref{app:tech} summarizes the truncation choices used for the $\Lambda\to\infty$ extrapolations of the numerical results presented in Sec.~\ref{sec:3}.

\FloatBarrier

\section{Bootstrap results}
\label{sec:3}
In this section, we report our numerical predictions for the BCFT data, including some scaling dimensions, coefficients in the conformal block decompositions, and other universal quantities such as $\a$, $C_t$ and $C_D$. 
These data are extracted from separate bootstrap analyses of the two-point functions involving the relevant bulk $\mathrm{O}(N)$ vector and singlet operators, as discussed in Sec.~\ref{sec:2}. 
The following three subsections present the results from the $\langle\phi_a\phi_b\rangle$, $\langle\phi_a S\rangle$, and $\langle SS\rangle$ bootstraps, respectively.
Below, we denote these correlators simply by $\langle \phi \phi\rangle$, $\langle \phi S\rangle$ and $\langle S S\rangle$.

\subsection{$\langle\phi \phi\rangle$}
\label{sec:3:1}
\FloatBarrier
We start with the $\langle\phi\phi\rangle$ bootstrap, whose results provide useful inputs for the subsequent analyses. 
Ref.~\cite{Hu:2025yrs} reported results for $N=1,2,3,4,5$.
Previous estimates suggest $N_c\simeq5$ \cite{Padayasi:2021sik,Hu:2025yrs}, making the case $N=6$ particularly useful for testing this prediction.
We therefore extend the calculation to $N=6$, using truncations up to $\Lambda=52$ for our main extrapolation.
This analysis yields estimates of several BCFT quantities such as $a_{\phi}$, $b_{\phi t}$, $b_{\phi D}$, $\lambda_{\phi\phi S}a_S$ and $\Delta_{T'}$.

\begin{table}[tbp]
\small
\centering
\begin{tabular}{c|ccccc}
\noalign{\hrule height 1.0pt}
 &$a_{\phi}$&$b_{\phi t}$&$b_{\phi D}$& $\lambda_{\phi\phi S}a_{S}$&$\Delta_{T'}$ \\
\hline
This work & $3.869^{+0.013}_{-0.015}$ & $0.5222_{-0.0020}^{+0.0024}$ & $0.236_{-0.008}^{+0.006}\,{}^{*}$&$1.967_{-0.032}^{+0.026}$&$ 3.31_{-0.11}^{+0.08}\,{}^{*}$\\
\hline
Large-$N$ expansion&3.714&0.5360&0.212&1.748&3.36\\
\noalign{\hrule height 1.0pt}
\end{tabular}
\caption{\label{ta:2}
Main $\langle\phi\phi\rangle$ bootstrap results at $N=6$ and the corresponding estimates from the large-$N$ expansion.
Data marked with an asterisk are obtained from linear fits, while the others are obtained from power-law fits. 
The large-$N$ comparison uses the leading-order expression for $b_{\phi D}$ and expressions up to order $1/N$ for the other quantities.
}
\end{table}

In Table~\ref{ta:2}, we summarize the main bootstrap results for $(a_\phi$, $b_{\phi t}$, $b_{\phi D}$, $\lambda_{\phi\phi S}a_S,\Delta_{T'})$. 
The table also lists the corresponding values obtained by evaluating the large-$N$ expressions at $N=6$.\footnote{For detailed discussions of the large-$N$ expansion, see Refs.~\cite{Metlitski:2020cqy,Padayasi:2021sik,Henriksson:2022rnm}. All quantities here except $b_{\phi D}$ are evaluated at $N=6$ using their large-$N$ expansions up to order $1/N$. For $b_{\phi D}$, only the leading-order result is available in Ref.~\cite{Padayasi:2021sik}. The expressions used here are given in Eqs.~\eqref{eq:a:7}, \eqref{eq:a:10}, \eqref{eq:a:13}--\eqref{eq:a:15}.}
The bootstrap results and the large-$N$ estimates are broadly comparable, especially given that the value $N=6$ considered here is not very large. 
Additional bootstrap results for $(\Delta_{S''},\hat{\Delta}_{\hat S},\hat{\Delta}_{\hat S'},\lambda_{\phi\phi S'}a_{S'},\lambda_{\phi\phi S''}a_{S''},\lambda_{\phi\phi T}a_T,b_{\phi\hat S})$ are listed in Table~\ref{ta:3} of Appendix~\ref{app:c:1}.

As a further comparison for the main results, we also evaluate the available $\epsilon=d-2$ and $\epsilon=4-d$ expansions at $d=3$ \cite{Padayasi:2021sik,Henriksson:2022rnm,Dey:2020lwp,Giombi:2025pxx}.\footnote{For coefficients given in squared form, we take the positive square root without re-expanding.}
For $a_\phi$, the $\epsilon=d-2$ expansion gives $a_\phi \approx 3.488$ at $N=6$, about $10\%$ below our result.
Applying the $[0,1]$ Pad\'e prescription of Ref.~\cite{Giombi:2025pxx} to the next-to-leading-order $\epsilon=4-d$ series for $\sqrt{\epsilon} a_\phi$ gives $a_\phi\approx3.820$ at $N=6$, about $1.3\%$ below our central estimate.
We also construct a two-sided $[2,2]$ Padé approximant for $a_\phi^2$ by matching the available expansions around $d=2$ and $d=4$.
Taking the positive square root gives $a_\phi\approx3.943$ at $N=6$, about $1.9\%$ above our central estimate.
For $(b_{\phi t},b_{\phi D},\Delta_{T'})$, the $\epsilon=4-d$ expansion through next-to-leading order gives $(0.523, 0.242, 3.571)$ at $N=6$.
Applying a two-sided $[2,2]$ Pad\'e approximant to $b_{\phi D}^2$ and then taking the positive square root gives $b_{\phi D}\approx0.227$.
This value lies close to the lower end of our bootstrap range.
Following the Pad\'e prescription of Ref.~\cite{Henriksson:2025hwi}, we apply a $[3,3]$ approximant to the six-loop $\epsilon=4-d$ expansion for $\Delta_{T'}$ \cite{Henriksson:2022rnm}. 
At $N=6$, we obtain $\Delta_{T'}\approx 3.38$, closer to our bootstrap result than the next-to-leading-order estimate.
For $\lambda_{\phi\phi S}a_S$, neither the direct estimates from the two $\epsilon$-expansions nor the Pad\'e estimates agree closely with our bootstrap result.
For some quantities, the available $\epsilon$-expansions are known only to very low orders.
This limits the accuracy of both direct and Pad\'e estimates at $\epsilon=1$.
Overall, these estimates from the $\epsilon$-expansions are broadly comparable with our bootstrap results.

The coefficients in Table~\ref{ta:2} allow us to compute other universal quantities, such as $\a$, $C_t$ and $C_D$.
Substituting the $N=6$ estimates of $(a_\phi,b_{\phi t})$ into Eq.~\eqref{eq:2:4} gives
\begin{equation}
\label{eq:3:2}
\a|_{N=6}=\frac{a_{\phi}^2}{32\pi b_{\phi t}^2}-\frac{N-2}{2\pi}=-0.090_{-0.010}^{+0.007}.
\end{equation}
While Refs.~\cite{Padayasi:2021sik,Hu:2025yrs} did not explicitly compute the $N=6$ case, the trends shown there suggest that $\alpha<0$ at $N=6$. Our result is consistent with this expectation. A more systematic study of the $N$-dependence and a more refined estimate of $N_c$ will be discussed later. 
Using the same values of $(a_\phi,b_{\phi t})|_{N=6}$, we estimate the Zamolodchikov norm $C_t$ from Eq.~\eqref{eq:2:24} with $\mathcal{O}=\phi$:
\begin{equation}
C_t|_{N=6,\mathcal{O}=\phi}=\(\frac{a_\phi}{4\pi b_{\phi t}}\)^2=0.348_{-0.007}^{+0.005}.
\end{equation}
Substituting the estimates for $(a_{\phi},b_{\phi D})|_{N=6}$ into Eq.~\eqref{eq:2:23} with $\mathcal{O}=\phi$ gives
\begin{equation}
\label{eq:3:3}
C_D|_{N=6,\mathcal{O}=\phi}=\(\frac{\Delta_{\phi}a_\phi}{4\pi b_{\phi D}}\)^2=0.451^{+0.028}_{-0.017}.
\end{equation}
The estimate of $C_D$ will be compared with that obtained using $\mathcal{O}=S$ in Sec.~\ref{sec:3:2}.

\begin{figure}[tbp]
\centering 
\includegraphics[width=0.68\textwidth]{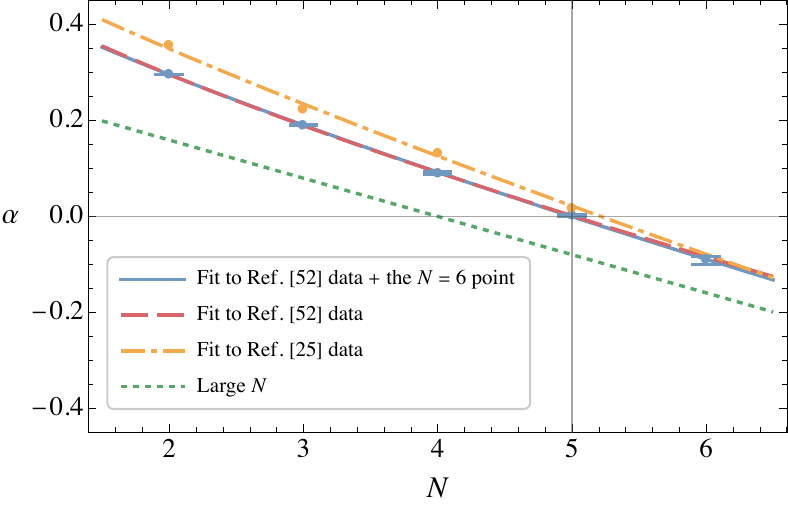}
\caption{\label{fig:3}Estimates of $\alpha$ and power-law fits to the central values. 
The blue points show the results for $N=2,3,4,5$ from Ref.~\cite{Hu:2025yrs} and our result for $N=6$; the error bars indicate the quoted uncertainties.
The solid blue and dashed red lines are power-law fits including and excluding the $N=6$ point, respectively.
The two fits nearly overlap and are difficult to distinguish at the scale of the plot.
The orange points show the results obtained at a relatively low truncation order in Ref.~\cite{Padayasi:2021sik}, and the dash-dotted line is a power-law fit to its results at $N=2,3,4,5,10,20$.
The green dashed line shows the large-$N$ prediction for $\alpha$ from Eq.~\eqref{eq:a:16}.
The gray vertical and horizontal lines mark $N=5$ and $\a=0$, respectively.}
\end{figure}

We now examine the $N$-dependence of $\a$ and estimate the critical value $N_c$. 
Assuming that $\alpha(N)$ varies monotonically, we determine $N_c$ by solving $\alpha(N_c)=0$. 
Using the data for $N=2,3,4,5,6$,\footnote{For $N=2,3,4,5,6$, the values used in the fit are $\a=0.2957(6)$, $0.1903(7)$, $0.0906(35)$, $0.002(2)$, $-0.090_{-0.010}^{+0.007}$, respectively. The first four values are taken from Ref.~\cite{Hu:2025yrs}, while the $N=6$ value is obtained in this work.}
we fit $\alpha(N)$ to the form $\alpha=p+qN^r$, so the condition $\alpha(N_c)=0$ gives $N_c=(-p/q)^{1/r}$, where $p$, $q$, and $r$ are fit parameters.
With the fitted parameters, we find
\begin{equation}
\label{eq:nc}
N_c=4.996_{-0.051}^{+0.045}.
\end{equation}
We estimate the sensitivity to the input ranges by refitting after shifting every data point to the upper endpoint of its quoted range, and then to the lower endpoint.
Our estimate is consistent with $N_c=5.016(28)$ from the power-law fit to the $N=2,3,4,5$ data in Ref.~\cite{Hu:2025yrs}. 
That study also used the truncated bootstrap with $\eta$-minimization.
The fit of Ref.~\cite{Hu:2025yrs} predicts $\alpha(6)\approx-0.084$, which lies within the uncertainty range of our bootstrap estimate $-0.090_{-0.010}^{+0.007}$ in Eq.~\eqref{eq:3:2}.
This agreement, as clearly shown in Fig.~\ref{fig:3}, lends support to the use of the power-law form in estimating $N_c$.
A power-law refit of the $N=2,3,4,5,10,20$ results of Ref.~\cite{Padayasi:2021sik}, obtained at a relatively low truncation order, gives $N_c=5.215$.
This value is outside our quoted sensitivity range but still near $5$.
Fig.~\ref{fig:3} shows the fits for $\alpha$ discussed above.
All these fitted curves cross the $\alpha=0$ line near $N=5$.
As a complementary check, we also fit $\alpha(N)$ by incorporating the analytic large-$N$ behavior summarized in Sec.~\ref{app:a}.
Specifically, we use the constrained form $\alpha(N)=-N/(4\pi)+1/\pi+qN^r$, with $q$ and $r$ as fit parameters. 
The first two terms are fixed by the large-$N$ expansion of $\alpha$ in Eq.~\eqref{eq:a:16}, while the remaining finite-$N$ correction is fitted by $qN^r$.
Using the $N=2,3,4,5,6$ data, the alternative fit with large-$N$ analytic input gives 
\begin{equation}
    N_c=5.003^{+0.046}_{-0.056},
\end{equation}
consistent with the estimate obtained in Eq.~\eqref{eq:nc}.
Taken together, these results support the robust conclusion that $N_c\approx 5$.

\FloatBarrier
\subsection{$\langle\phi S\rangle$}
\label{sec:3:2}
The mixed-correlator bootstrap yields new BCFT data over a broad range of $N$. When combined with the $\langle\phi\phi\rangle$ results, it allows us to disentangle several bulk OPE and BOE coefficients for $N=2,3,4,5,6$. 
While Ref.~\cite{Hu:2025yrs} studied the mixed correlator at $N=1$, here we extend the analysis to $N=2,3,4,5,6,8,10,12$.

\begin{figure}[tbp]
\centering 
\includegraphics[width=0.68\textwidth]{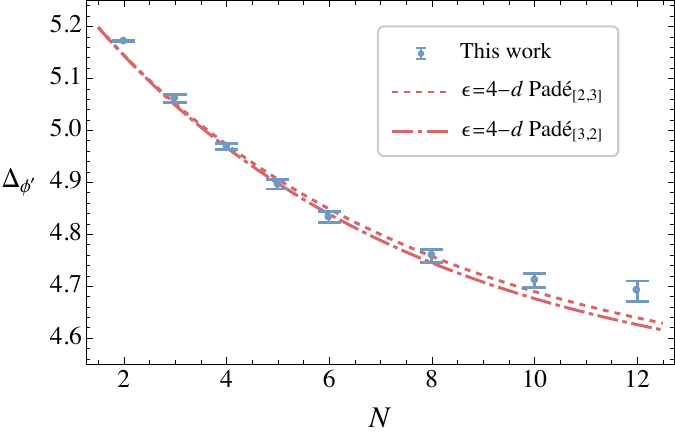}
\caption{\label{fig:Deltaphi:e}The bootstrap estimates of $\Delta_{\phi'}$ from the $\<\phi S\>$ bootstrap, compared with the $[2,3]$ and $[3,2]$ Pad\'e resummations applied to the five-loop $\epsilon=4-d$ series \cite{Henriksson:2025hwi}.}
\end{figure}

Below, we present the extrapolation results obtained from power-law fits for $N=2,3,4,5,6,8,10,12$.
As representative examples of the fits, we show the results for $\Delta_{\phi'}$ and $\hat{\Delta}_{\hat{S}}$ in Figs.~\ref{fig:Deltaphi2}--\ref{fig:deltas} of Appendix~\ref{app:tech}.
The fits give
\begin{equation}
\begin{split}
\label{eq:phiS:Deltaphi2}
\Delta_{\phi'} ={}& 5.1725(13),5.061(8),4.969(6),4.896(9),
4.833(11),\\
& 4.759_{-0.013}^{+0.011},4.712_{-0.015}^{+0.012},
4.692_{-0.022}^{+0.018}
\end{split}
\end{equation}
and
\begin{equation}
\begin{split}
\label{eq:phiS:deltaS}
\hat{\Delta}_{\hat{S}}={}& 4.916_{-0.016}^{+0.014},4.982(38),5.052_{-0.022}^{+0.019},5.116_{-0.027}^{+0.024},5.179_{-0.033}^{+0.028},\\
& 5.292_{-0.043}^{+0.034},5.372_{-0.055}^{+0.041},5.479_{-0.075}^{+0.054}
\end{split}
\end{equation}
for $N=2,3,4,5,6,8,10,12$.
For $N=2$, a fuzzy-sphere study gives the finite-size estimate $\Delta_{\phi'}=5.296350$ \cite{Dey:2026cso}, about $2.4\%$ above our estimate.
Figure~\ref{fig:Deltaphi:e} compares our estimates of $\Delta_{\phi'}$ with the $[2,3]$ and $[3,2]$ Pad\'e resummations applied to the five-loop $\epsilon=4-d$ series reported in Ref.~\cite{Henriksson:2025hwi}.
Both Pad\'e approximants reproduce the overall decreasing trend and show reasonable overall agreement with the bootstrap estimates.
The separation between the two Pad\'e estimates becomes more noticeable as $N$ increases, suggesting some sensitivity to the choice of approximant at larger $N$.
The Pad\'e results at larger $N$ should therefore be interpreted with caution, and their growing discrepancies with the bootstrap estimates are understandable.
The scaling dimension $\hat{\Delta}_{\hat S}$ can also be extracted from the $\langle\phi\phi\rangle$ and $\langle SS\rangle$ bootstraps, and we compare the available estimates in Sec.~\ref{sec:3:3}.
A comparison of our estimates with the analytic large-$N$ expansion is presented separately in Sec.~\ref{sec:largen:2}.

Results for the bulk scaling dimensions $(\Delta_{\phi'},\Delta_{\phi''},\Delta_{\phi'''})$ and the boundary scaling dimensions $(\hat{\Delta}_{\hat S},\hat{\Delta}_{\hat S'})$ are summarized in Table~\ref{ta:5} of Appendix~\ref{app:c:2}.
Table~\ref{ta:8} of the same appendix presents the ratio $a_S/\lambda_{\phi\phi S}$, together with coefficient combinations involving $D$, $\hat S$, $\phi'$, and $\phi''$.
Permutation symmetry of scalar three-point functions implies $\lambda_{\phi S\phi}=\lambda_{\phi\phi S}$, and we therefore denote $\lambda_{\phi S\phi}$ by $\lambda_{\phi\phi S}$ throughout this work.

\begin{figure}[tbp]
\centering
\includegraphics[width=0.7\textwidth]{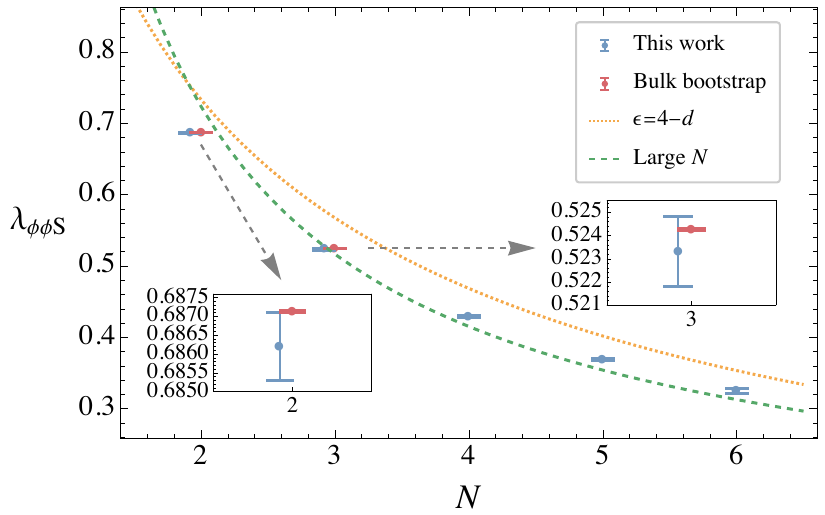}
\caption{\label{fig:lambda:compare}Comparison of the $\lambda_{\phi\phi S}$ estimates in Eq.~\eqref{eq:3:lambda} with bulk-bootstrap results \cite{Chester:2019ifh,Chester:2020iyt}, the $[1,2]$ Pad\'e estimate from the $\epsilon=4-d$ expansion \cite{Henriksson:2022rnm}, and the large-$N$ approximation in Eq.~\eqref{eq:a:9}.
The bars on the data points indicate the estimated uncertainties.
The points are horizontally shifted for clarity.}
\end{figure}

We can extract some bulk OPE and BOE coefficients for $N=2,3,4,5,6$ by combining the data in Table~\ref{ta:8} with the boundary bootstrap results obtained from $\langle\phi\phi\rangle$.
Following a commonly used sign convention, we take $\lambda_{\phi\phi S}>0$.
Since our bootstrap result gives $\lambda_{\phi\phi S}a_S>0$, it follows that $a_S>0$ in this convention.
For $N=2,3,4,5,6$, we combine $\lambda_{\phi\phi S} a_{S}$ from $\langle\phi\phi\rangle$ and $a_{S}/\lambda_{\phi\phi S}$ from $\langle\phi S\rangle$ to obtain
\begin{equation}
\begin{split}
\label{eq:as:1}
a_S^{\mathrm{this\ work(1)}}&=\sqrt{\(\lambda_{\phi\phi S} a_{S}\)\(\frac{a_{S}}{ \lambda_{\phi\phi S}}\)}\\
&=5.578(7),5.379(12),5.488_{-0.017}^{+0.023},5.756(20),6.059_{-0.062}^{+0.053}.
\end{split}
\end{equation}
For $N=2,3$, our results are in good agreement with the previous estimates \cite{Hu:2025yrs},  
\begin{equation}
\label{eq:as:3}
a_S^{\mathrm{prev}}=\frac{\lambda_{\phi\phi S} a_{S}}{\lambda_{\phi\phi S}^\text{bulk}}=5.571(12),5.369(12), 
\end{equation}
which were obtained by combining $\lambda_{\phi\phi S}a_S$ from $\langle\phi\phi\rangle$ with the bulk bootstrap estimates $\lambda_{\phi\phi S}^{\text{bulk}}$ reported in Refs.~\cite{Chester:2019ifh,Chester:2020iyt}.
We will later derive alternative estimates of $a_S$ by combining $\lambda_{\phi\phi S}^{\text{bulk}}$ with $a_S/\lambda_{\phi\phi S}$ from $\langle\phi S\rangle$, and compare them with the above estimates.
The same boundary bootstrap data also provide estimates of the bulk OPE coefficient,
\begin{equation}
\begin{split}
\label{eq:3:lambda}
\lambda_{\phi\phi S}^{\mathrm{this\ work}}&=\sqrt{\(\lambda_{\phi\phi S} a_{S}\)\(\frac{a_{S}}{ \lambda_{\phi\phi S}}\)^{-1}}\\
&=0.6862(9),0.5233(15),0.4290_{-0.0016}^{+0.0021},0.3689(18),0.3247_{-0.0037}^{+0.0032},
\end{split}
\end{equation}
for $N=2,3,4,5,6$.
The results for $N=2,3$ are in good agreement with the previous bulk bootstrap results $\lambda_{\phi\phi S}^{\mathrm{bulk}}=0.687126(27)$ \cite{Chester:2019ifh}, $0.524261(59)$ \cite{Chester:2020iyt}.
For $N=2$, the fuzzy-sphere result $\lambda_{\phi\phi S}=0.6909(7)$ \cite{Feng:2026iii} lies about $0.7\%$ above ours.
The $N=4,5,6$ estimates are new predictions.
For comparison, we apply the $[1,2]$ Pad\'e approximant to the $\epsilon=4-d$ series for $\lambda_{\phi\phi S}^{2}$ collected in Ref.~\cite{Henriksson:2022rnm}. 
Taking the positive square root at $\epsilon=1$ gives $\lambda_{\phi\phi S}=0.4684,0.4017,0.3533$ for $N=4,5,6$, about $9\%$ above our central estimates.
Evaluating the large-$N$ approximation in Eq.~\eqref{eq:a:9} at the same values of $N$ gives $0.4151,0.3539,0.3126$, about $3$--$4\%$ below our central estimates.
Figure~\ref{fig:lambda:compare} compares our estimates with the bulk-bootstrap results, the Pad\'e estimates, and the large-$N$ prediction in Eq.~\eqref{eq:a:9}.
Our new numerical results for $N=4,5,6$ lie between the Pad\'e and large-$N$ predictions, while remaining reasonably close to both. 
This provides a useful consistency check.
A more detailed analysis of the large-$N$ behavior is presented in Sec.~\ref{sec:largen:2}.

Combining $\lambda_{\phi\phi S}^{\mathrm{this\ work}}$ in Eq.~\eqref{eq:3:lambda} with the other $\langle\phi S\rangle$ and $\langle\phi\phi\rangle$ bootstrap results allows us to extract several additional bulk OPE and BOE coefficients.
The resulting estimates of $(a_S,\lambda_{\phi\phi S},b_{SD},b_{S\hat S})$ are summarized in Table~\ref{ta:7}.
For some other quantities, the available data determine only coefficient products.
The products $\lambda_{\phi S\phi'}a_{\phi'}$ and $\lambda_{\phi S\phi''}a_{\phi''}$ are listed in Table~\ref{ta:add:extract} of Appendix~\ref{app:c:2}.

\begin{figure}[tbp]
\centering 
\includegraphics[width=0.7\textwidth]{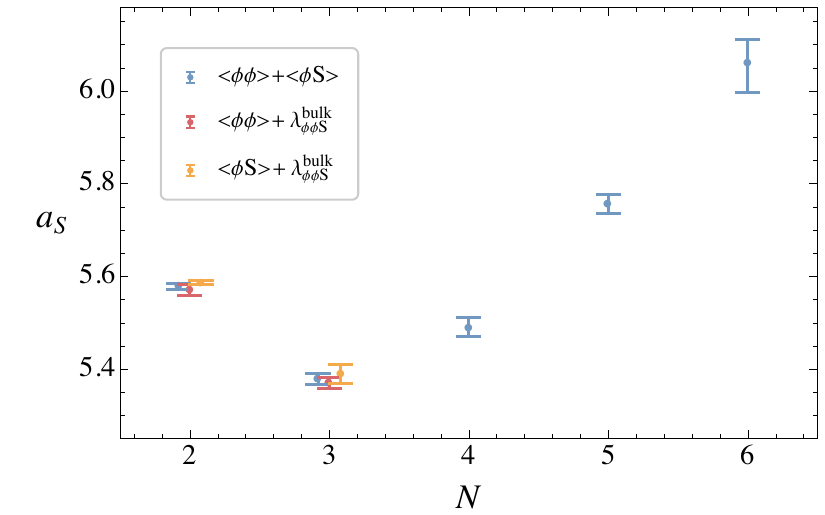}
\caption{\label{fig:29}Estimates of $a_S$ from three extraction procedures.
Blue points combine the $\langle\phi\phi\rangle$ and $\langle\phi S\rangle$ bootstrap results.
Red and orange points combine the bulk-bootstrap estimate $\lambda_{\phi\phi S}^{\mathrm{bulk}}$ with the $\langle\phi\phi\rangle$ and $\langle\phi S\rangle$ results, respectively. 
The three procedures are defined in Eqs.~\eqref{eq:as:1}, \eqref{eq:as:3}, and \eqref{eq:as:2}.
The bars indicate the quoted uncertainty ranges, and the points are shifted horizontally for clarity.}
\end{figure}

For $N=2,3$, we also perform a complementary calculation of $(a_S,b_{SD},b_{S\hat S})$ using the more precise bulk-bootstrap estimates $\lambda_{\phi\phi S}^{\mathrm{bulk}}$ from Refs.~\cite{Chester:2019ifh,Chester:2020iyt}. 
The results obtained in this way are summarized in Table~\ref{ta:6}. 
The quoted ranges in Tables~\ref{ta:7} and \ref{ta:6} overlap for all three coefficients.
For $a_S$, using $\lambda_{\phi\phi S}^{\mathrm{bulk}}$ gives
\begin{equation}
\label{eq:as:2}
a_S^{\mathrm{this\ work(2)}}=\lambda^\text{bulk}_{\phi\phi S}\(\frac{a_S}{\lambda_{\phi\phi S}}\)=5.586(4),5.389(20)\ \mathrm{for}\ N=2,3.
\end{equation} 
Although the numerical estimates obtained from different extraction methods differ slightly, they are compatible within the quoted uncertainties, as shown in Fig.~\ref{fig:29}. 
We now comment on the uncertainties in our estimates of $a_S$.\footnote{To identify the dominant sources of uncertainty, we consider the
following first-order signed variations of the extracted quantities:
\begin{align}
&\frac{\delta a_S^{\mathrm{prev}}}{a_S^{\mathrm{prev}}}\approx\frac{\delta(\lambda_{\phi\phi S}^\text{bulk})^{-1}}{(\lambda_{\phi\phi S}^\text{bulk})^{-1}}+\frac{\delta(\lambda_{\phi\phi S} a_{S})}{(\lambda_{\phi\phi S} a_{S})}=-\frac{\delta(\lambda_{\phi\phi S}^\text{bulk})}{(\lambda_{\phi\phi S}^\text{bulk})}+\frac{\delta(\lambda_{\phi\phi S} a_{S})}{(\lambda_{\phi\phi S} a_{S})},\\
&\frac{\delta a_S^{\mathrm{this\ work(1)}}}{a_S^{\mathrm{this\ work(1)}}}\approx\frac{\delta\sqrt{(\lambda_{\phi\phi S} a_{S})}}{\sqrt{(\lambda_{\phi\phi S} a_{S})}}+\frac{\delta\sqrt{(a_{S}/\lambda_{\phi\phi S})}}{{\sqrt{(a_{S}/\lambda_{\phi\phi S})}}}=\frac{1}{2}\(\frac{\delta(\lambda_{\phi\phi S} a_{S})}{(\lambda_{\phi\phi S} a_{S})}+\frac{\delta(a_{S}/\lambda_{\phi\phi S})}{(a_{S}/\lambda_{\phi\phi S})}\),\\
&\frac{\delta a_S^{\mathrm{this\ work(2)}}}{a_S^{\mathrm{this\ work(2)}}}\approx\frac{\delta(\lambda_{\phi\phi S}^\text{bulk})}{(\lambda_{\phi\phi S}^\text{bulk})}+\frac{\delta(a_{S}/\lambda_{\phi\phi S})}{(a_{S}/\lambda_{\phi\phi S})}.
\end{align}
The above relative-error analysis should only be regarded as approximate, since $\lambda_{\phi\phi S} a_{S}$ and $a_{S}/\lambda_{\phi\phi S}$ from the boundary bootstrap are not independent.
Nevertheless, it is useful for identifying the dominant source of uncertainty when the relative uncertainties of $\lambda_{\phi\phi S}a_S$ and $a_S/\lambda_{\phi\phi S}$ differ significantly.}
The bulk-bootstrap estimates $\lambda_{\phi\phi S}^{\text{bulk}}$ are relatively precise, so the final uncertainties are dominated by those of $\lambda_{\phi\phi S}a_S$ from $\langle\phi\phi\rangle$ and $a_S/\lambda_{\phi\phi S}$ from $\langle\phi S\rangle$.
For $N=2$, the ratio $a_{S}/\lambda_{\phi\phi S}$ has a smaller relative uncertainty, and $a_{S}^{\mathrm{this\ work(2)}}$ in Eq.~\eqref{eq:as:2} accordingly gives the narrowest quoted range.
For $N=3$, it is less straightforward to identify the method with the smallest uncertainty, since the inputs entering the $\<\phi S\>$ and $\<\phi\phi\>$ analyses have comparable precision. 
In this case, $a_S^{\mathrm{this\ work(1)}}$ in Eq.~\eqref{eq:as:1} has the smallest relative uncertainty.

For $b_{SD}$ and $b_{S\hat S}$, the only difference between the calculations for the results in Tables~\ref{ta:7} and \ref{ta:6} is the choice of $\lambda_{\phi\phi S}$.
The resulting estimates are very close. 
This is because the differences caused by using different estimates of $\lambda_{\phi\phi S}$ are smaller than the uncertainties from the other quantities entering the extraction.
In particular, the two tables report identical values for $b_{S\hat S}$ after rounding.

\begin{figure}[tbp]
\centering 
\includegraphics[width=0.7\textwidth]{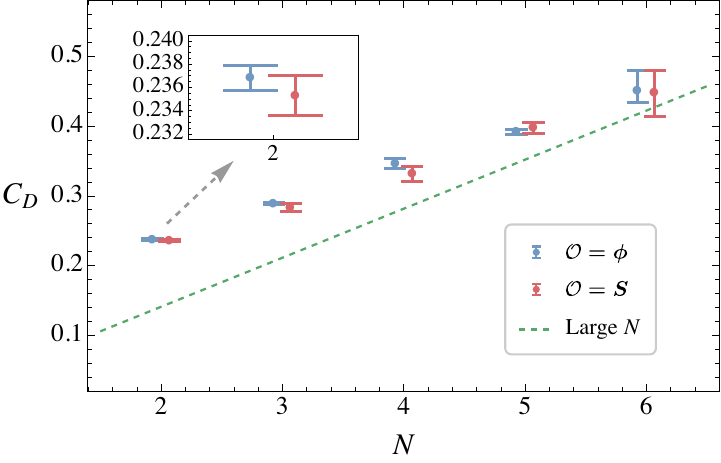}
\caption{\label{fig:17}Estimates of $C_D$ obtained from different operators, together with their uncertainties.
The large-$N$ results are from Eq.~\eqref{eq:a:17}.
The results for $N=2,3,4,5$ with $\mathcal{O}=\phi$ are taken from Ref.~\cite{Hu:2025yrs}.
The $N=6$ result for $\mathcal O=\phi$ and the $N=2,3,4,5,6$ results for $\mathcal O=S$ are given in Eqs.~\eqref{eq:3:3} and \eqref{eq:3:11}, respectively.
The points are horizontally shifted for clarity.
In addition to the estimates shown here, $C_D$ can be obtained directly from the $\langle\phi S\rangle$ bootstrap as given in Eq.~\eqref{eq:cd:mixed}, including for $N=8,10,12$.}
\end{figure}

Using the results above, we can also compute $C_D$ from Eq.~\eqref{eq:2:23} by setting $\mathcal{O}=S$.
Following Refs.~\cite{Gliozzi:2015qsa,Hu:2025yrs}, we compare this result with that obtained using $\mathcal O=\phi$.
Since the value of $C_D$ is independent of the choice of $\mathcal{O}$, the resulting numerical estimates should agree within their quoted uncertainties.
Expressing this formula in terms of the quantities obtained directly from the bootstrap gives
\begin{equation}
\label{eq:phiS:cd}
    C_D|_{\mathcal{O}=S}=\(\frac{\Delta_{S}a_{S}}{4\pi b_{SD}}\)^2=\(\frac{\Delta_S}{4\pi}\)^2\(\frac{b_{\phi D}}{a_{\phi}}\)^2 \left(\frac{a_S}{\lambda_{\phi\phi S}}\right)^2 \left(\frac{b_{\phi D}b_{SD}}{\lambda_{\phi\phi S}a_\phi}\right)^{-2}.
\end{equation}
Here $a_{\phi}$ and $b_{\phi D}$ are direct bootstrap results from $\langle \phi\phi\rangle$, while $a_S/\lambda_{\phi\phi S}$ and $(b_{\phi D} b_{SD})/(\lambda_{\phi\phi S}a_{\phi})$ are direct results from the $\langle \phi S\rangle$ bootstrap in Table~\ref{ta:8}. 
For $N=2,3,4,5,6$, Eq.~\eqref{eq:phiS:cd} gives
\begin{equation}
\label{eq:3:11}
C_D|_{\mathcal{O}=S}=0.2353(17),0.2828(57),0.331(11),0.397(8),0.448_{-0.035}^{+0.032}.
\end{equation}
For $N=1$, Ref.~\cite{Hu:2025yrs} found close agreement between the values of
$C_D$ extracted using $\mathcal{O}=S$ and $\mathcal{O}=\phi$. 
We examine this consistency for $N=2,3,4,5,6$.
For comparison, $C_D|_{\mathcal{O}=\phi}$ takes the values $0.2368(11), 0.2883(16),0.346(7),0.3914(32)$ for $N=2,3,4,5$ \cite{Hu:2025yrs}, and $0.451^{+0.028}_{-0.017}$ for $N=6$ from Eq.~\eqref{eq:3:3}. 
To visualize this more clearly, we present these $C_D$ estimates in Fig.~\ref{fig:17}. 
The two determinations overlap within their quoted uncertainties. 
Although they share some input data, they combine results from separate $\eta$-minimizations for $\langle\phi\phi\rangle$ and $\langle\phi S\rangle$ in different ways.
Their agreement therefore provides a nontrivial self-consistency check.

The agreement above is consistent with the operator independence of $C_D$. 
We can also use this property to extract $C_D$ directly from the mixed-correlator bootstrap:
\begin{align}
C_D^{\mathrm{mixed}}
&=\sqrt{\(C_D|_{\mathcal O=\phi}\)\(C_D|_{\mathcal O=S}\)}
 =\frac{\Delta_\phi\Delta_S}{16\pi^2}
 \left(\frac{a_S}{\lambda_{\phi\phi S}}\right)
 \left(\frac{b_{\phi D}b_{SD}}{\lambda_{\phi\phi S}a_\phi}\right)^{-1}
 \notag\\
&=0.189680(2),0.2361(4),0.2856(23),0.3387(17),
  0.3942_{-0.0023}^{+0.0025},0.4503_{-0.0038}^{+0.0040},
\notag\\
&\mathrel{\phantom{=}}0.571(7),0.686_{-0.008}^{+0.010},
  0.805_{-0.020}^{+0.025}
\label{eq:cd:mixed}
\end{align}
for $N=1,2,3,4,5,6,8,10,12$.
We compute these results using the $\langle\phi S\rangle$ data from Ref.~\cite{Hu:2025yrs} for $N=1$ and from Table~\ref{ta:8} for the remaining values of $N$.
Since $C_D^{\mathrm{mixed}}$ is defined as the geometric mean of $C_D|_{\mathcal O=\phi}$ and $C_D|_{\mathcal O=S}$,
the mixed estimates lie between the other two estimates.
For $N=1$, our estimate of $C_D^{\mathrm{mixed}}$ lies entirely within the ranges of both $C_D|_{\mathcal O=\phi}=0.18966(10)$ and $C_D|_{\mathcal O=S}=0.18970(10)$ reported in Ref.~\cite{Hu:2025yrs}.
The quoted relative uncertainties of $C_D^{\mathrm{mixed}}$ are generally smaller than those of $C_D|_{\mathcal O=\phi}$ and $C_D|_{\mathcal O=S}$, except for $C_D|_{\mathcal O=\phi}$ at $N=3$.
The particularly small uncertainty at $N=1$ reflects the high precision of
the Ising mixed-correlator results of Ref.~\cite{Hu:2025yrs}.
For $N=8,10,12$, the leading large-$N$ predictions $(0.563,0.703,0.844)$ from Eq.~\eqref{eq:a:17} differ from our corresponding central estimates by less than $5\%$.
This agreement provides support for these new estimates.

\begin{table}[tbp]
\centering
\small
\begin{tabular}{c|cccc}
\noalign{\hrule height 1.0pt}
$N$ & $a_S$ &$\lambda_{\phi\phi S}$ & $b_{SD}$ & $b_{S\hat{S}}$    \\
\hline
2&5.578(7)	&0.6862(9)&1.383(6)&	0.44(4) \\
\hline
3&5.379(12)	&0.5233(15)&	1.284(12)&	0.49(6)\\
\hline
4&$5.488_{-0.017}^{+0.023}$	&$0.4290_{-0.0016}^{+0.0021}$&	1.262(23)	&0.48(11)\\
\hline
5&5.756(20)	&0.3689(18)&	1.249(16)&	$0.73_{-0.07}^{+0.04}$\\
\hline
6&$6.059_{-0.062}^{+0.053}$&$0.3247_{-0.0037}^{+0.0032}$&$1.270_{-0.050}^{+0.056}$&$0.72_{-0.33}^{+0.42}$ \\
\noalign{\hrule height 1.0pt}
\end{tabular}
\caption{\label{ta:7}Bulk one-point, OPE and BOE coefficients extracted by combining the $\langle\phi S\rangle$ and $\langle\phi\phi\rangle$ bootstrap results for $N=2,3,4,5,6$.
The uncertainty is obtained by first determining a range for each input choice and then taking the envelope of these ranges.}
\end{table}

\begin{table}[tbp]
\small
\centering
\begin{tabular}{c|ccc}
\noalign{\hrule height 1.0pt}
$N$ & $a_S$ & $b_{SD}$ & $b_{S\hat{S}}$ \\
\hline
2&5.586(4)	&1.385(4)&	0.44(4) \\
\hline
3&5.389(20) &	1.286(9)&	0.49(6) \\
\noalign{\hrule height 1.0pt}
\end{tabular}
\caption{\label{ta:6}Bulk one-point and BOE coefficients for $N=2,3$, obtained using the $\langle\phi S\rangle$ bootstrap results and the bulk-bootstrap input $\lambda_{\phi\phi S}^{\mathrm{bulk}}$ \cite{Chester:2019ifh,Chester:2020iyt}.}
\end{table}

\FloatBarrier

\subsection{$\< S S\>$}
\label{sec:3:3}
We next study the bootstrap of $\langle S S\rangle$, which provides additional information on the singlet sector and allows for consistency checks with results from the $\langle \phi \phi\rangle$ and $\langle \phi S\rangle$ analyses. 
Combined with the $\langle\phi\phi\rangle$ results, it also determines the bulk coefficient ratio $\lambda_{SSS'}/\lambda_{\phi\phi S'}$.

The $\eta$-function for $\langle SS \rangle$ is constructed from Eq.~\eqref{eq2:22} with $(\Delta_{S},\Delta_{S'},\lambda_{SSS},a_{S})$ taken as inputs.
Unlike the analyses of $\langle\phi\phi\rangle$ and $\langle\phi S\rangle$, where only scaling dimensions are fixed as inputs, the present calculation also requires coefficient inputs in both the bulk and boundary channels.
The reason is that using $(\Delta_{S},\Delta_{S'})$ alone leads to solutions with negative boundary-channel coefficients, such as $a_S^2<0$ or $b_{SD}^2<0$, which is inconsistent with the real BOE coefficients from the $\langle\phi\phi\rangle$ and $\langle\phi S\rangle$ crossing equations. 
We therefore fix the leading coefficients $(a_{S}\lambda_{SSS},a_{S}^2)$ by
\begin{equation}
 \lambda_{SSS}^{\mathrm{input}}=1.53244304(58)\text{\cite{Chang:2024whx}},0.830914(32) \text{\cite{Chester:2019ifh}},0.5055(11) \text{\cite{Chester:2020iyt}}\ \mathrm{for}\ N=1,2,3,
\end{equation}
and
\begin{equation}
 a_{S}^{\mathrm{input}}=6.677424(16)\text{\cite{Hu:2025yrs}},5.586(4),5.389(20) \ \mathrm{for}\ N=1,2,3.
\end{equation}
For $N=2,3$, we take $a_S$ from Eq.~\eqref{eq:as:2}.
As a stability check, we also repeat the bootstrap analysis using the other two estimates of $a_S$ in Eqs.~\eqref{eq:as:1} and \eqref{eq:as:3}.
We find only minor changes in the results, as summarized in Tables~\ref{ta:22}--\ref{ta:23} in Appendix~\ref{app:c}.

We now present the bootstrap results obtained from the $\langle SS\rangle$ two-point function.
Table~\ref{ta:10} summarizes the bootstrap estimates of $(\Delta_{S''},\hat{\Delta}_{\hat S},\hat{\Delta}_{\hat S'},b_{SD},b_{S\hat S})$ that are central to the discussion below.\footnote{For $N=1$, the normal boundary condition breaks $\mathrm{O}(1)\simeq\mathbb{Z}_2$ and leaves no residual internal symmetry.
Accordingly, $\hat S,\hat S',\hat S'',\ldots$ do not denote singlets under a residual internal symmetry but simply label different boundary scalar operators.
In Ref.~\cite{Hu:2025yrs}, these operators were denoted by $\hat N,\hat N',\hat N'',\ldots$.}
Supplementary estimates of $(\Delta_{S'''},b_{S\hat S'},\lambda_{SSS'}a_{S'},\lambda_{SSS''}a_{S''})$ are listed in Table~\ref{ta:add:ss} of Appendix~\ref{app:c:3}.
By combining $\lambda_{SSS'}a_{S'}$ from $\<SS\>$ with $\lambda_{\phi\phi S'}a_{S'}$ from $\langle\phi\phi\rangle$ \cite{Hu:2025yrs}, we eliminate the common factor $a_{S'}$ and obtain
\begin{equation}
\frac{\lambda_{SSS'}}{\lambda_{\phi\phi S'}}=28.962(23),33.17(25),38.63(56)\ \mathrm{for}\ N=1,2,3.
\end{equation}
For $N=1$, our result is consistent with $28.975(61)$ and $28.963(111)$, derived respectively from the bulk-bootstrap OPE-coefficient ranges reported in Ref.~\cite{Simmons-Duffin:2016wlq} and Ref.~\cite{Reehorst:2021hmp}.
The latter estimate has rigorous bounds, and its central value is remarkably close to ours.
To our knowledge, the $N=2,3$ results are the first numerical estimates of this ratio.
For comparison, evaluating the leading large-$N$ expression in Eq.~\eqref{eq:a:S2:ratio} gives $29.61$ and $44.41$ for $N=2,3$.
These values differ from our estimates by approximately $11\%$ and $15\%$.
Given the small values of $N$, this level of agreement is reasonable.

\begin{figure}[tbp]
\captionsetup[subfigure]{skip=2pt}
\captionsetup[subfigure]{
  margin={20pt,0pt},
  justification=centering,
  singlelinecheck=false
}
\setlength{\lineskip}{1em}
\centering
\begin{subfigure}[t]{0.48\textwidth}
\includegraphics[width=\textwidth]{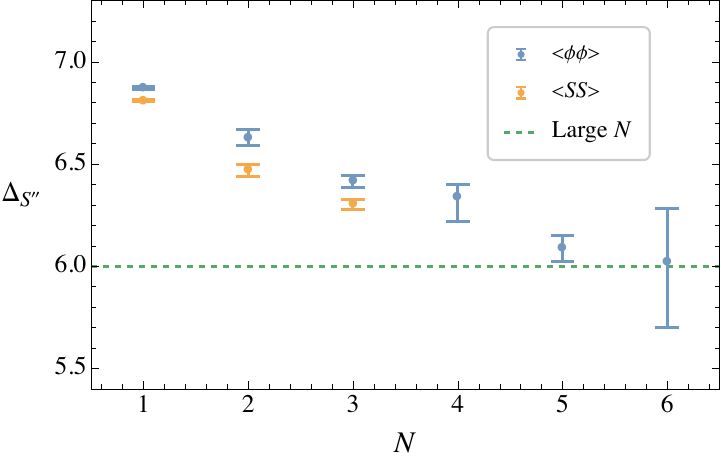}
\caption{$\Delta_{S''}$}
\end{subfigure}
\par\medskip
\begin{subfigure}[t]{0.48\textwidth}
\includegraphics[width=\textwidth]{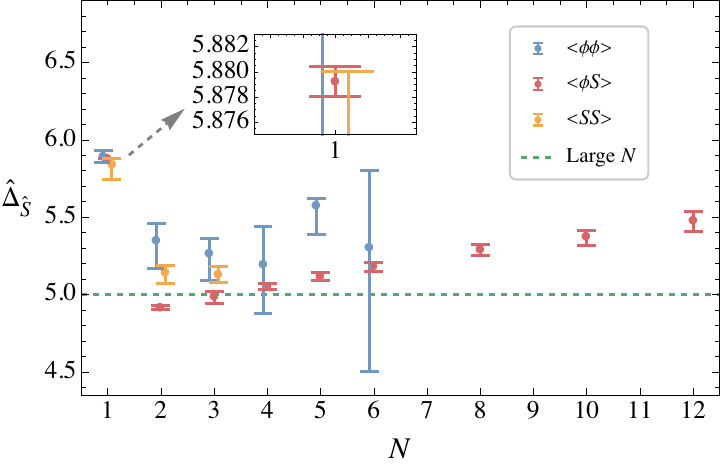}
\caption{$\hat{\Delta}_{\hat{S}}$}
\end{subfigure}
\hfill
\begin{subfigure}[t]{0.48\textwidth}
\includegraphics[width=\textwidth]{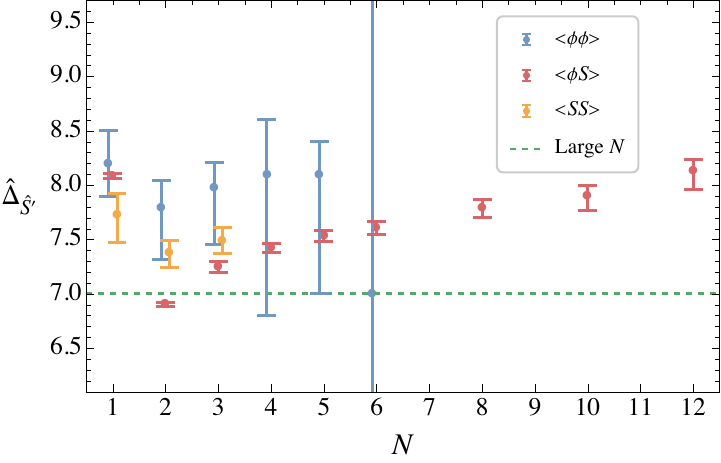}
\caption{$\hat{\Delta}_{\hat{S}'}$}
\end{subfigure}
\caption{\label{fig:diffcorr:dimension}Bootstrap estimates of scaling dimensions from different correlators.
The green dashed lines show the large-$N$ results from Eqs.~\eqref{eq:a} and~\eqref{eq:a:hatS}.
Bootstrap results are collected in Tables~\ref{ta:a2:7}--\ref{ta:a2:8} of Appendix~\ref{app:b}.
Error estimates include only input-induced and extrapolation uncertainties.
The error estimate for the $N=6$ $\langle\phi\phi\rangle$ result $\hat{\Delta}_{\hat S'}=7(8)$ is clipped to the plot range.
Points are horizontally shifted for clarity.}
\end{figure}

\begin{figure}[tbp]
\captionsetup[subfigure]{skip=2pt}
\captionsetup[subfigure]{
  margin={20pt,0pt},
  justification=centering,
  singlelinecheck=false
}
\setlength{\lineskip}{1em}
\centering
\begin{subfigure}[t]{0.48\textwidth}
\includegraphics[width=\textwidth]{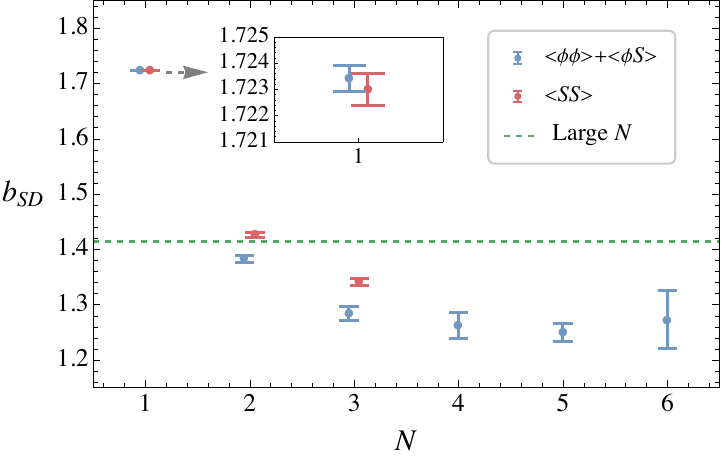}
\caption{$b_{SD}$}
\end{subfigure}
\hfill
\begin{subfigure}[t]{0.48\textwidth}
\includegraphics[width=\textwidth]{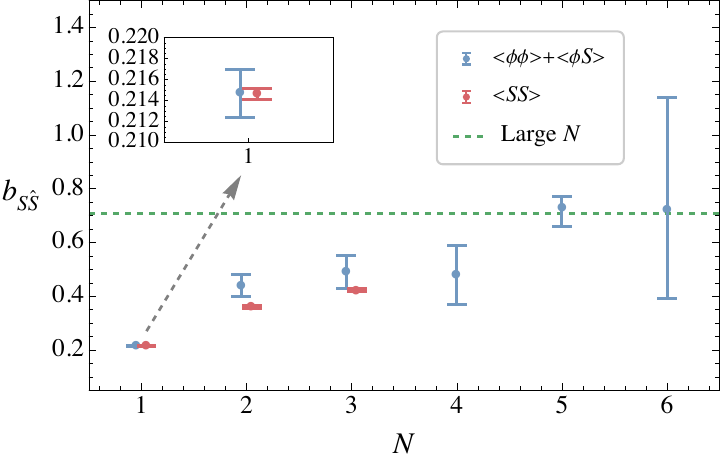}
\caption{$b_{S\hat{S}}$}
\end{subfigure}
\caption{\label{fig:diffcorr:coeff}Bootstrap estimates of BOE coefficients from different correlators.
The green dashed lines show the large-$N$ results from Eqs.~\eqref{eq:a:bsd}--\eqref{eq:a:bss}.
Bootstrap results are summarized in Tables~\ref{ta:a2:9}--\ref{ta:a2:10} of Appendix~\ref{app:b}.
Error estimates include only input-induced and extrapolation uncertainties.
Here, ``$\langle\phi\phi\rangle+\langle\phi S\rangle$'' denotes BOE coefficients disentangled using results from these two bootstraps.
Points are horizontally shifted for clarity.}
\end{figure}

We next compare scaling dimensions and boundary coefficients extracted from different two-point functions.
Fig.~\ref{fig:diffcorr:dimension} shows the comparisons for $\Delta_{S''}$, $\hat{\Delta}_{\hat S}$, and $\hat{\Delta}_{\hat{S}'}$, while Fig.~\ref{fig:diffcorr:coeff} shows those for $b_{SD}$ and $b_{S\hat S}$.
The $\langle\phi\phi\rangle$ results for $N=1,2,3,4,5$ and the $\langle\phi S\rangle$ results for $N=1$ are based on the $\eta$-minimization study of Ref.~\cite{Hu:2025yrs}, while the remaining bootstrap results are from our work.
The numerical values underlying these comparisons are collected in Tables~\ref{ta:a2:7}--\ref{ta:a2:10} in Appendix~\ref{app:b}.
Our quoted uncertainties include only the input-induced uncertainty and the extrapolation uncertainty. 
For several quantities, the uncertainty intervals from bootstrap analyses of different correlators do not overlap, suggesting that the present uncertainty prescription may not capture all systematic effects.
The better agreement at $N=1$ may be due to the more precise input data and the sparser low-lying spectrum.

We discuss the $N=1$ case in more detail below. 
The two-point functions studied here have simpler spectra at $N=1$ than at $N>1$. 
This makes the low-lying contributions easier to resolve and their estimates more reliable.
For the bulk singlet $S''$, the $\langle\phi\phi\rangle$ and $\langle SS\rangle$
analyses give $\Delta_{S''}=6.873(7)$~\cite{Hu:2025yrs} and $6.808(5)$ in Table~\ref{ta:10}, respectively.
Both values lie below the bulk bootstrap estimate of $6.8956(43)$~\cite{Simmons-Duffin:2016wlq}.
The next operator resolved in the truncated spectrum is estimated to have scaling dimension $10.111_{-0.034}^{+0.028}$ from the $\langle\phi\phi\rangle$ bootstrap \cite{Hu:2025yrs} and $10.080(17)$ from the $\langle SS\rangle$ bootstrap.
The latter value is listed in Table~\ref{ta:add:ss} of Appendix~\ref{app:c:3}.
There is no clearly resolved state corresponding to the operator at $7.2535(51)$ reported in Ref.~\cite{Simmons-Duffin:2016wlq}.
This suggests that the truncated bootstrap employed here may not fully resolve the spectrum in this region and that additional uncertainty may be introduced by operators with very close scaling dimensions.

For the boundary scaling dimension $\hat{\Delta}_{\hat S}$, our $N=1$ estimate $5.84_{-0.10}^{+0.04}$ from $\langle SS\rangle$ agrees within the quoted uncertainties with the $\langle\phi\phi\rangle$ estimate $5.89(4)$ and the $\langle\phi S\rangle$ estimate $5.8792(12)$ from Ref.~\cite{Hu:2025yrs}.
Its central value differs from these two estimates by less than $0.9\%$ and from the fuzzy-sphere estimate $5.858$ \cite{Dedushenko:2024nwi} by about $0.3\%$.
For $\hat{\Delta}_{\hat{S}'}$, the discrepancy is more pronounced.
The estimates obtained from $\langle S S\rangle$ and $\langle \phi S\rangle$ do not overlap within their quoted uncertainties, as shown in Fig.~\ref{fig:diffcorr:dimension}.
The larger discrepancy for $\hat{\Delta}_{\hat{S}'}$ may reflect the denser spectrum at higher scaling dimensions, where nearby operators are harder to distinguish in each correlator analysis.
Since the contributions of the nearby operators vary among two-point functions, the resulting estimates may differ between correlators.

Turning to the boundary coefficients, we also find particularly good agreement among the different correlator analyses for $N=1$.
The $\langle S S\rangle$ bootstrap gives
\begin{equation}
(b_{SD},b_{S\hat{S}})=(1.7230(6),0.2146(5)).
\end{equation}
These values agree well with the estimates $(1.7234(5),0.2147(23))$ \cite{Hu:2025yrs} obtained by combining results from $\langle \phi\phi\rangle$ and $\langle\phi S\rangle$.
Our value of $b_{SD}$ is also consistent with the fuzzy-sphere result $1.74(22)$ \cite{Zhou:2024dbt}.
It is close to the boundary bootstrap estimate $1.742(6)$ obtained using relatively low-order truncations \cite{Gliozzi:2015qsa} and the Monte Carlo estimate $1.69(1)$ \cite{Przetakiewicz:2025gzi}.

For $N>1$, estimates of the same scaling dimensions and boundary coefficients extracted from different correlators generally differ more than at $N=1$.
Some of these estimates do not overlap within their quoted uncertainties, as can be seen in Figs.~\ref{fig:diffcorr:dimension}--\ref{fig:diffcorr:coeff}.
As $N$ increases, operators whose scaling dimensions become degenerate
in the $N\to\infty$ limit may become harder to distinguish.
Their combined contribution may vary across correlators, which could help explain the larger discrepancies among the estimates.

Fig.~\ref{fig:diffcorr:dimension} also shows a marked decrease in the central values of the boundary scaling dimension $\hat{\Delta}_{\hat S}$ from $N=1$ to $N=2$ in all three correlator analyses. 
To explain this apparently discontinuous behavior at $N=1$, we interpret the leading nonprotected boundary singlet $\hat S$ as a composite operator built from the protected operator.
In the Ising case, the lowest nonidentity primary is the displacement operator $D$ with $\hat{\Delta}_D=3$, and there is no tilt operator because the $Z_2$ symmetry is discrete. 
As a result, the leading nonprotected boundary operator at $N=1$ is expected to be the composite $[DD]$. 
For $N>1$, the presence of the tilt operator $t_i$ with
$\hat{\Delta}_t=2$ allows for additional scalar singlets built from it. 
In particular, the composite $[tt]$ is expected to have a lower scaling dimension than $[DD]$.
This difference in the leading boundary composite operator can account for the lower estimates for $\hat\D_{\hat S}$ at $N>1$.\footnote{In the $\epsilon=4-d$ expansion, there is a mixing between $DD$ and $t\Box_{\parallel} t$ for $N>1$ as they are of dimension 8 at $d=4$. Therefore, the next nonprotected boundary scaling dimension $\hat{\Delta}_{\hat S'}$ may be an averaged dimension of nearby operators for $N>1$. 
In the Ising case, the next composite $D\Box_{\parallel} D$ is more isolated, so the mixing effects should be milder for $\hat{\Delta}_{\hat S'}$ at $N=1$. }

The differences between the correlator analyses are also reflected in the estimates of $C_D$.
Using $b_{SD}$ obtained directly from the $\langle SS\rangle$ bootstrap gives 
\begin{equation}
\left.C_D\right|_{\mathcal O=S}^{\langle SS\rangle}=0.18978_{-0.00011}^{+0.00015},0.2217_{-0.0016}^{+0.0019},0.260(5),
\end{equation}
for $N=1,2,3$. 
The $N=1$ result is consistent with the previous $\mathcal O=S$ estimate $0.18970(10)$ and $\mathcal O=\phi$ estimate $0.18966(10)$ from Ref.~\cite{Hu:2025yrs}.
Its quoted uncertainty range includes the full range of the mixed-correlator estimate $0.189680(2)$ in Eq.~\eqref{eq:cd:mixed}.
The Monte Carlo estimate $C_D=0.198(3)$ \cite{Przetakiewicz:2025gzi} is about $4\%$ above our $N=1$ result.
For $N=2,3$, these estimates from $\langle SS\rangle$ are appreciably lower than the corresponding $\mathcal O=S$ estimates $(0.2353(17),0.2828(57))$ obtained by combining the $\langle\phi\phi\rangle$ and $\langle\phi S\rangle$ results in Eq.~\eqref{eq:3:11}.
They are also lower than the $\mathcal O=\phi$ estimates $(0.2368(11),0.2883(16))$ from the $\langle\phi\phi\rangle$ bootstrap \cite{Hu:2025yrs} and the mixed-correlator estimates $(0.2361(4),0.2856(23))$ in Eq.~\eqref{eq:cd:mixed}.
These deviations suggest that the $\langle SS\rangle$ calculation may be more sensitive to systematic effects not captured by the present uncertainty prescription.

\begin{table}[tbp]
\small
\centering
\begin{tabular}{c|ccccc}
\noalign{\hrule height 1.0pt}
$N$ &$\Delta_{S''}$ &$\hat{\Delta}_{\hat{S}}$ & $\hat{\Delta}_{\hat{S}'}$&$b_{SD}$ & $b_{S\hat{S}}$ \\
\hline
1&6.808(5)&$5.84_{-0.10}^{+0.04}$&$7.73_{-0.26}^{+0.19}\,{}^{*}$&1.7230(6)&$0.2146(5)^{*}$\\
\hline
2&$6.470_{-0.031}^{+0.028}$&$5.14_{-0.07}^{+0.05}$&$7.38_{-0.14}^{+0.11}$&$1.4267_{-0.0049}^{+0.0043}$&0.359(6)\\
\hline
3&$6.303_{-0.028}^{+0.024}$&5.13(5)&7.49(12)&$1.3413_{-0.0070}^{+0.0061}$&0.422(6)\\
\noalign{\hrule height 1.0pt}
\end{tabular}
\caption{\label{ta:10}Main results obtained directly from the $\langle SS\rangle$ bootstrap. Data marked with an asterisk are obtained from linear fits, while the others are obtained from power-law fits.}
\end{table}

\section{Large-$N$ behavior}
\label{sec:largen}
With bootstrap results extending to $N=6$ and even to $N=12$ for some quantities, we explore the asymptotic behavior of selected BCFT data as $N$ increases.
In Sec.~\ref{app:a}, we summarize the analytic large-$N$ expressions in $d=3$. 
In Sec.~\ref{sec:largen:2}, we extrapolate our finite-$N$ bootstrap results to the $N\to\infty$ limit.
The comparison between the extrapolated limits and the analytic predictions provides a consistency check on the bootstrap results obtained at finite $N$.
\subsection{Analytic large-$N$ expressions}
\label{app:a}
Here, we collect the analytic large-$N$ expressions in $d=3$ needed for comparison with the numerical bootstrap results.
For the boundary quantities, we consider the normal boundary condition.
We review the known results for bulk and boundary CFT data. 
Using these results, we obtain large-$N$ expressions for several additional quantities such as $a_S/\lambda_{\phi\phi S}$, $a_S$, $b_{SD}$, $b_{S\hat S}$, and $C_D$.

We begin with the bulk data, including several low-lying scaling dimensions and OPE coefficients.
In the $N\to\infty$ limit, the low-lying bulk vector spectrum is sparse compared with the singlet spectrum, with operator dimensions approaching $1/2,9/2,13/2,\ldots$.
The degeneracies of these levels increase with dimension.
For the two lowest vector operators $\phi$ and $\phi'$, the next-to-leading-order (NLO) expansions of the scaling dimensions are \cite{Henriksson:2022rnm}
\begin{equation}
\Delta_{\phi}=\frac{1}{2}+\frac{4}{3\pi^2 N}+O(N^{-2}),
\quad
\Delta_{\phi'}=\frac{9}{2}+\frac{4}{3\pi^2 N}+O(N^{-2}).
\label{eq:a:2}
\end{equation}
The bulk singlet spectrum contains operators whose dimensions approach the even integers $2,4,6,8,\ldots$ as $N\to\infty$.
The scaling dimensions of the two lowest primaries have the following large-$N$ expansions \cite{Henriksson:2022rnm}:
\begin{equation}
\Delta_{S}=2-\frac{32}{3\pi^2 N}+O(N^{-2}),
\quad
\Delta_{S'}=4-\frac{64}{3\pi^2 N}+O(N^{-2}).
\label{eq:a:4}
\end{equation}
The next bulk singlet level contains multiple operators that are degenerate at leading order, with
\begin{equation}
\Delta_{S''}=6+O(N^{-1}).
\label{eq:a}
\end{equation}
The bulk traceless symmetric spectrum has levels at $1,3,5,\ldots$ in the $N\to\infty$ limit.
The scaling dimensions of the first two operators are given by \cite{Henriksson:2022rnm}
\begin{equation}
\Delta_{T}=1+\frac{32}{3\pi^2 N}+O(N^{-2}),\quad
\Delta_{T'}=3+\frac{64}{3\pi^2 N}+O(N^{-2}).
\label{eq:a:7}
\end{equation}
The square of the OPE coefficient $\lambda_{\phi\phi S}$ has the large-$N$ expansion \cite{Henriksson:2022rnm}:
\begin{equation}
\label{eq:a:8}
\lambda_{\phi\phi S}^2=\frac{4}{\pi^2 N}+\frac{96}{\pi^4 N^2}+O(N^{-3}).
\end{equation}
To maintain a consistent order, we take the square root and re-expand the result in powers of $1/N$:
\begin{equation}
\label{eq:a:9}
\lambda_{\phi\phi S}=\sqrt{\frac{1}{N}}\(\frac{2}{\pi}+\frac{24}{\pi^3 N}+O(N^{-2})\).
\end{equation}
Substituting the $1/N$ correction to $\Delta_S$ into the large-$N$ expansion of $\langle\phi\phi\rangle$ \cite{Padayasi:2021sik} gives
\begin{equation}
\label{eq:a:10}
\lambda_{\phi\phi S}a_{S}=\frac{3}{2}+\frac{44}{3\pi^2 N}+O(N^{-2}).
\end{equation}
For the OPE coefficients involving the next-to-leading bulk singlet $S'$, the large-$N$ expansions give \cite{Henriksson:2022rnm}
\begin{equation}
 \lambda_{\phi\phi S'}=\frac{2\sqrt{2}}{3\pi^2N}+O(N^{-2}), \quad 
 \lambda_{SSS'}=\sqrt{2}\left(1-\frac{1}{2N}\right)+O(N^{-2}).
\label{eq:a:S2}
\end{equation}
Their ratio therefore behaves as
\begin{equation}
\label{eq:a:S2:ratio}
\frac{\lambda_{SSS'}}{\lambda_{\phi\phi S'}}
=\frac{3\pi^2}{2}N+O(N^{0}).
\end{equation}

The large-$N$ boundary data are obtained by matching the boundary-channel conformal block expansion to the large-$N$ expansion of $\langle\phi\phi\rangle$ in Refs.~\cite{Metlitski:2020cqy,Padayasi:2021sik}.\footnote{It is not known whether multiple boundary primaries share a given leading scaling dimension at large $N$.
If they do, the large-$N$ correlators determine only combinations of their contributions, rather than the data of individual operators.
When using these large-$N$ results, we assume no degeneracy within each boundary symmetry sector.}
Apart from the protected operators $D$ and $t_i$, the singlet and vector spectra approach the odd-integer levels $5,7,9,\ldots$ as $N\to\infty$.
The first two boundary singlets above $D$ have the scaling dimensions:
\begin{equation}
\hat{\Delta}_{\hat S}=5+O(N^{-1}),
\qquad
\hat{\Delta}_{\hat S'}=7+O(N^{-1}).
\label{eq:a:hatS}
\end{equation}
For later use, we record the following coefficient expressions from Ref.~\cite{Padayasi:2021sik}:
\begin{align}
&a_{\phi}^2=2N\(1+\(1-\frac{4}{3\pi^2}\)\frac{1}{N}+O(N^{-2})\),\label{eq:a:11}\\
&b_{\phi t}^2=\frac{1}{4}\(1+\(1-\frac{4}{3\pi^2}\)\frac{1}{N}+O(N^{-2})\),\label{eq:a:12}\\
&b_{\phi D}^2=\frac{4}{9\pi^2}+O(N^{-1})\label{eq:a:13}.
\end{align}
For the quantities in Eqs.~\eqref{eq:a:11}--\eqref{eq:a:12}, we take the square root and further expand the results in powers of $1/N$.
In particular, we divide $a_{\phi}$ by $\sqrt{N}$ so that the resulting quantity has a finite large-$N$ limit. This gives
\begin{align}
\frac{a_{\phi}}{\sqrt{N}}&=\sqrt{2}+\frac{\sqrt{2}}{2}\(1-\frac{4}{3\pi^2}\)\frac{1}{N}+O(N^{-2}),\label{eq:a:14}\\
b_{\phi t}&=\frac{1}{2}+\frac{1}{4}\(1-\frac{4}{3\pi^2}\)\frac{1}{N}+O(N^{-2}).\label{eq:a:15}
\end{align}
Substituting Eqs.~\eqref{eq:a:11} and \eqref{eq:a:12} into the expression for $\alpha$ in Eq.~\eqref{eq:2:4} yields \cite{Metlitski:2020cqy}
\begin{equation}
\label{eq:a:16}
\alpha=\frac{1}{32\pi}\frac{a_{\phi}^2}{b_{\phi t}^2}-\frac{N-2}{2\pi}=-\frac{N}{4\pi}+\frac{1}{\pi}+O(N^{-1}).
\end{equation}

Using the results above, we obtain large-$N$ expressions for several other universal quantities studied in this work.
Combining Eqs.~\eqref{eq:a:8} and \eqref{eq:a:10} gives
\begin{equation}
\label{eq:a:ratio}
\frac{a_S}{\lambda_{\phi\phi S}}=\frac{\lambda_{\phi\phi S}a_S}{\lambda_{\phi\phi S}^2}=\frac{3\pi^2}{8}N-\frac{16}{3}+O(N^{-1}).
\end{equation}
This ratio is the coefficient of the boundary identity block upon normalizing the $\langle\phi S\rangle$ crossing equation in Eq.~\eqref{eq2:21} by its leading bulk-channel coefficient.
Equations~\eqref{eq:a:9} and \eqref{eq:a:10} imply
\begin{equation}
\label{eq:a:as}
\frac{a_S}{\sqrt{N}}=\frac{3\pi}{4}-\frac{5}{3\pi N}+O(N^{-2}).
\end{equation}
Combining Eqs.~\eqref{eq:a:2}, \eqref{eq:a:11} and \eqref{eq:a:13}, we find
 \begin{equation}
  \label{eq:a:17}
 C_D|_{\mathcal{O}=\phi}=\(\frac{\Delta_{\phi}a_{\phi}}{4\pi b_{\phi D}}\)^2=\frac{9 N}{128}+O(N^{0}).
 \end{equation}
The $O(N^0)$ term is not determined by the presently available large-$N$ results, because it also depends on the unknown $O(N^{-1})$ correction to $b_{\phi D}^2$.
Since $C_D$ is independent of the choice of bulk operator, the two evaluations of Eq.~\eqref{eq:2:23} should agree: $C_D|_{\mathcal{O}=\phi}= C_D|_{\mathcal{O}=S}$.
Substituting the large-$N$ expressions in Eqs.~\eqref{eq:a:4}, \eqref{eq:a:as}, and \eqref{eq:a:17} into Eq.~\eqref{eq:2:23} for $\mathcal O=S$, we obtain
\begin{equation}
b_{SD}
=\frac{\Delta_S a_S}{4\pi\sqrt{C_D}}
=\sqrt{2}+O(N^{-1}),
\label{eq:a:bsd}
\end{equation}
where we choose the positive sign convention for $b_{SD}$.
Assuming that the contribution at $\hat{\Delta}=5$ is carried by a single operator $\hat S$, the boundary-block decomposition of the leading large-$N$ singlet two-point function in Ref.~\cite{Giombi:2020rmc} gives $b_{S\hat S}^2/b_{SD}^2=1/4$ in $d=3$.
Matching the normalization via Eq.~\eqref{eq:a:bsd} then gives
\begin{equation}
b_{S\hat S}=\frac{1}{\sqrt{2}}+O(N^{-1}).
\label{eq:a:bss}
\end{equation}
These analytic expressions can provide the reference values for the large-$N$ extrapolations of our finite-$N$ bootstrap results.
\FloatBarrier

\subsection{Numerical large-$N$ extrapolations}
\label{sec:largen:2}
We now examine the $N$-dependence of selected BCFT data extracted from the $\langle\phi\phi\rangle$ and $\langle\phi S\rangle$ bootstraps.
We fit these finite-$N$ data, extrapolate to $N\to\infty$, and compare the resulting limits with the analytic predictions summarized in Sec.~\ref{app:a}.

We use the following fitting procedure throughout this subsection unless stated otherwise.
We fit the numerical data to the three-parameter power-law form $p+qN^{-c}$, where $p$, $q$, and $c$ are fit parameters.
Here $p$ gives the extrapolated $N\to\infty$ value when $c>0$. 
When the three-parameter fitting procedure fails to converge, we fix $c=1$, corresponding to a linear fit in $1/N$.
Extrapolated values obtained from fits linear in $1/N$ are marked with an asterisk ``$*$''.
We estimate the uncertainties from two additional fits, with all data points set to the upper or lower endpoints of their quoted ranges.
Unlike in Sec.~\ref{sec:3}, the extrapolation uncertainties are not included in the errors quoted in this section.
Instead, they are encoded in the differences of the extrapolation results associated with different fit ranges.
For some important quantities, we also present plots to assess how well the fitted curves describe the finite-$N$ bootstrap data and to compare the bootstrap data with the analytic large-$N$ approximations.

\begin{figure}[tbp]
\includegraphics[width=0.48\textwidth]{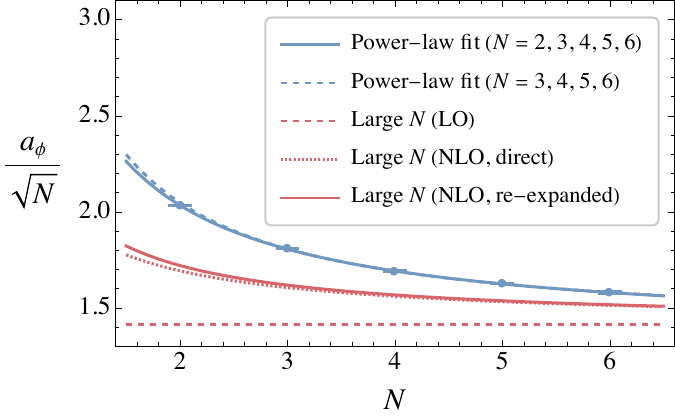}
\hfill
\includegraphics[width=0.48\textwidth]{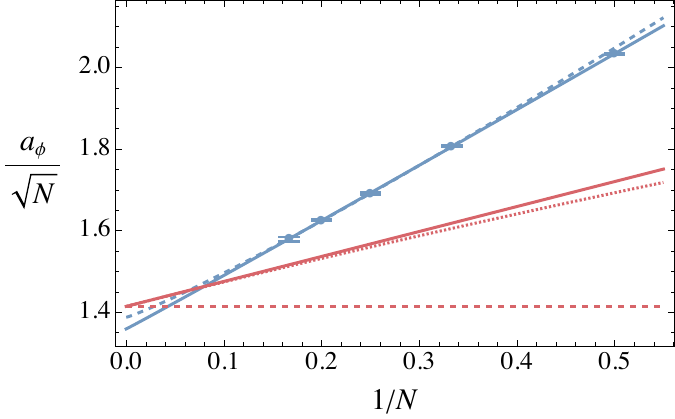}
\caption{\label{fig:4}The $N$-dependence of $\frac{a_\phi}{\sqrt{N}}$ in Table~\ref{ta:a2:1} from $\langle\phi\phi\rangle$ (blue), compared with the large-$N$ approximations (red).
Here ``LO'' denotes the leading large-$N$ result. 
``NLO, direct'' is obtained by taking the square root of Eq.~\eqref{eq:a:11} and dividing by $\sqrt{N}$, while ``NLO, re-expanded'' uses its expansion in powers of $1/N$ in Eq.~\eqref{eq:a:14}.
Blue lines are three-parameter power-law fits, and the bars indicate the estimated uncertainties.
The extrapolated $N\to\infty$ values are given in Eq.~\eqref{eq:aphi:largeN}.}
\end{figure}

We first consider the results from $\langle\phi\phi\rangle$ collected in Table~\ref{ta:a2:1}.
The results for $N=2,3,4,5$ are taken from the $\eta$-minimization study of Ref.~\cite{Hu:2025yrs}, while those for $N=6$ are computed in this work.
For the one-point coefficient $a_\phi$, the analytic expression in Eq.~\eqref{eq:a:14} shows that it scales as $\sqrt{N}$ at large $N$. 
We therefore fit $a_\phi/\sqrt{N}$ to estimate its finite large-$N$ limit.
The fitted curves in Fig.~\ref{fig:4} closely follow the bootstrap data points, providing a good description of their $N$-dependence. 
The difference between the bootstrap estimates and the large-$N$ approximations decreases with $N$. 
Motivated by this trend, we further extract the large-$N$ extrapolations from our power-law fits:
\begin{equation}
\label{eq:aphi:largeN}
 \lim_{N\to\infty}\frac{a_{\phi}}{\sqrt{N}}\approx
\begin{cases}
1.36(2)\ &\mathrm{full\ dataset}\ (N=2,3,4,5,6),\\
1.39_{-0.03}^{+0.02}\ &\mathrm{last\ 4\ points}\ (N=3,4,5,6).
\end{cases}
\end{equation}
Excluding the $N=2$ point moves the extrapolated value closer to the large-$N$ prediction $\sqrt{2}\approx1.41$. 
The remaining discrepancy may reflect sizable finite-$N$ corrections, since the available data extend only to $N=6$.

\begin{figure}[tbp]
\includegraphics[width=0.48\textwidth]{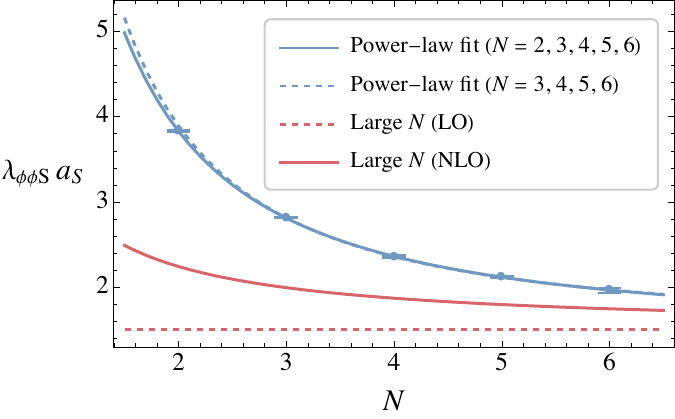}
\hfill
\includegraphics[width=0.48\textwidth]{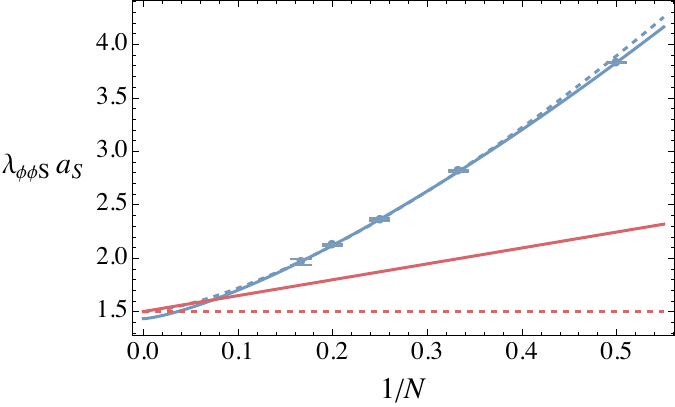}
\caption{\label{fig:24}The $N$-dependence of $\lambda_{\phi\phi S}a_S$ in Table~\ref{ta:a2:1} from $\langle\phi\phi\rangle$ (blue), compared with the large-$N$ approximations from Eq.~\eqref{eq:a:10} (red). 
Here ``LO'' and ``NLO'' denote the leading-order and next-to-leading-order large-$N$ approximations, respectively.
Blue lines are three-parameter power-law fits, and the bars indicate the estimated uncertainties.
The extrapolated $N\to\infty$ values are given in Eq.~\eqref{eq:lambda1:largeN}.}
\end{figure}

Similarly, we analyze the $N$-dependence of $\lambda_{\phi\phi S}a_S$.
This quantity provides the necessary input for the extraction of the bulk OPE and BOE coefficients. 
The fits to $\lambda_{\phi\phi S}a_S$ give
\begin{equation}
\label{eq:lambda1:largeN}
 \lim_{N\to\infty}\lambda_{\phi \phi S}a_S\approx
\begin{cases}
1.43_{-0.09}^{+0.06}\ &\mathrm{full\ dataset}\ (N=2,3,4,5,6),\\
1.50^{+0.05}_{-0.16}\ &\mathrm{last\ 4\ points}\ (N=3,4,5,6).
\end{cases}
\end{equation}
Both central values are close to the analytic large-$N$ limit of $1.50$, especially the one from the restricted fit.
The fitted curves agree very well with the bootstrap data points, as shown in Fig.~\ref{fig:24}.
The bootstrap estimates differ substantially from the large-$N$ approximation at small $N$, but move toward the analytic large-$N$ limit as $N$ increases.

\begin{figure}[tbp]
\includegraphics[width=0.47\textwidth]{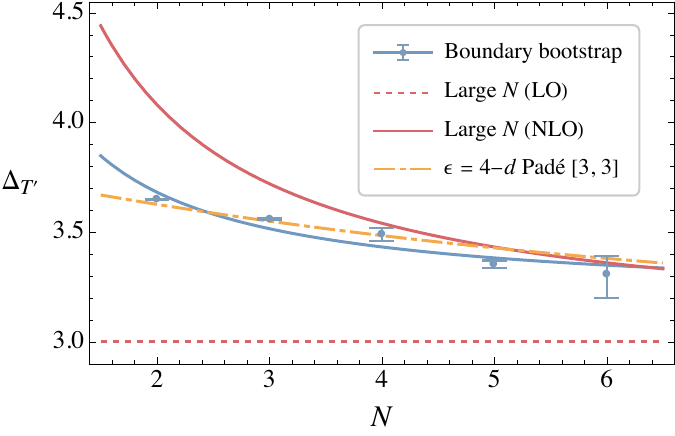}
\hfill
\includegraphics[width=0.47\textwidth]{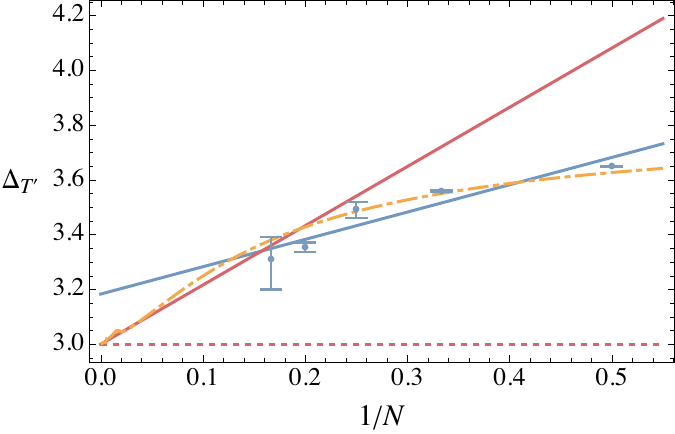}
\caption{\label{fig:5}The $N$-dependence of $\Delta_{T'}$ in Table~\ref{ta:a2:1} from $\langle\phi\phi\rangle$ (blue), compared with the large-$N$ approximations from Eq.~\eqref{eq:a:7} (red) and the $[3,3]$ Pad\'e resummation of the $\epsilon=4-d$ series \cite{Henriksson:2025hwi} (orange).
Here ``LO'' and ``NLO'' denote the leading-order and next-to-leading-order large-$N$ approximations, respectively.
The blue line is a linear fit in $1/N$, and the bars indicate the estimated uncertainties.
The extrapolated $N\to\infty$ values are given in Eq.~\eqref{eq:other:largeN}.
The $[3,3]$ Pad\'e estimates are close to the bootstrap central values for $N=2,3,4$, with relative differences below $0.6\%$.}
\end{figure}

\begin{figure}[tbp]
\includegraphics[width=0.48\textwidth]{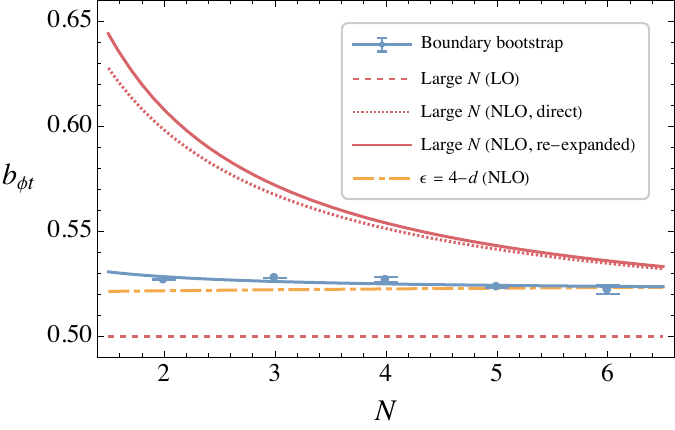}
\hfill
\includegraphics[width=0.48\textwidth]{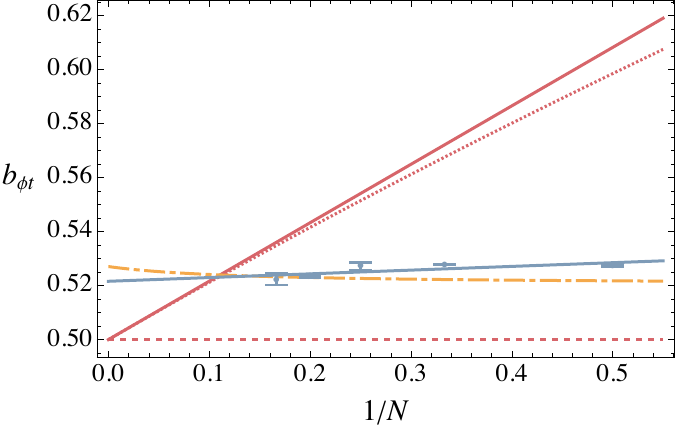}
\caption{\label{fig:6}The $N$-dependence of $b_{\phi t}$ in Table~\ref{ta:a2:1} from $\langle\phi\phi\rangle$ (blue), compared with the large-$N$ approximations (red) and the $\e=4-d$ expansions \cite{Dey:2020lwp,Padayasi:2021sik} (orange). 
Here ``LO'' denotes the leading large-$N$ result. 
``NLO, direct'' is obtained by taking the square root of Eq.~\eqref{eq:a:12}, while ``NLO, re-expanded'' uses its expansion in powers of $1/N$ in Eq.~\eqref{eq:a:15}.
The blue line is a linear fit in $1/N$, and the bars indicate the estimated uncertainties.
The extrapolated $N\to\infty$ values are given in Eq.~\eqref{eq:other:largeN}.}
\end{figure}

\begin{figure}[tbp]
\includegraphics[width=0.48\textwidth]{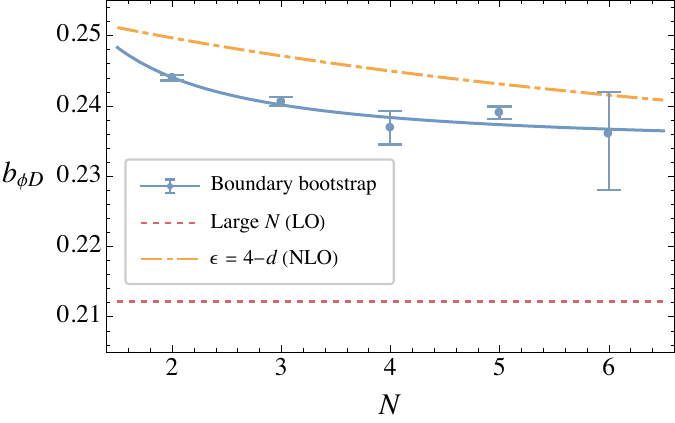}
\hfill
\includegraphics[width=0.48\textwidth]{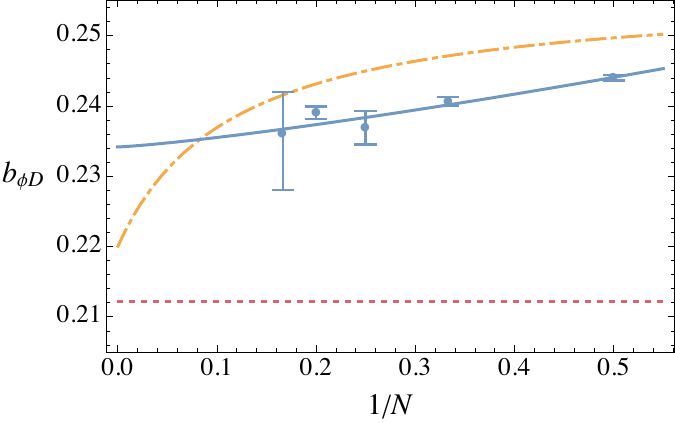}
\caption{\label{fig:7}The $N$-dependence of $b_{\phi D}$ in Table~\ref{ta:a2:1} from $\langle\phi\phi\rangle$ (blue), compared with the large-$N$ approximation from Eq.~\eqref{eq:a:13} (red) and the $\e=4-d$ expansions \cite{Dey:2020lwp,Padayasi:2021sik} (orange).
Here ``LO'' and ``NLO'' denote the leading-order and next-to-leading-order approximations.
The blue line is a three-parameter power-law fit, and the bars indicate the estimated uncertainties.
The extrapolated $N\to\infty$ values are given in Eq.~\eqref{eq:other:largeN}.}
\end{figure}

Figs.~\ref{fig:5}--\ref{fig:7} show the variation of $\Delta_{T'}$, $b_{\phi t}$, and $b_{\phi D}$ with $N$, together with the large-$N$ approximations in Eqs.~\eqref{eq:a:7}, \eqref{eq:a:12}--\eqref{eq:a:13}, and \eqref{eq:a:15}.
Agreement with the large-$N$ predictions generally improves as $N$ increases.
The fits to the bootstrap results give
\begin{equation}
\label{eq:other:largeN}
 \lim_{N\to\infty}(\Delta_{T'}, b_{\phi t}, b_{\phi D})
\approx (3.2^{*}, 0.52^{*}, 0.23).
\end{equation}
The fitted curves shown in Figs.~\ref{fig:5}--\ref{fig:7} capture the overall trends of the bootstrap data, although there are some deviations from individual data points.
The large uncertainties in the finite-$N$ data hinder precise extrapolation, so we report only the extrapolation results based on the central values.
Before rounding, the extrapolated central values differ from the corresponding large-$N$ predictions $(\Delta_{T'}, b_{\phi t}, b_{\phi D})=(3,0.5,2/(3\pi)\approx 0.21)$ by about $6\%$, $4\%$, and $10\%$, respectively.
As shown in Fig.~\ref{fig:5}, the $[3,3]$ Pad\'e approximant to the $\epsilon=4-d$ expansion of $\Delta_{T'}$ \cite{Henriksson:2025hwi} is closer to the bootstrap central values than the large-$N$ approximation at small $N$, with relative differences below $0.6\%$ for $N=2,3,4$.\footnote{However, its pole at $N\approx 192$ ($1/N\approx 0.0052$) warrants caution in interpreting its large-$N$ behavior.}
The weak $N$-dependence of $b_{\phi t}$ and $b_{\phi D}$ is consistent with the $\epsilon=4-d$ expansion \cite{Dey:2020lwp,Padayasi:2021sik}, as shown in Figs.~\ref{fig:6} and \ref{fig:7}.

\FloatBarrier

\begin{figure}[tbp]
\includegraphics[width=0.48\textwidth]{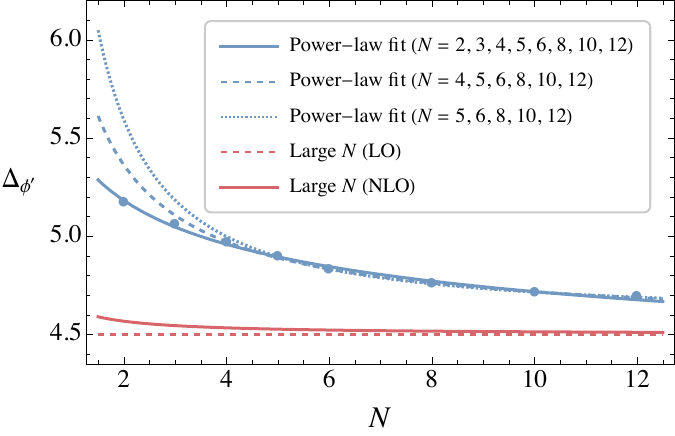}
\hfill
\includegraphics[width=0.48\textwidth]{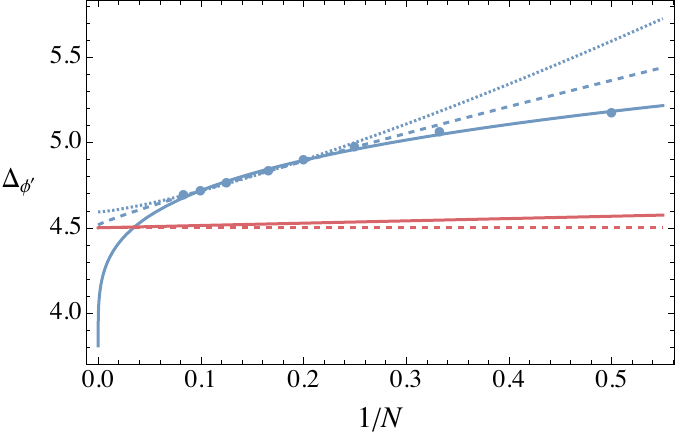}
\caption{\label{fig:16}The $N$-dependence of $\Delta_{\phi'}$ in Table~\ref{ta:5} from the $\langle\phi S\rangle$ bootstrap (blue), compared with predictions from the large-$N$ expansion from Eq.~\eqref{eq:a:2} (red). 
Here ``LO'' and ``NLO'' denote the leading-order and next-to-leading-order large-$N$ approximations, respectively.
The estimated uncertainties of the bootstrap data points are negligible on the scale of the plot and are omitted for clarity.
The blue lines are three-parameter power-law fits.
The extrapolated $N\to\infty$ values from these fits are given in Eqs.~\eqref{eq:deltaphiN:1}--\eqref{eq:deltaphiN:2}.}
\end{figure}

We next turn to the results from the $\langle\phi S\rangle$ bootstrap listed in Table~\ref{ta:5}.
We first fit our bootstrap estimates of $\Delta_{\phi'}$ as a function of $N$ and compare them with the large-$N$ expansion in Eq.~\eqref{eq:a:2}, as shown in Fig.~\ref{fig:16}.
Our bootstrap estimates move closer to the analytic large-$N$ predictions as $N$ increases.
Although the fitted curves agree well with the data points used in each fit, they vary with the choice of data points to some extent.
The fit using the full dataset $N=2,3,4,5,6,8,10,12$ gives
\begin{equation}
\label{eq:deltaphiN:1}
\Delta_{\phi'}|_{N\to\infty}^{\mathrm{full\ dataset}}\approx 3.79_{-0.18}^{+0.08}.
\end{equation}
The result shows a noticeable discrepancy from the analytic large-$N$ limit $9/2=4.50$. 
To examine the dependence of the extrapolation on the choice of data points, we perform several fits using different subsets of the available data. We obtain
\begin{equation}
\label{eq:deltaphiN:2}
 \lim_{N\to\infty}\Delta_{\phi'}\approx
\begin{cases}
4.36_{-0.12}^{+0.06}&\mathrm{fit\ with}\ N=3,4,5,6,8,10,12,\\
4.52_{-0.06}^{+0.03}&\mathrm{fit\ with}\ N=4,5,6,8,10,12,\\
4.59_{-0.06}^{+0.04}&\mathrm{fit\ with}\ N=5,6,8,10,12.
\end{cases}
\end{equation}
Excluding the $N=2,3$ data points brings the extrapolated values closer to the large-$N$ prediction.\footnote{We also repeat the fits after excluding the largest-$N$ point ($N=12$). 
Fits using $N=3,4,5,6,8,10$, $N=4,5,6,8,10$, and $N=5,6,8,10$ give extrapolated values of $4.11(13)$, $4.41(3)$, and $4.54(3)$, respectively.
Excluding $N=12$ improves agreement with the analytic limit only in the last case.}
However, this tendency does not persist if the $N=4$ point is omitted, as the fit then relies more on the less precise data at larger $N$.
The differences among these extrapolated values indicate an appreciable dependence on the fit range. 
In the following analysis, we compare the full-dataset fits with fits restricted to high-$N$ subsets.

For $\Delta_{\phi''}$ and $\Delta_{\phi'''}$, we obtain
\begin{equation}
\lim_{N\to\infty}(\Delta_{\phi''},\Delta_{\phi'''})\approx
\begin{cases}
(7.12_{-0.07}^{+0.05},10.11_{-0.13}^{+0.10})^{*} &\mathrm{full\ dataset}\ (N=2,3,4,5,6,8,10,12),\\
(7.03_{-0.09}^{+0.06},10.04_{-0.18}^{+0.11})^{*} &\mathrm{last\ 6\ points}\ (N=4,5,6,8,10,12).
\end{cases}
\end{equation}
These values should be regarded as effective scaling dimensions for groups of operators around the large-$N$ levels at $6.5$ and $8.5$.
The deviations of these extrapolated values from the corresponding large-$N$ levels increase at higher scaling dimensions.
These deviations may reflect contributions from closely spaced operators that are not separately resolved.

\begin{figure}[tbp]
\centering 
\captionsetup[subfigure]{skip=1pt}
\setlength{\lineskip}{1em}
\begin{subfigure}[t]{\textwidth}
\includegraphics[width=0.48\textwidth]{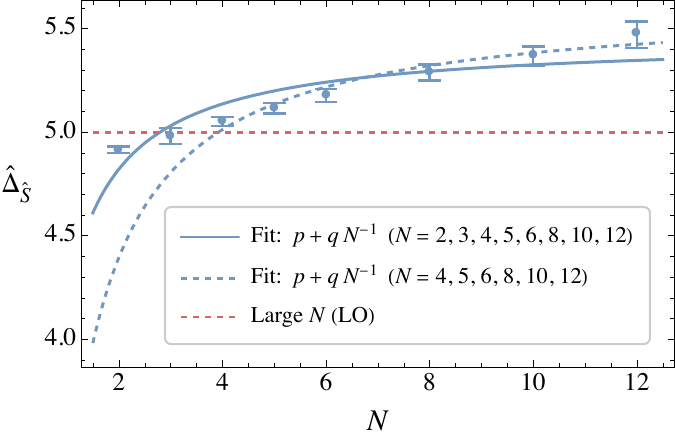}
\hfill
\includegraphics[width=0.48\textwidth]{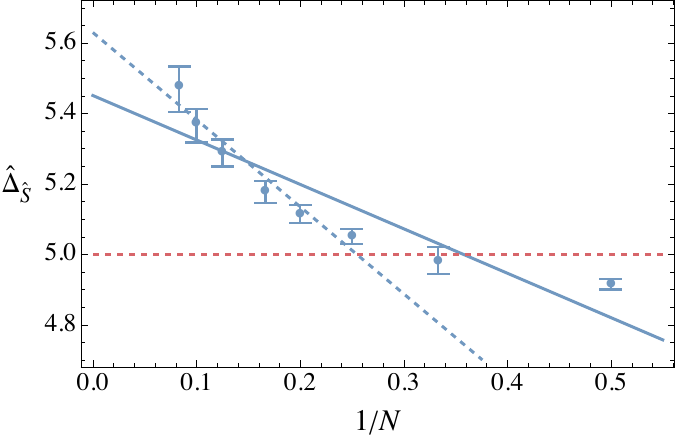}
\caption{$\hat{\Delta}_{\hat{S}}$}
\end{subfigure}
\hfill
\begin{subfigure}[t]{\textwidth}
\includegraphics[width=0.48\textwidth]{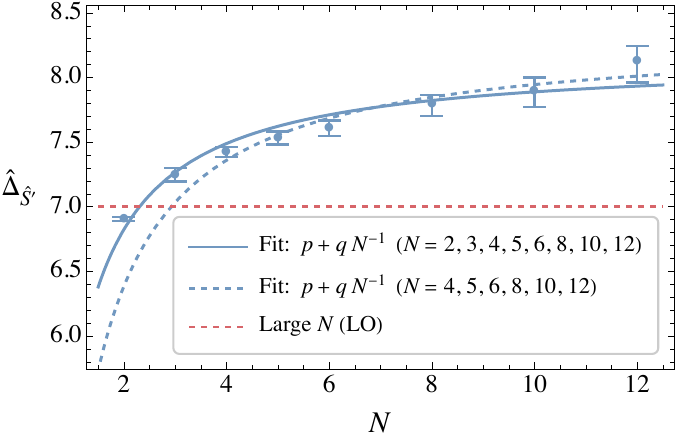}
\hfill
\includegraphics[width=0.48\textwidth]{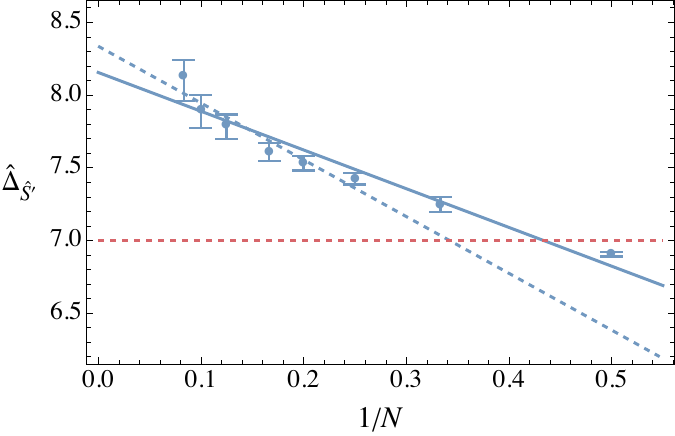}
\caption{$\hat{\Delta}_{\hat{S}'}$}
\end{subfigure}
\caption{\label{fig:delta:N}The $N$-dependence of $\hat{\Delta}_{\hat{S}}$ and $\hat{\Delta}_{\hat{S'}}$ in Table~\ref{ta:5} from the $\langle\phi S\rangle$ bootstrap (blue), compared with predictions from the large-$N$ expansion in Eq.~\eqref{eq:a:hatS} (red).
The blue lines are linear fits in $1/N$, and the bars indicate the estimated uncertainties.
The extrapolated $N\to\infty$ values from these fits are given in Eq.~\eqref{eq:largeN:delta}.}
\end{figure}

The extrapolations of $\hat{\Delta}_{\hat S}$ and $\hat{\Delta}_{\hat S'}$ give the estimates
\begin{equation}
\label{eq:largeN:delta} 
\lim_{N\to\infty}(\hat{\Delta}_{\hat{S}},\hat{\Delta}_{\hat{S}'})\approx 
\begin{cases}
(5.45(6),8.15_{-0.14}^{+0.11})^{*} &\mathrm{full\ dataset}\ (N=2,3,4,5,6,8,10,12),\\
(5.63(9),8.33_{-0.20}^{+0.15})^{*} &\mathrm{last\ 6\ points}\ (N=4,5,6,8,10,12).
\end{cases}
\end{equation}
The estimates all lie above the leading large-$N$ values $5$ and $7$ in Eq.~\eqref{eq:a:hatS}.
As shown in Fig.~\ref{fig:delta:N}, the fits to the full dataset show some deviations from the data points, 
which is more clearly seen in the plots against $1/N$.\footnote{We have not found better power-law fits that both agree well with most of the data points and approach finite limits as $N\to\infty$.}
Over the range of $N$ studied, both $\hat{\Delta}_{\hat S}$ and $\hat{\Delta}_{\hat S'}$ tend to move further from their leading large-$N$ values as $N$ increases.
This unusual behavior may indicate a non-monotonic dependence on $N$.
If so, the extrapolations would be particularly sensitive to the fit range. 
More reliable extrapolations may require data at larger values of $N$.

\begin{figure}[tbp]
\includegraphics[width=0.48\textwidth]{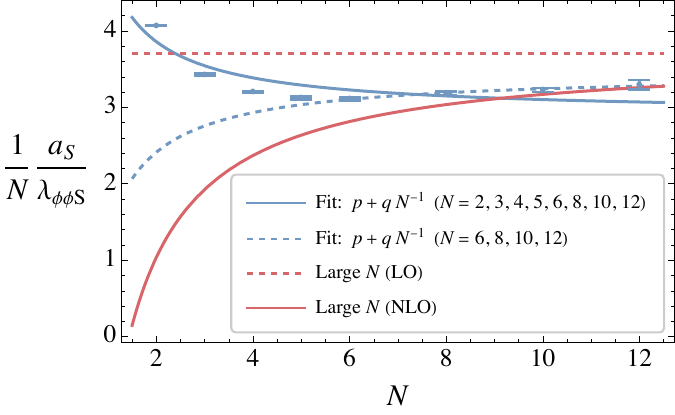}
\hfill
\includegraphics[width=0.48\textwidth]{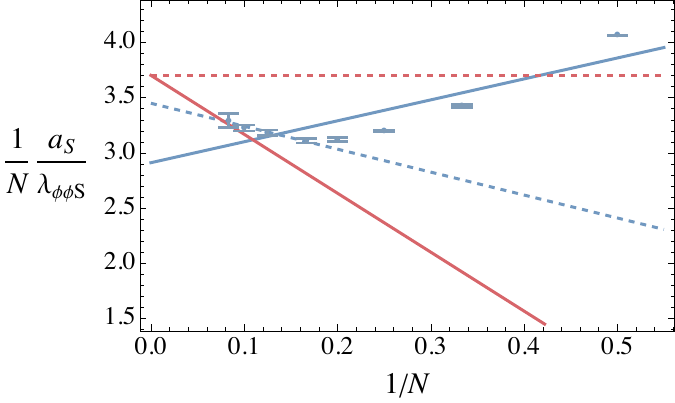}
\caption{\label{fig:largen:ratio}The $N$-dependence of $a_S/(N\lambda_{\phi\phi S})$ obtained from
the $\langle\phi S\rangle$ bootstrap (blue), compared with predictions from the large-$N$ approximations in Eq.~\eqref{eq:a:ratio} (red).
The bootstrap results for $a_S/\lambda_{\phi\phi S}$ are listed in Table~\ref{ta:8}.
Here ``LO'' and ``NLO'' denote the leading-order and next-to-leading-order large-$N$ approximations, respectively.
The blue lines are linear fits in $1/N$, and the bars indicate the estimated uncertainties.
The extrapolated $N\to\infty$ values are given in Eq.~\eqref{eq:largeN:ratio}.}
\end{figure}

We next consider the ratio $a_S/\lambda_{\phi\phi S}$, which is listed in Table~\ref{ta:8}.
Its large-$N$ expansion in Eq.~\eqref{eq:a:ratio} predicts linear growth with $N$.
We therefore consider the rescaled ratio $a_S/(N\lambda_{\phi\phi S})$, which has a finite large-$N$ limit. 
The fits give
\begin{equation}
\label{eq:largeN:ratio} 
\lim_{N\to\infty}\frac{1}{N}\frac{a_S}{\lambda_{\phi\phi S}}\approx 
\begin{cases}
2.91(5)^{*} &\mathrm{full\ dataset}\ (N=2,3,4,5,6,8,10,12),\\
3.45(9)^{*} &\mathrm{last\ 4\ points}\ (N=6,8,10,12).
\end{cases}
\end{equation}
As shown in Fig.~\ref{fig:largen:ratio}, the bootstrap estimates of $a_S/(N\lambda_{\phi\phi S})$ show a non-monotonic dependence on $N$, which is more apparent in the plot against $1/N$.
Even so, the last four central values move toward the leading large-$N$ prediction.
We therefore also perform a restricted fit using these four data points at $N=6,8,10,12$.
The restricted fit lies about $7\%$ below the analytic large-$N$ limit $3\pi^2/8\approx 3.70$ in Eq.~\eqref{eq:a:ratio}.
Relative to the full-dataset fit, the restricted fit agrees more closely with the analytic prediction. 
This indicates sizable finite-$N$ corrections at small $N$.

\FloatBarrier

\begin{figure}[tbp]
\centering 
\captionsetup[subfigure]{skip=1pt}
\setlength{\lineskip}{1em}
\begin{subfigure}[t]{\textwidth}
\includegraphics[width=0.48\textwidth]{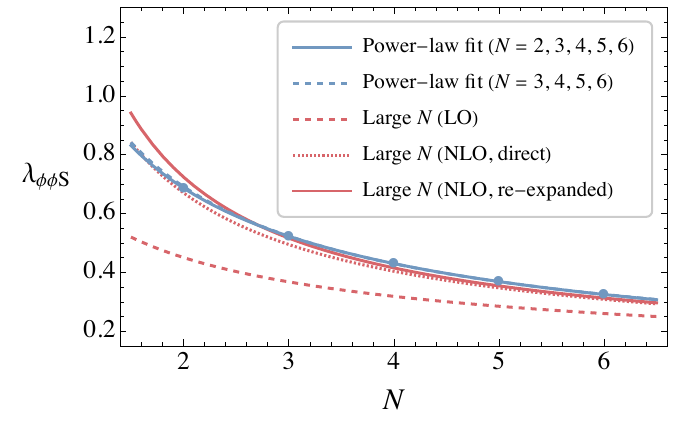}
\hfill
\includegraphics[width=0.48\textwidth]{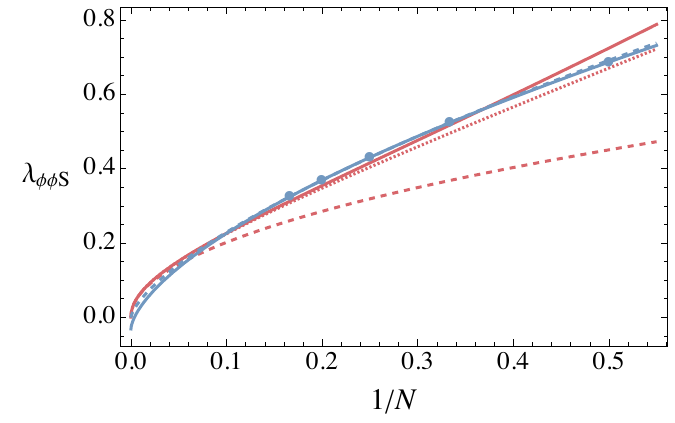}
\caption{\label{fig:25:1}$\lambda_{\phi\phi S}$}
\end{subfigure}
\hfill
\begin{subfigure}[t]{\textwidth}
\includegraphics[width=0.48\textwidth]{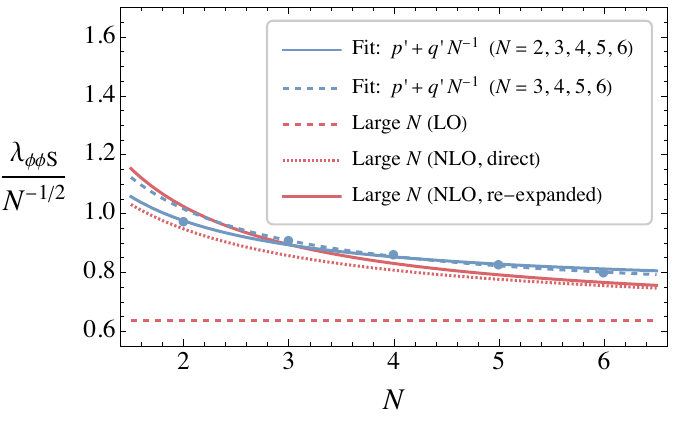}
\hfill
\includegraphics[width=0.48\textwidth]{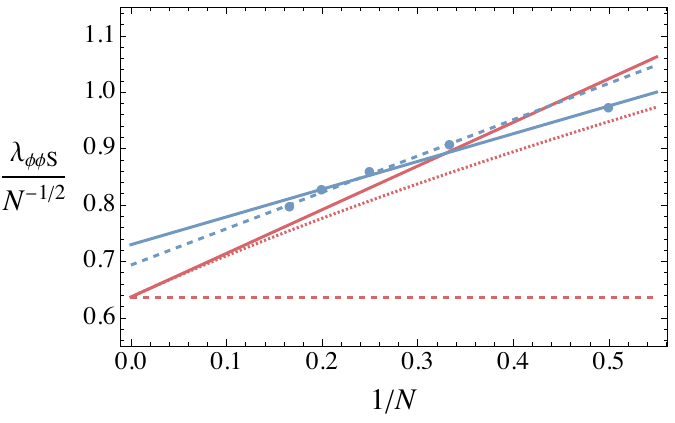}
\caption{$\lambda_{\phi\phi S}/(N^{-1/2})$}
\label{fig:25:2}
\end{subfigure}
\caption{\label{fig:25}The $N$-dependence of $\lambda_{\phi\phi S}$ in Table~\ref{ta:7} (blue), compared with predictions from the large-$N$ expansion (red). 
Here ``LO'' denotes the leading large-$N$ result. 
``NLO, direct'' is obtained by taking the square root of Eq.~\eqref{eq:a:8}, while ``NLO, re-expanded'' uses its expansion in powers of $1/N$ in Eq.~\eqref{eq:a:9}.
For the three-parameter power-law fits in (a), the extrapolated $N\to\infty$ values and the fitted exponents are given in Eqs.~\eqref{eq:largeN:lambda:1}--\eqref{eq:largeN:lambda:2}.
For (b), the blue lines are linear fits in $1/N$, and the extrapolated $N\to\infty$ values are given in Eq.~\eqref{eq:largeN:lambda:3}.
The estimated uncertainties of the data points are negligible on the scale of the plot and are omitted for clarity.}
\end{figure}

Finally, we examine the $N$-dependence of several expansion coefficients obtained by combining the results from the $\langle\phi\phi\rangle$ and $\langle\phi S\rangle$ bootstrap analyses. 
The numerical results are listed in Table~\ref{ta:7}.
We first consider the bulk OPE coefficient $\lambda_{\phi\phi S}$, which decreases with $N$ over the range studied here.
As shown in Fig.~\hyperref[fig:25:1]{\ref*{fig:25}(a)}, the fitted curves lie very close to the NLO large-$N$ approximations.
The fits to $\lambda_{\phi\phi S}$ give
\begin{equation}
\label{eq:largeN:lambda:1}
\lim_{N\to\infty}\lambda_{\phi\phi S}\approx
\begin{cases}
-0.032_{-0.022}^{+0.012}&\mathrm{full\ dataset}\ (N=2,3,4,5,6),\\
-0.001_{-0.055}^{+0.022}&\mathrm{last\ 4\ points}\ (N=3,4,5,6).
\end{cases}
\end{equation}
Both central estimates are close to zero, as expected from the absence of a constant term in the large-$N$ expansion. The quoted uncertainty interval for the latter result includes zero.
The corresponding fitted exponents associated with $\lambda_{\phi\phi S}$ are
\begin{equation}
\begin{split}
\label{eq:largeN:lambda:2}
c
&=-\lim_{N\to\infty}\log_{N}{\(\lambda_{\phi\phi S}-\lim_{N\to\infty}\lambda_{\phi\phi S}\)}\\
&\approx
\begin{cases}
0.638_{-0.021}^{+0.009}&\mathrm{full\ dataset}\ (N=2,3,4,5,6),\\
0.685_{-0.071}^{+0.030}&\mathrm{last\ 4\ points}\ (N=3,4,5,6).
\end{cases}
\end{split}
\end{equation}
The fitted exponents are somewhat larger than the leading large-$N$ exponent $1/2$, possibly due to subleading finite-$N$ corrections.
To incorporate the analytic large-$N$ scaling explicitly, we also fit $\lambda_{\phi\phi S}$ to the form $(p'+q'N^{-1})N^{-1/2}$.
Here $p'$ and $q'$ are fit parameters, with $p'$ giving the coefficient of the leading $N^{-1/2}$ term.
Although such fits shown in Fig.~\hyperref[fig:25:2]{\ref*{fig:25}(b)} do not describe the data as well as those in Fig.~\hyperref[fig:25:1]{\ref*{fig:25}(a)}, they allow us to estimate the coefficient of the leading $N^{-1/2}$ term.
We obtain
\begin{equation}
\label{eq:largeN:lambda:3}
\lim_{N\to\infty}\frac{\lambda_{\phi\phi S}}{N^{-1/2}}\approx
\begin{cases}
0.729(9)^{*} &\mathrm{full\ dataset}\ (N=2,3,4,5,6),\\
0.693(11)^{*} &\mathrm{last\ 4\ points}\ (N=3,4,5,6).
\end{cases}
\end{equation}
The fit excluding the $N=2$ point gives a leading amplitude closer to the analytic large-$N$ coefficient $2/\pi\approx0.637$ predicted in Eq.~\eqref{eq:a:9}.

\begin{figure}[tbp]
\includegraphics[width=0.48\textwidth]{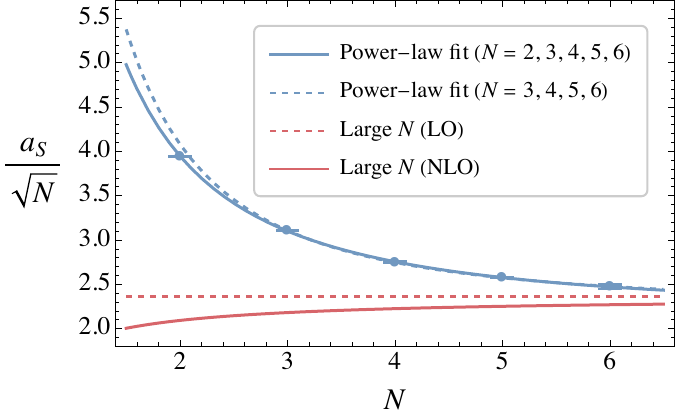}
\hfill
\includegraphics[width=0.48\textwidth]{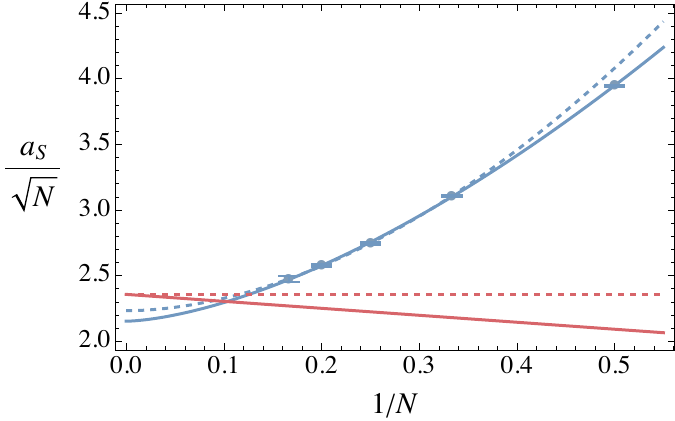}
\caption{\label{fig:add:as}The $N$-dependence of $a_S/\sqrt{N}$ in Table~\ref{ta:7} (blue), compared with predictions from the large-$N$ expansion from Eq.~\eqref{eq:a:as} (red). 
Here ``LO'' and ``NLO'' denote the leading-order and next-to-leading-order large-$N$ approximations, respectively.
The blue lines are three-parameter power-law fits, and the bars indicate the estimated uncertainties.
The extrapolated $N\to\infty$ values from these fits are given in Eq.~\eqref{eq:largeN:as}.} 
\end{figure}

We now turn to the one-point function coefficient $a_S$.
Although our estimates of $a_S$ in Table~\ref{ta:7} exhibit a non-monotonic dependence on $N$, the rescaled quantity $a_S/\sqrt{N}$ displays a more regular $N$-dependence, as shown in Fig.~\ref{fig:add:as}.
At small $N$, our bootstrap estimates are much higher than the large-$N$ predictions, but the difference decreases considerably for larger $N$.
Fitting the rescaled quantity $a_S/\sqrt{N}$, we obtain
\begin{equation}
\label{eq:largeN:as}
\lim_{N\to\infty}\frac{a_S}{\sqrt{N}}\approx
\begin{cases}
2.15(5)&\mathrm{full\ dataset}\ (N=2,3,4,5,6),\\
2.23_{-0.09}^{+0.05}&\mathrm{last\ 4\ points}\ (N=3,4,5,6).
\end{cases}
\end{equation}
The extrapolated central values are reasonably close to the leading large-$N$ prediction $3\pi/4\approx 2.36$, lying about $5\%$--$9\%$ below it.

\begin{figure}[tbp]
\includegraphics[width=0.48\textwidth]{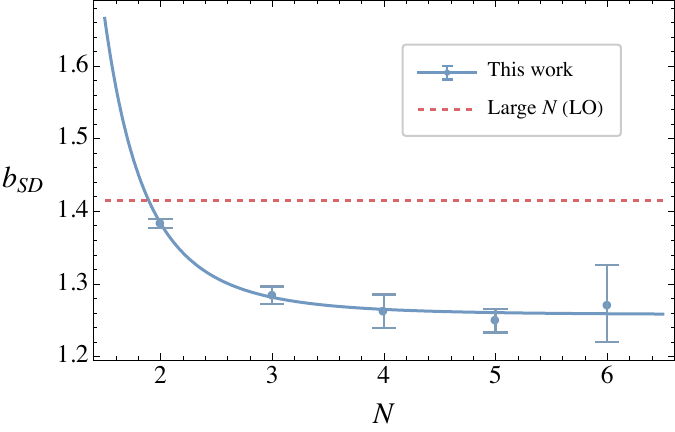}
\hfill\includegraphics[width=0.48\textwidth]{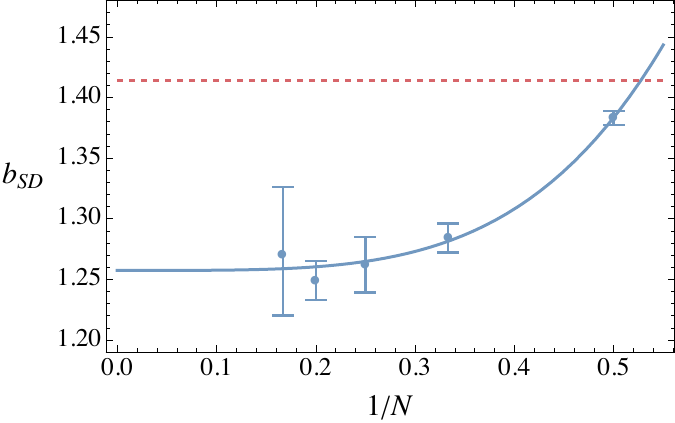}
\caption{\label{fig:add:bsd}The $N$-dependence of $b_{SD}$ in Table~\ref{ta:7} (blue), compared with the leading large-$N$ prediction from Eq.~\eqref{eq:a:bsd} (red).
The blue line is a three-parameter power-law fit, and the bars indicate the estimated uncertainties.}
\end{figure}

For $b_{SD}$, we fit the central estimates at $N=2,3,4,5,6$ to examine the overall trend.
This fit gives a finite and nonzero extrapolated value
\begin{equation}
\lim_{N\to\infty}b_{SD}\approx
1.257.
\end{equation}
The uncertainty in the $N=6$ data is too large to reliably estimate the error in this extrapolated value.
The extrapolated central value is about $11\%$ below the analytic prediction $\sqrt{2}\approx1.414$ in Eq.~\eqref{eq:a:bsd}.
This difference and the trend in the data shown in Fig.~\ref{fig:add:bsd} suggest that $b_{SD}$ may vary non-monotonically with $N$.
A more reliable analysis of the large-$N$ behavior would require extending the calculation to larger $N$.

Taken together, the broad agreement with the expected large-$N$ trends provides indirect support for the reliability of our finite-$N$ bootstrap results. 
The remaining deviations indicate sizable finite-$N$ corrections and sensitivity to the fit range.

\FloatBarrier
\section{Discussion}
\label{sec:4}
In this paper, we performed separate bootstrap analyses of bulk correlators $\langle\phi\phi\rangle$, $\langle\phi S\rangle$, and $\langle SS\rangle$ at the normal transition of the three-dimensional $\mathrm{O}(N)$ model.
We numerically estimated the corresponding low-lying bulk and boundary spectra, together with the coefficients entering those channels.
We extracted bulk one-point, BOE, and bulk OPE coefficients for several low-lying operators. 
These include estimates of $a_S$, $\lambda_{\phi\phi S}$, $b_{SD}$, and $b_{S\hat S}$ for $N=2,3,4,5,6$.
Some of the bulk OPE coefficients are not currently accessible through conventional studies of four-point functions in CFTs without a boundary.
For $N=2,3$, the results for $a_S$ and $\lambda_{\phi\phi S}$ provide useful comparisons with existing estimates.
We also obtained the bulk ratio $\lambda_{SSS'}/\lambda_{\phi\phi S'}$ for $N=1,2,3$, where the $N=1$ result agrees closely with the estimates in the literature.
We evaluated $C_t$ in Eq.~\eqref{eq:2:24} and $\alpha$ in Eq.~\eqref{eq:2:4} for $N=6$.
Using our $N=6$ result for $\alpha$ and the $N=2,3,4,5$ results of Ref.~\cite{Hu:2025yrs}, we re-estimated $N_c$ and also obtained a value close to 5.
Stability across fitting forms and agreement with previous estimates \cite{Padayasi:2021sik,Hu:2025yrs} further support the conclusion that $N_c\approx5$.
Our new $N=6$ estimate of $C_D|_{\mathcal O=\phi}$ in Eq.~\eqref{eq:3:3} extends the results reported in Ref.~\cite{Hu:2025yrs} for $N=2,3,4,5$.
Combining the $\langle\phi\phi\rangle$ and $\langle\phi S\rangle$ results gave the estimates of $C_D|_{\mathcal O=S}$ in Eq.~\eqref{eq:3:11}.
The agreement between the two extractions provides a self-consistency check of the extracted BCFT data.
Using the operator independence of $C_D$, we also obtained new estimates for $N=8,10,12$ directly from the mixed-correlator bootstrap via Eq.~\eqref{eq:cd:mixed}.
Finally, we analyzed the $N$-dependence of some quantities and compared their large-$N$ extrapolations with analytic large-$N$ predictions.

A remaining limitation of the present analysis is that some sources of uncertainty were not fully quantified.
The quoted uncertainty ranges accounted only for variations in the input data and the extrapolation procedure.
As discussed in Sec.~\ref{sec:3:3}, the quoted uncertainty ranges for several quantities did not overlap across the different correlator analyses.
This suggests that the present procedure may underestimate the total uncertainties in these quantities.
As the operator spectrum becomes denser at higher scaling dimensions, the separate minimization performed for each correlator may not resolve individual operators. 
This difficulty in distinguishing nearby operators may contribute to an underestimation of the uncertainties.
Additionally, the choice of truncation family may also have some effect on our estimates (see Sec.~\ref{sec:2:3} and Appendix~\ref{app:c:2}).
Despite this limitation, the low-lying BCFT data obtained here may provide useful input for further studies.
One possible application is to extend the test of conformal invariance proposed in Ref.~\cite{Podo:2026hfh} to the normal boundary universality class.

While the present results support the effectiveness of the truncated bootstrap with $\eta$-minimization in BCFTs, there remain several interesting directions for future work. 
\begin{itemize}
\item
A natural extension is to solve the crossing equations for different two-point functions simultaneously.
This could provide stronger constraints on the extracted BCFT data.
We could also study additional two-point functions involving the traceless symmetric scalar and the antisymmetric spin-one current operator. Along this line, we may further consider the correlators involving the spin-two stress tensor.
\item
Another direction is to study other boundary universality classes, such as the ordinary and special transitions.
Previous work on these transitions includes numerical boundary bootstrap analyses in Refs.~\cite{Liendo:2012hy,Gliozzi:2015qsa,Gliozzi:2016cmg,Toldin:2023fny}.
More recently, Ref.~\cite{Kangsabanik:2026tdm} applied the truncated Polyakov bootstrap to the special transition of the 3d Ising model and reported more precise estimates.
These studies may provide data for comparison with results from such an extension.
\item
It would also be worthwhile to study the anisotropic special transition \cite{Diehl:1996kd,Diehl:1984}, which is relevant for systems with anisotropic surface interactions.
The anisotropic special boundary condition breaks $\mathrm{O}(N)$ to $\mathrm{O}(M)\times \mathrm{O}(N-M)$, where some components exhibit special-like behavior and others exhibit ordinary-like behavior \cite{Diehl:1984}.
\item
In addition, the method can be generalized to situations involving fermionic degrees of freedom. 
A representative example is boundary criticality associated with Gross-Neveu-Yukawa (GNY) theories \cite{Giombi:2021cnr,Herzog:2022jlx,Shen:2024pve,Jiang:2025sfb,Ge:2025cmk,Diatlyk:2026gab,Diatlyk:2026eta}.
Existing estimates of low-lying CFT data in Refs.~\cite{Erramilli:2022kgp,Atanasov:2022bpi,Mitchell:2024hix} could provide useful input for such an extension.
\end{itemize}

Further developments of $\eta$-minimization could open the way to a broader range of applications, particularly in bootstrap problems where positivity may be absent.
Possible extensions of $\eta$-minimization include incorporating artificial-intelligence algorithms \cite{Kantor:2021kbx,Kantor:2021jpz,Kantor:2022epi,Niarchos:2023lot,Niarchos:2025cdg,Niarchos:2026vow}, analytic input for high-lying contributions \cite{Li:2023tic,Poland:2023bny,Barrat:2025wbi,Poland:2025ide}, random weights \cite{Poland:2023bny,Barrat:2025wbi,Poland:2025ide,Poland:2023vpn},
or more sophisticated weights \cite{Li:2021uki,Laio:2022ayq,Benjamin:2026lbj}. 
The reliability of such residual-minimization approaches is supported by the stabilization of low-lying numerical estimates with increasing truncation order, as also observed in related truncated-bootstrap studies \cite{Li:2024ggr,Kangshabanik:2026onk}.
A particularly promising application is the study of non-unitary CFTs, where positivity constraints are generally unavailable.
Such theories can be used to study the critical behavior of the $\mathrm{O}(N)$ loop model with noninteger or nonpositive $N$, the Yang-Lee edge singularity, percolation, and disordered systems \cite{Gliozzi:2013ysa,Gliozzi:2014jsa,Shimada:2015gda,Hogervorst:2016itc,Hikami:2017hwv,Hikami:2017sbg,Leclair:2018trn,Hikami:2018qpz,Padayasi:2023hpd,Shimamori:2025afk}.

\begin{acknowledgments}
We thank Yongwei Guo and Runzhe Hu for helpful discussions.
We are grateful to David Poland for suggesting the extension of the $\langle\phi\phi\rangle$ analysis to $N=6$.
This work was supported by the National Natural Science Foundation of China (Grant Nos. 12522504 and 12205386).
The authors thank the Yukawa Institute for Theoretical Physics at Kyoto University, where part of this work was initiated during the workshop ``Progress of Theoretical Bootstrap''.
\end{acknowledgments}

\appendix
\section{Supplementary technical details}
\label{app:tech}
This appendix summarizes the input values and truncation choices used in our bootstrap.
In the main text, Sec.~\ref{sec:2:3} presents concrete examples of the extrapolation step for the $N=6$ $\langle\phi\phi\rangle$ bootstrap. 
Here we provide additional examples for the $\langle\phi S\rangle$ and $\langle SS\rangle$ bootstraps.

\begin{figure}[tbp]
\captionsetup[subfigure]{
  margin={15pt,0pt},
  justification=centering,
  singlelinecheck=false
}
\captionsetup[subfigure]{skip=2pt}
\setlength{\lineskip}{0.5 em}
\begin{subfigure}[t]{0.48\textwidth}
\centering
\includegraphics[width=\textwidth]{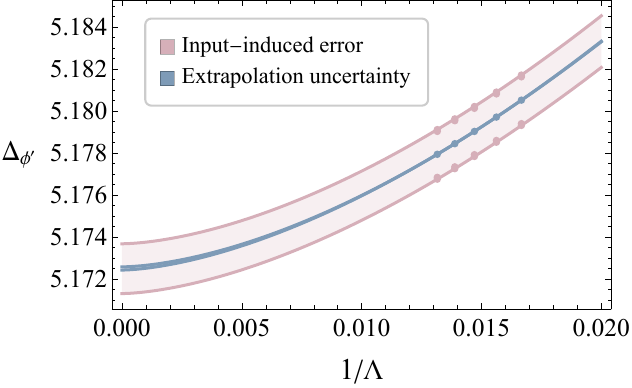}
\caption{$N=2$}
\end{subfigure}
\hfill
\begin{subfigure}[t]{0.48\textwidth}
\centering
\includegraphics[width=\textwidth]{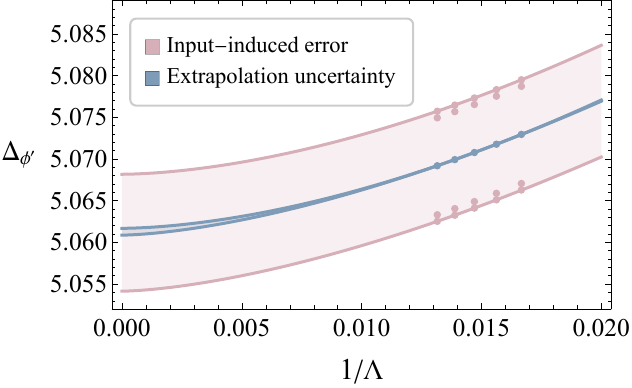}
\caption{$N=3$}
\end{subfigure}
\hfill
\begin{subfigure}[t]{0.48\textwidth}
\centering
\includegraphics[width=\textwidth]{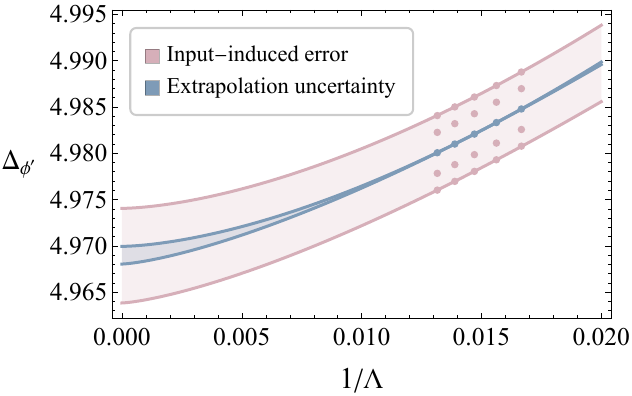}
\caption{$N=4$}
\end{subfigure}
\hfill
\begin{subfigure}[t]{0.48\textwidth}
\centering
\includegraphics[width=\textwidth]{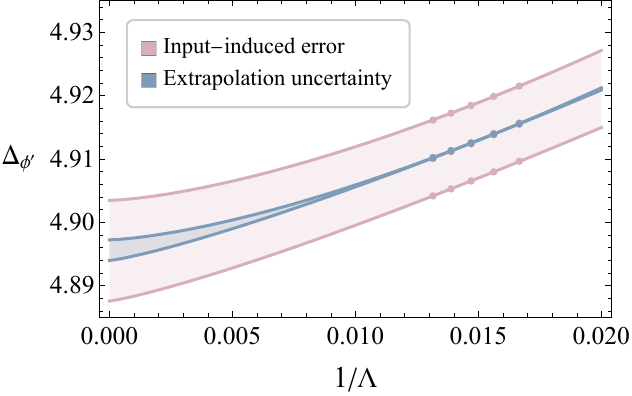}
\caption{$N=5$}
\end{subfigure}
\hfill
\begin{subfigure}[t]{0.48\textwidth}
\centering
\includegraphics[width=\textwidth]{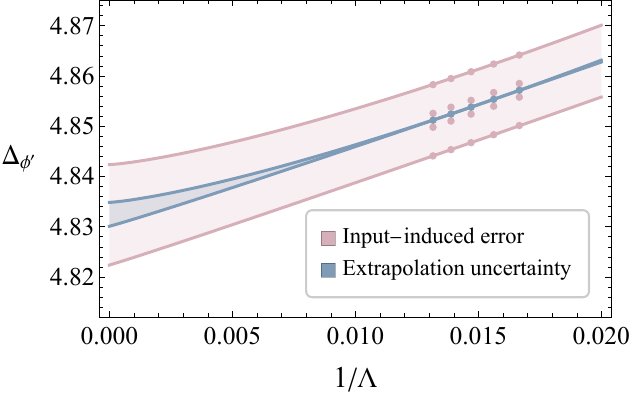}
\caption{$N=6$}
\end{subfigure}
\hfill
\begin{subfigure}[t]{0.48\textwidth}
\centering
\includegraphics[width=\textwidth]{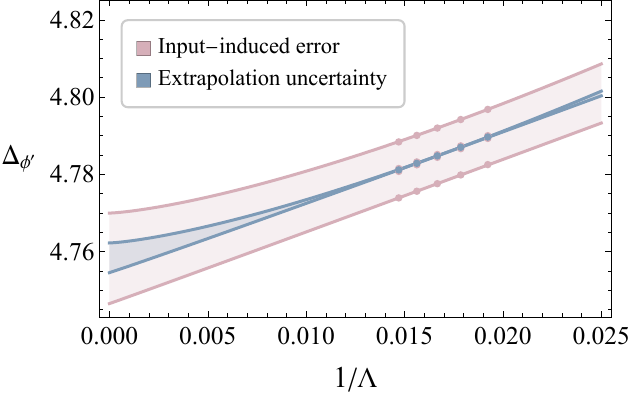}
\caption{$N=8$}
\end{subfigure}
\hfill
\begin{subfigure}[t]{0.48\textwidth}
\centering
\includegraphics[width=\textwidth]{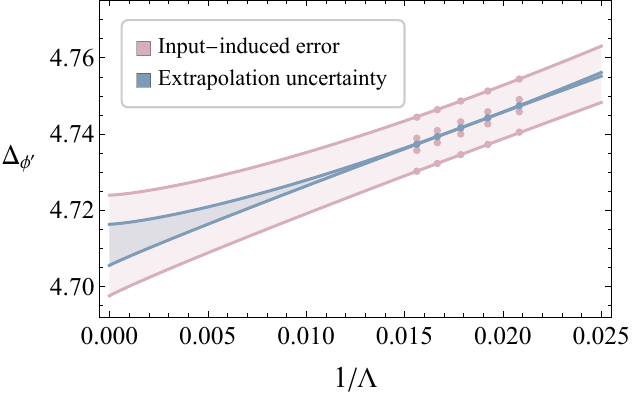}
\caption{$N=10$}
\end{subfigure}
\hfill
\begin{subfigure}[t]{0.48\textwidth}
\centering
\includegraphics[width=\textwidth]{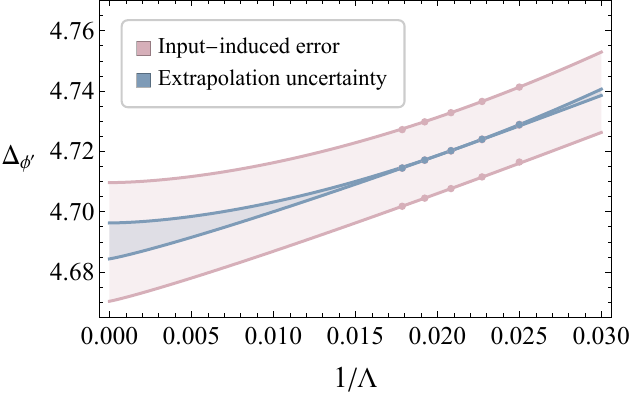}
\caption{$N=12$}
\end{subfigure}
\caption{\label{fig:Deltaphi2}Power-law fits of $\Delta_{\phi'}$ from the $\langle\phi S\rangle$ bootstrap for various $N$.
}
\end{figure}

\begin{figure}[tbp]
\captionsetup[subfigure]{skip=2pt}
\captionsetup[subfigure]{
  margin={15pt,0pt},
  justification=centering,
  singlelinecheck=false
}
\setlength{\lineskip}{0.5 em}
\begin{subfigure}[t]{0.48\textwidth}
\centering
\includegraphics[width=\textwidth]{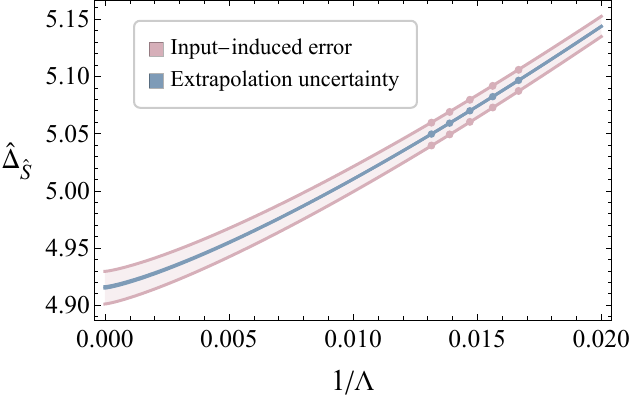}
\caption{$N=2$}
\end{subfigure}
\hfill
\begin{subfigure}[t]{0.48\textwidth}
\centering
\includegraphics[width=\textwidth]{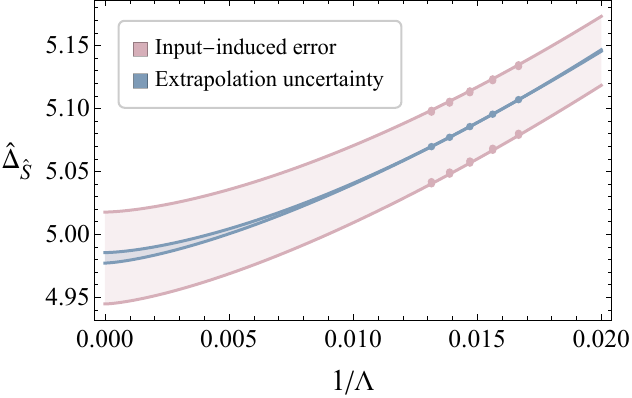}
\caption{$N=3$}
\end{subfigure}
\hfill
\begin{subfigure}[t]{0.48\textwidth}
\centering
\includegraphics[width=\textwidth]{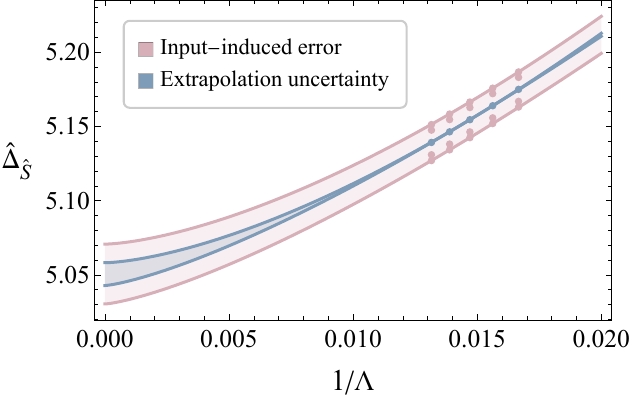}
\caption{$N=4$}
\end{subfigure}
\hfill
\begin{subfigure}[t]{0.48\textwidth}
\centering
\includegraphics[width=\textwidth]{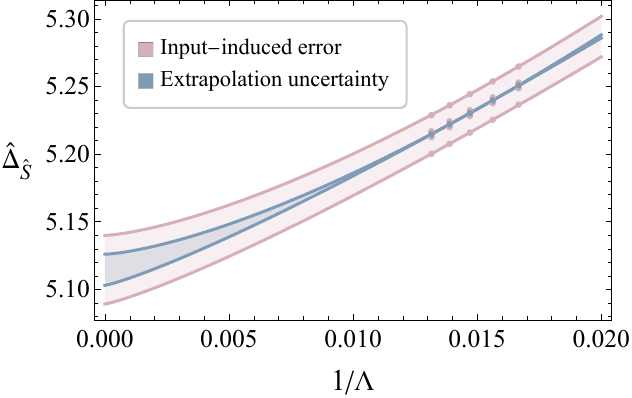}
\caption{$N=5$}
\end{subfigure}
\hfill
\begin{subfigure}[t]{0.48\textwidth}
\centering
\includegraphics[width=\textwidth]{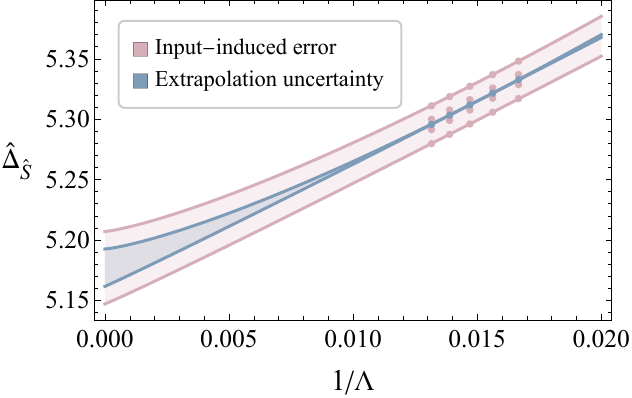}
\caption{$N=6$}
\end{subfigure}
\hfill
\begin{subfigure}[t]{0.48\textwidth}
\centering
\includegraphics[width=\textwidth]{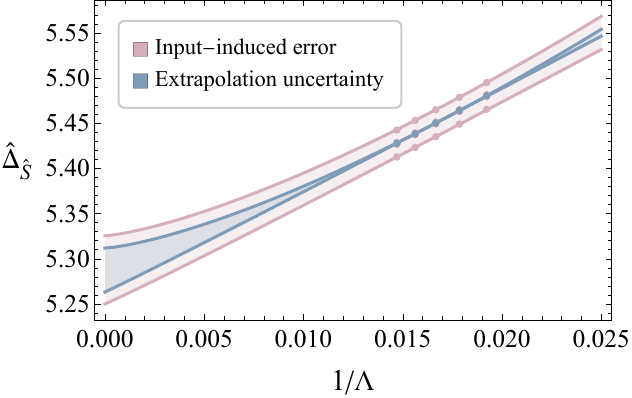}
\caption{$N=8$}
\end{subfigure}
\hfill
\begin{subfigure}[t]{0.48\textwidth}
\centering
\includegraphics[width=\textwidth]{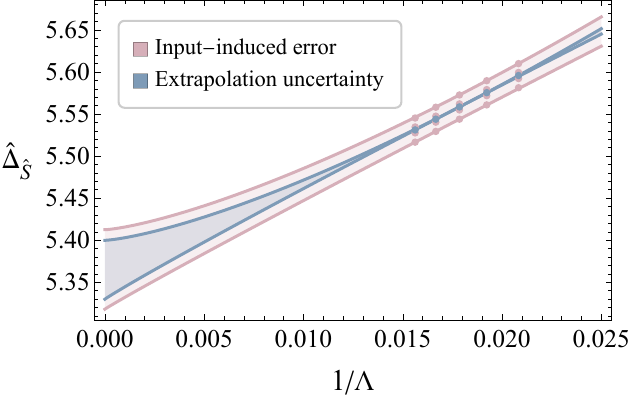}
\caption{$N=10$}
\end{subfigure}
\hfill
\begin{subfigure}[t]{0.48\textwidth}
\centering
\includegraphics[width=\textwidth]{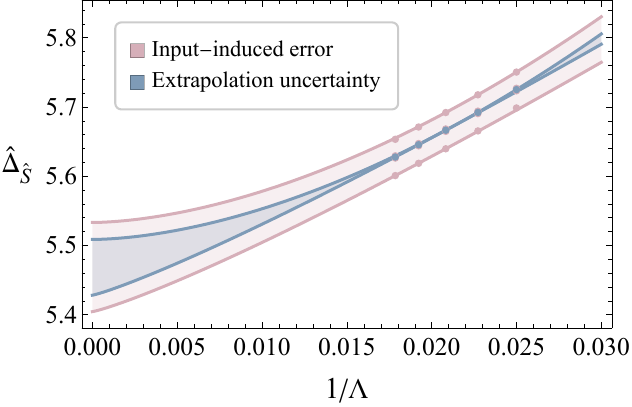}
\caption{$N=12$}
\end{subfigure}
\caption{\label{fig:deltas}Power-law fits of $\hat{\Delta}_{\hat{S}}$ from the $\langle\phi S\rangle$ bootstrap for various $N$.
}
\end{figure}

\begin{figure}[tbp]
\captionsetup[subfigure]{skip=1pt}
\captionsetup[subfigure]{
  margin={30pt,0pt},
  justification=centering,
  singlelinecheck=false
}
\setlength{\lineskip}{1em}
\centering
\begin{subfigure}[t]{0.49\textwidth}
\includegraphics[width=\textwidth]{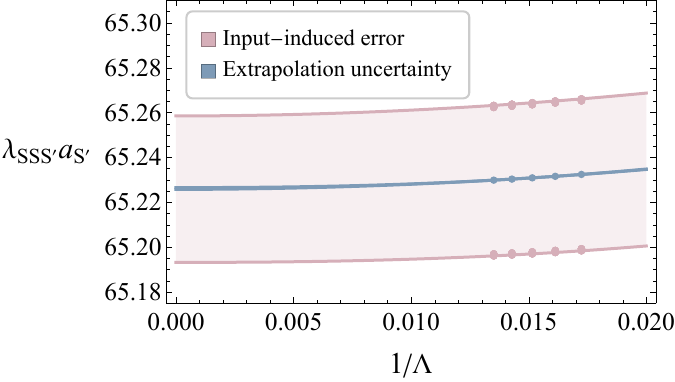}
\caption{$N=1$}
\end{subfigure}
\hfill
\begin{subfigure}[t]{0.49\textwidth}
\includegraphics[width=\textwidth]{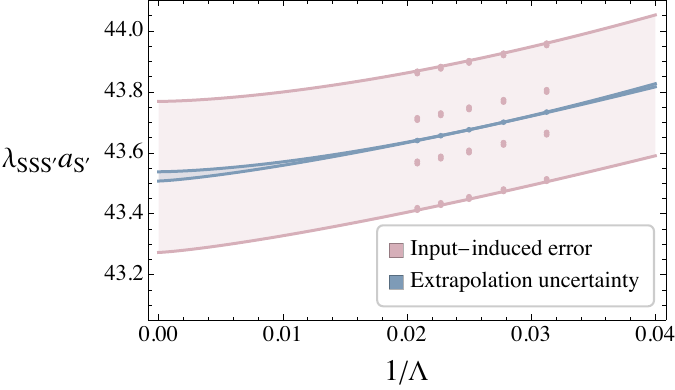}
\caption{$N=2$}
\end{subfigure}
\hfill
\begin{subfigure}[t]{0.49\textwidth}
\includegraphics[width=\textwidth]{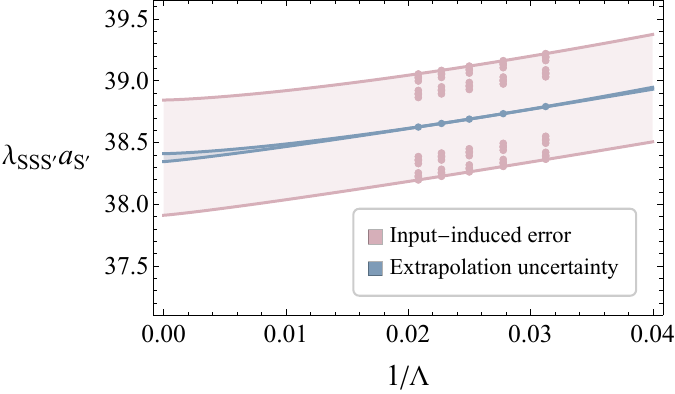}
\caption{$N=3$}
\end{subfigure}
\caption{\label{fig:SS:l2}Power-law fits of $\lambda_{SSS'}a_{S'}$ from the $\langle S S\rangle$ bootstrap.
}
\end{figure}

The input values and the truncation choices used for the extrapolations are summarized in Tables~\ref{ta:1}--\ref{ta:truncation}.\footnote{For $N=5$, the differing decimal precision in the estimate $\Delta_T=1.1568^{+0.009}_{-0.010}$ reported in Ref.~\cite{Kos:2013tga} may have led to its interpretation as $\Delta_T=1.1568(10)$ in Refs.~\cite{Padayasi:2021sik,Hu:2025yrs}.
We use the same input uncertainty for $\Delta_T$ at $N=5$ as in Refs.~\cite{Padayasi:2021sik,Hu:2025yrs}.
The resulting uncertainties may therefore be underestimated.}
Figs.~\ref{fig:Deltaphi2} and \ref{fig:deltas} show the $\Lambda\to\infty$ extrapolations of $\Delta_{\phi'}$ and $\hat{\Delta}_{\hat S}$ from the $\langle\phi S\rangle$ bootstrap for $N=2,3,4,5,6,8,10,12$. Fig.~\ref{fig:SS:l2} shows the corresponding extrapolation of $\lambda_{SSS'}a_{S'}$ from the $\langle SS\rangle$ bootstrap for $N=1,2,3$.
For a given truncation family, both the input-induced uncertainty and the extrapolation uncertainty tend to increase with $N$.
The former reflects the decreasing precision of the input values as $N$ grows.
The latter may be due to the lower truncation orders accessible within the same family at larger $N$.

\begin{table}[tbp]
\small
\setlength{\tabcolsep}{2pt}
\centering
\begin{tabular}{c|ccccc}
\noalign{\hrule height 1.0pt}
$N$& $\Delta_{\phi}$ & $\Delta_{S}$ & $\Delta_{S'}$ &$\Delta_{T}$&Method\\
\hline
1&$0.518148806(\mathbf{24})$\cite{Chang:2024whx}&$1.41262528(\mathbf{29})$\cite{Chang:2024whx}&$3.82951(\mathbf{61})$\cite{Reehorst:2021hmp}&-&CB \\
\hline
2&0.51908(1)\cite{Hasenbusch:2025yrl}&1.51128(5)\cite{Hasenbusch:2025yrl}&3.789(4)\cite{Hasenbusch:2019jkj}&1.23629($\mathbf{11}$)\cite{Chester:2019ifh}&CB\cite{Chester:2019ifh},MC\cite{Hasenbusch:2025yrl,Hasenbusch:2019jkj}\\
\hline
3&0.518936($\mathbf{67}$)\cite{Chester:2020iyt}&1.5948(2)\cite{Hasenbusch:2020pwj}&3.759(2)\cite{Hasenbusch:2020pwj}&1.20954($\mathbf{32}$)\cite{Chester:2020iyt}&CB\cite{Chester:2020iyt},MC\cite{Hasenbusch:2020pwj}\\
\hline
4&0.51812(4)\cite{Hasenbusch:2021rse}&1.66340(35)\cite{Hasenbusch:2021rse}&3.755(5)\cite{Hasenbusch:2021rse}&$1.1864^{+0.0024}_{-0.0034}$\cite{Kos:2013tga}&CB\cite{Kos:2013tga},MC\cite{Hasenbusch:2021rse}\\
\hline
5&0.516985(45)\cite{Hasenbusch:2021rse}&1.7182(10)\cite{Hasenbusch:2021rse}&3.754(7)\cite{Hasenbusch:2021rse}&$1.1568(10)$\cite{Kos:2013tga}&CB\cite{Kos:2013tga},MC\cite{Hasenbusch:2021rse}\\
\hline
6&0.515785(70)\cite{Hasenbusch:2021rse}&1.7625(9)\cite{Hasenbusch:2021rse}&3.758(14)\cite{Hasenbusch:2021rse}&$1.1401^{+0.0085}_{-0.0095}$\cite{Kos:2013tga}&CB\cite{Kos:2013tga},MC\cite{Hasenbusch:2021rse}\\
\hline
8&0.513375(75)\cite{Hasenbusch:2021rse}&1.8248(10)\cite{Hasenbusch:2021rse}&-&-&MC\\
\hline
10&0.51151(6)\cite{Hasenbusch:2021rse}&1.8632(12)\cite{Hasenbusch:2021rse}&-&-&MC\\
\hline
12&0.50995(15)\cite{Hasenbusch:2021rse}&1.8892(17)\cite{Hasenbusch:2021rse}&-&-&MC\\
\hline
$\infty$&1/2&2&4&1& \\
\noalign{\hrule height 1.0pt}
\end{tabular}
\caption{\label{ta:1}Input values $(\Delta_{\phi},\Delta_{S},\Delta_{S'},\Delta_{T})$ from recent studies.
``MC'' and ``CB'' denote Monte Carlo simulations and the conformal bootstrap, respectively.
Parentheses indicate uncertainties in the final quoted digits, while boldface uncertainties are rigorous.
A dash ``-'' indicates that the corresponding quantity is not used in the subsequent analysis, rather than necessarily being unavailable.}
\end{table}

\begin{table}[tbp]
\small
\begin{subtable}{1\textwidth}
\centering
\begin{tabular}{c|cc}
\noalign{\hrule height 1.0pt}
$N$ & $(n_{\mathrm{bulk}},n_{\mathrm{bdy}})$ &$\Lambda_{\mathrm{max}}$\\
\hline
6&(6,2),(7,3),(8,4),(9,5)&52\\
\noalign{\hrule height 1.0pt}
\end{tabular}
\caption{$\langle\phi\phi\rangle$ bootstrap.
For the $\Delta_T^{-}$ input, the $(9,5)$ truncation is not included.}
\end{subtable}

\medskip

\begin{subtable}{1\textwidth}
\centering
\begin{tabular}{c|cc}
\noalign{\hrule height 1.0pt}
$N$ & $(n_{\mathrm{bulk}},n_{\mathrm{bdy}})$ &$\Lambda_{\mathrm{max}}$\\
\hline
2, 3, 4, 5, 6&(18,13), (19,14), (20,15), (21,16), (22,17)& 76\\
\hline
8&(16,11), (17,12), (18,13), (19,14), (20,15)& 68\\
\hline
10&(15,10), (16,11), (17,12), (18,13), (19,14)&64\\
\hline
12&(13,8), (14,9), (15,10), (16,11), (17,12)&56\\
\noalign{\hrule height 1.0pt}
\end{tabular}
\caption{$\langle\phi S\rangle$ bootstrap.}
\end{subtable}

\medskip

\begin{subtable}{1\textwidth}
\centering
\begin{tabular}{c|cc}
\noalign{\hrule height 1.0pt}
$N$ & $(n_{\mathrm{bulk}},n_{\mathrm{bdy}})$&$\Lambda_{\mathrm{max}}$\\
\hline
1&(18,13), (19,14), (20,15), (21,16), (22,17)& 74\\
\hline
2,3&(11,7), (12,8), (13,9), (14,10), (15,11)& 48\\
\noalign{\hrule height 1.0pt}
\end{tabular}
\caption{$\langle S S\rangle$ bootstrap.}
\end{subtable}
\caption{\label{ta:truncation}Truncation choices used for the $\Lambda\to\infty$ extrapolations reported in the main text.}
\end{table}

\section{Data for comparisons}
\label{app:b}
To facilitate direct comparison among the results, we collect the numerical data used in the analysis presented in the main text.
For $\langle\phi\phi\rangle$ at $N=1,2,3,4,5$ and $\langle\phi S\rangle$ at $N=1$, we use results reported in Ref.~\cite{Hu:2025yrs} or obtained from the accompanying data. The remaining bootstrap results are obtained in this work.

In Table~\ref{ta:a2:1}, we collect $(a_\phi,b_{\phi t},b_{\phi D},\Delta_{T'},\lambda_{\phi\phi S}a_S)$ from the $\langle\phi\phi\rangle$ bootstrap for $N=2,3,4,5,6$. These quantities are used in the discussion of their $N$-dependence.
In Tables~\ref{ta:a2:7}--\ref{ta:a2:10}, we summarize the data underlying the comparison of estimates from different correlators considered in the main text.
Table~\ref{ta:a2:7} gives the results for $\Delta_{S''}$, Table~\ref{ta:a2:8} gives those for $\hat{\Delta}_{\hat S}$ and $\hat{\Delta}_{\hat{S}'}$, and Tables~\ref{ta:a2:9}--\ref{ta:a2:10} give those for $b_{SD}$ and $b_{S\hat S}$.

\begin{table}[htbp]
\centering
\small
\begin{tabular}{c|ccccc}
\noalign{\hrule height 1.0pt}
$N$ & $a_\phi$ &$b_{\phi t}$ & $b_{\phi D}$ &$\Delta_{T'}$ &$\lambda_{\phi\phi S}a_S$\\
\hline
2&2.875(2)	&0.5272(2)&0.2440(4)& 3.6484(24)&$3.8278_{-0.0070}^{+0.0079}$\\
\hline
3&3.129(3)	&0.5278(2)&0.2406(6)&3.559(4)&$2.8148_{-0.0051}^{+0.0060}$\\
\hline
4&3.380(7)	&0.5272(13)&0.2369(24)&3.49(3)&$2.354_{-0.010}^{+0.017}$\\
\hline
5&3.634(5)	&0.5235(5)&0.2390(9)&3.354(17)&$2.1235_{-0.0091}^{+0.0094}$\\
\hline
6& $3.869^{+0.013}_{-0.015}$ & $0.5222_{-0.0020}^{+0.0024}$ & $0.236_{-0.008}^{+0.006}$&$3.31_{-0.11}^{+0.08}$&$1.967_{-0.032}^{+0.026}$\\
\noalign{\hrule height 1.0pt}
\end{tabular}
\caption{\label{ta:a2:1}Results from $\langle\phi\phi\rangle$ used in the $N$-dependence analysis in Sec.~\ref{sec:largen:2}.
Data for $N=2,3,4,5$ are taken from Ref.~\cite{Hu:2025yrs}, while the $N=6$ data are from this work (see Table~\ref{ta:2}).}
\end{table}

\begin{table}[htbp]
\centering
\small
\setlength{\tabcolsep}{4pt}
\begin{tabular}{c|cccccc}
\noalign{\hrule height 1.0pt}
Correlator&$N=1$&$N=2$&$N=3$&$N=4$&$N=5$&$N=6$  \\
\hline
$\langle\phi\phi\rangle$&6.873(7)\cite{Hu:2025yrs}&6.63(4)\cite{Hu:2025yrs}&$6.420_{-0.037}^{+0.024}$\cite{Hu:2025yrs}&$6.34_{-0.12}^{+0.06}$\cite{Hu:2025yrs}&$6.088_{-0.067}^{+0.063}$\cite{Hu:2025yrs}&$6.02_{-0.32}^{+0.26}$\\
\hline
$\langle SS\rangle$&6.808(5)&$6.470_{-0.031}^{+0.028}$&$6.303_{-0.028}^{+0.024}$&-&-&-\\
\noalign{\hrule height 1.0pt}
\end{tabular}
\caption{\label{ta:a2:7}Estimates of $\Delta_{S''}$ extracted from the bootstrap analyses of different correlators.
The $\langle\phi\phi\rangle$ results for $N=1,2,3,4,5$ are taken from Ref.~\cite{Hu:2025yrs}, while the $N=6$ result is obtained in this work and reported in Table~\ref{ta:3}.
The $N=1,2,3$ $\langle SS\rangle$ results are from Table~\ref{ta:10}.}
\end{table}

\begin{table}[htbp]
\centering
\small
\begin{tabular}{c|ccc|ccc}
\noalign{\hrule height 1.0pt}
$N$&\multicolumn{3}{c|}{$\hat{\Delta}_{\hat{S}}$}&\multicolumn{3}{c}{$\hat{\Delta}_{\hat{S}'}$}\\
\hline
&$\langle\phi\phi\rangle$&$\langle\phi S\rangle$&$\langle S S\rangle$&$\langle\phi\phi\rangle$&$\langle\phi S\rangle$&$\langle S S\rangle$\\
\hline
1&5.89(4)\cite{Hu:2025yrs}&5.8792(12)\cite{Hu:2025yrs}&$5.84_{-0.10}^{+0.04}$&8.2(3)\cite{Hu:2025yrs}&8.086(24)\cite{Hu:2025yrs}&$7.73_{-0.26}^{+0.19}$\\
\hline
2&$5.35_{-0.18}^{+0.11}$\cite{Hu:2025yrs}&$4.916_{-0.016}^{+0.014}$&$5.14_{-0.07}^{+0.05}$&$7.79_{-0.47}^{+0.25}$\cite{Hu:2025yrs}&6.905(17)&$7.38_{-0.14}^{+0.11}$\\
\hline
3&$5.26_{-0.17}^{+0.10}$\cite{Hu:2025yrs}&4.982(38)&5.13(5)&$7.98_{-0.53}^{+0.23}$\cite{Hu:2025yrs}&$7.247_{-0.053}^{+0.051}$&7.49(12)\\
\hline
4&$5.19_{-0.31}^{+0.25}$\cite{Hu:2025yrs}&$5.052_{-0.022}^{+0.019}$&-&$8.1_{-1.3}^{+0.5}$\cite{Hu:2025yrs}&$7.425_{-0.042}^{+0.036}$&-\\
\hline
5&$5.57_{-0.18}^{+0.05}$\cite{Hu:2025yrs}&$5.116_{-0.027}^{+0.024}$&-&$8.1_{-1.1}^{+0.3}$\cite{Hu:2025yrs}&$7.536_{-0.056}^{+0.046}$&-\\
\hline
6&$5.3_{-0.8}^{+0.5}$&$5.179_{-0.033}^{+0.028}$&-&$7(8)$&$7.612_{-0.068}^{+0.055}$&-\\
\hline
8&-&$5.292_{-0.043}^{+0.034}$&-&-&$7.795_{-0.096}^{+0.071}$&-\\
\hline
10&-&$5.372_{-0.055}^{+0.041}$&-&-&$7.90_{-0.13}^{+0.10}$&-\\
\hline
12&-&$5.479_{-0.075}^{+0.054}$&-&-&$8.13_{-0.17}^{+0.11}$&-\\
\noalign{\hrule height 1.0pt}
\end{tabular}
\caption{\label{ta:a2:8}Estimates of $\hat{\Delta}_{\hat{S}}$ and $\hat{\Delta}_{\hat{S}'}$ extracted from the bootstrap analyses of different correlators. 
The $N=1$ results from $\langle\phi S\rangle$ are taken from Ref.~\cite{Hu:2025yrs}.
The $\langle\phi\phi\rangle$ estimates for $N=1,2,3,4,5$ are obtained from the numerical data accompanying Ref.~\cite{Hu:2025yrs}.
The remaining results are obtained in this work: the $\langle\phi\phi\rangle$ result for $N=6$ is reported in Table~\ref{ta:3}, the $\langle\phi S\rangle$ results for $N=2,3,4,5,6,8,10,12$ in Table~\ref{ta:5}, and the $\langle SS\rangle$ results for $N=1,2,3$ in Table~\ref{ta:10}.
}
\end{table}

\begin{table}[htbp]
\centering
\small
\begin{tabular}{c|cccccc}
\noalign{\hrule height 1.0pt}
Correlator&$N=1$&$N=2$&$N=3$&$N=4$&$N=5$&$N=6$  \\
\hline
$\langle\phi\phi\rangle$+$\langle\phi S\rangle$&1.7234(5)\cite{Hu:2025yrs}&1.383(6)&1.284(12)&1.262(23)&1.249(16)&$1.270_{-0.050}^{+0.056}$\\
\hline
$\langle SS\rangle$&1.7230(6)&$1.4267_{-0.0049}^{+0.0043}$&$1.3413_{-0.0070}^{+0.0061}$&-&-&-\\
\noalign{\hrule height 1.0pt}
\end{tabular}
\caption{\label{ta:a2:9}Estimates of $b_{SD}$ extracted from the bootstrap analyses of different correlators.
For the combined $\langle\phi\phi\rangle$ and $\langle\phi S\rangle$ analysis, the $N=1$ result is taken from Ref.~\cite{Hu:2025yrs}, while the results for $N=2,3,4,5,6$ are obtained in this work and reported in Table~\ref{ta:7}.
The $\langle SS\rangle$ results for $N=1,2,3$ are from Table~\ref{ta:10}.}
\end{table}

\begin{table}[htbp]
\centering
\small
\begin{tabular}{c|cccccc}
\noalign{\hrule height 1.0pt}
Correlator&$N=1$&$N=2$&$N=3$&$N=4$&$N=5$&$N=6$  \\
\hline
$\langle\phi\phi\rangle$+$\langle\phi S\rangle$&0.2147(23)\cite{Hu:2025yrs}&0.44(4)&0.49(6)&0.48(11)&$0.73_{-0.07}^{+0.04}$&$0.72_{-0.33}^{+0.42}$\\
\hline
$\langle SS\rangle$&0.2146(5)&0.359(6)&0.422(6)&-&-&-\\
\noalign{\hrule height 1.0pt}
\end{tabular}
\caption{\label{ta:a2:10}Estimates of $b_{S\hat{S}}$ extracted from the bootstrap analyses of different correlators.
For the combined $\langle\phi\phi\rangle$ and $\langle\phi S\rangle$ analysis, the $N=1$ result is taken from Ref.~\cite{Hu:2025yrs}, while the results for $N=2,3,4,5,6$ are obtained in this work and reported in Table~\ref{ta:7}.
The $\langle SS\rangle$ results for $N=1,2,3$ are from Table~\ref{ta:10}.}
\end{table}

\section{Additional bootstrap results}
\label{app:c}
This appendix collects additional direct results from the bootstrap analyses presented in Sec.~\ref{sec:3}.
We also report how the estimates change under alternative truncation families and input choices.

\subsection{$\<\phi\phi\>$}
\label{app:c:1}
In addition to the main results presented in Table~\ref{ta:2}, we report  in Table~\ref{ta:3} the estimates of $(\Delta_{S''},\hat{\Delta}_{\hat S},\hat{\Delta}_{\hat S'},\lambda_{\phi\phi S'}a_{S'},\lambda_{\phi\phi S''}a_{S''},\lambda_{\phi\phi T}a_T,b_{\phi\hat S})$ obtained from $\langle\phi\phi\rangle$.

Our present $\langle\phi\phi\rangle$ analysis is limited to relatively small values of $n_{\mathrm{bdy}}$.
For some input choices, we obtain results for $\hat{\Delta}_{\hat S'}$ at only two truncation orders, so its uncertainty can only be estimated roughly.
A more reliable assessment of its uncertainty would require extending the calculations to larger values of $n_{\mathrm{bulk}}$ and $n_{\mathrm{bdy}}$.
This is technically challenging for the $\langle\phi\phi\rangle$ bootstrap, since its two crossing equations involve more symmetry sectors.
Increasing $n_{\mathrm{bulk}}$ and $n_{\mathrm{bdy}}$ therefore introduces more unknown parameters than in the bootstrap analyses of the other correlators considered in this work.
We obtain a more precise estimate of $\hat{\Delta}_{\hat S'}$ from the $\langle\phi S\rangle$ bootstrap.
Estimating BCFT data for higher-dimensional operators beyond those listed in the tables would also require calculations at larger values of $n_{\mathrm{bulk}}$ and $n_{\mathrm{bdy}}$.
In addition, the spectrum in a given symmetry sector may become denser at higher scaling dimensions.
The individual contributions of closely spaced operators can be difficult to disentangle in our analysis and the resulting estimates should be interpreted with caution.

\begin{table}[htbp]
\small
\centering
\begin{tabular}{ccccccc}
\noalign{\hrule height 1.0pt}
$\Delta_{S''}$&$\hat{\Delta}_{\hat{S}}$& $\hat{\Delta}_{\hat{S'}}$ & $\lambda_{\phi\phi S'} a_{S'}$ & $\lambda_{\phi\phi S''} a_{S''}$ &$\lambda_{\phi\phi T}a_{T}$& $b_{\phi \hat{S}}$\\
\hline
$6.02_{-0.32}^{+0.26}$&$5.3_{-0.8}^{+0.5}\,{}^{*}$&$7(8)^{*}$&$0.657_{-0.016}^{+0.013}$ &$0.106(7)$ &$13.45_{-0.14}^{+0.09}$ & $0.014_{-0.005}^{+0.012}\,{}^{*}$\\
\noalign{\hrule height 1.0pt}
\end{tabular}
\caption{\label{ta:3}Direct bootstrap results from $\langle \phi\phi\rangle$ for $N=6$. Data marked with an asterisk are obtained from linear fits, while the other data are obtained from power-law fits.
The estimates in this table use the $n_{\mathrm{diff}}=4$ family listed in Table~\ref{ta:truncation}. 
For $\hat{\Delta}_{\hat S'}$, only a few truncation orders in this family are available for extrapolation.
Its quoted uncertainty is a rough symmetric estimate based on the largest absolute deviation from the tabulated central value as the inputs vary.
Using the $n_{\mathrm{diff}}=5$ family with $(n_{\mathrm{bulk}},n_{\mathrm{bdy}})=\{(10,5),(11,6),(12,7),(13,8)\}$ gives a central estimate $\hat{\Delta}_{\hat{S'}}\approx 8$.}
\end{table}
\FloatBarrier

\subsection{$\<\phi S\>$}
\label{app:c:2}
The scaling dimensions and coefficients obtained directly from the $\langle\phi S\rangle$ bootstrap are listed in Tables~\ref{ta:5}--\ref{ta:8}.
By combining the results of the $\langle\phi S\rangle$ and $\langle\phi\phi\rangle$ bootstrap analyses, we extract several bulk OPE and BOE coefficients, as discussed in the main text.
This analysis also provides the products $\lambda_{\phi S\phi'}a_{\phi'}$ and $\lambda_{\phi S\phi''}a_{\phi''}$, which are listed in Table~\ref{ta:add:extract}.

\begin{table}[htbp]
\small
\centering
\begin{tabular}{c|ccccc}
\noalign{\hrule height 1.0pt}
$N$ &$\Delta_{\phi'}$ & $\Delta_{\phi''}$ & $\Delta_{\phi'''}$ & $\hat{\Delta}_{\hat{S}}$ & $\hat{\Delta}_{\hat{S}'}$\\
\hline
2 &$5.1725(13)$& 8.154(5)&$11.384_{-0.010}^{+0.008}$ &$4.916_{-0.016}^{+0.014}$ & $6.905(17)$ \\
\hline
3 &$5.061(8)$	&$7.908(25)$	&$11.060_{-0.047}^{+0.041}$	&4.982(38)&	$7.247_{-0.053}^{+0.051}$\\
\hline
4 & 4.969(6)	&$7.727_{-0.020}^{+0.017}$	&$10.838_{-0.041}^{+0.033}$	&$5.052_{-0.022}^{+0.019}$	&$7.425_{-0.042}^{+0.036}$\\
\hline
5 & $4.896(9)$	&$7.593_{-0.029}^{+0.025}$	&$10.674_{-0.056}^{+0.046}$	&$5.116_{-0.027}^{+0.024}$	&$7.536_{-0.056}^{+0.046}$\\
\hline
6 &$4.833(11)$	&$7.477_{-0.035}^{+0.031}$	&$10.525_{-0.067}^{+0.057}$	&$5.179_{-0.033}^{+0.028}$	&$7.612_{-0.068}^{+0.055}$\\
\hline
8 & $4.759_{-0.013}^{+0.011}$	&$7.369_{-0.043}^{+0.035}$	&$10.414_{-0.086}^{+0.065}$	&$5.292_{-0.043}^{+0.034}$	&$7.795_{-0.096}^{+0.071}$\\
\hline
10 &$4.712_{-0.015}^{+0.012}$	&$7.292_{-0.052}^{+0.039}$	&$10.32_{-0.11}^{+0.08}$	&$5.372_{-0.055}^{+0.041}$	&$7.90_{-0.13}^{+0.10}$\\
\hline
12& $4.692_{-0.022}^{+0.018}$	&$7.285_{-0.069}^{+0.049}$	&$10.35_{-0.14}^{+0.09}$	&$5.479_{-0.075}^{+0.054}$	&$8.13_{-0.17}^{+0.11}$\\
\noalign{\hrule height 1.0pt}
\end{tabular}
\caption{\label{ta:5} The $\langle\phi S\rangle$ bootstrap results for scaling dimensions.}
\end{table}

\begin{table}[htbp]
\centering
\small
\begin{tabular}{c|ccccc}
\noalign{\hrule height 1.0pt}
$N$ & $\frac{a_S}{\lambda_{\phi\phi S}}$ & $\frac{b_{\phi D} b_{SD}}{\lambda_{\phi \phi S}a_{\phi}}$ & $\frac{b_{\phi \hat{S}} b_{S\hat{S}}}{\lambda_{\phi \phi S}a_{\phi}}$& $\frac{\lambda_{\phi S \phi'}a_{\phi'}}{\lambda_{\phi\phi S}a_{\phi}}$ & $\frac{\lambda_{\phi S \phi''}a_{\phi''}}{\lambda_{\phi\phi S}a_{\phi}}$ \\
\hline
2 &8.1296(47)& 0.17104(16)&$3.31(5)\times 10^{-3}$ &3.1918(23)&0.5836(8)  \\
\hline
3 &10.28(4)	&0.1886(10)&$5.23(22)\times 10^{-3}$&4.339(20)	&0.918(7)\\
\hline
4 &	12.79(4)	&0.2061(8)&$6.58(14)\times 10^{-3}$& 5.700(22)&	1.349(7)  \\
\hline
5 	&$15.60(9)$	&0.2226(14)	&$7.50_{-0.18}^{+0.20}\times 10^{-3}$&7.234(48)	&$1.859(19)$ \\
\hline
6 &	18.66(13)&	$0.2386(19)$	&$8.06_{-0.24}^{+0.27}\times 10^{-3}$& $8.909(76)$	&2.451(24)\\
\hline
8 &	25.44(21)&$0.2643(26)$&	$8.64_{-0.29}^{+0.37}\times 10^{-3}$& 12.65(14)	&$3.743(45)$ \\
\hline
10 	&32.25(27)&	$0.2836(34)$&	$8.82_{-0.36}^{+0.48}\times 10^{-3}$& 16.39(20)&	5.078(80)\\
\hline
12&39.51(76)&0.2992(60)	&$8.52_{-0.49}^{+0.71}\times 10^{-3}$& 20.44(47)&	6.44(17)\\
\noalign{\hrule height 1.0pt}
\end{tabular}
\caption{\label{ta:8}Selected direct results for combinations of OPE and BOE coefficients from $\langle\phi S\rangle$.}
\end{table}

\begin{table}[htbp]
\centering
\small
\setlength{\tabcolsep}{3.5pt}
\begin{tabular}{c|ccccc}
\noalign{\hrule height 1.0pt}
&$N=2$&$N=3$&$N=4$&$N=5$&$N=6$\\
\hline
$\lambda_{\phi S\phi'}a_{\phi'}$&6.297(16),6.305(8)&7.105(52),7.119(34)&$8.265_{-0.039}^{+0.055}$&$9.70(12)$&$11.19_{-0.18}^{+0.15}$\\
\hline
$\lambda_{\phi S\phi''}a_{\phi''}$&1.1513(36),1.1529(22)&1.505(15),1.506(11)&$1.956_{-0.016}^{+0.020}$&$2.493_{-0.035}^{+0.037}$&$3.079_{-0.062}^{+0.058}$\\
\noalign{\hrule height 1.0pt}
\end{tabular}
\caption{\label{ta:add:extract} Coefficient products obtained by combining the $\langle\phi S\rangle$ and $\langle\phi\phi\rangle$ bootstrap results.
For entries with two estimates, the first is obtained using only the $\langle\phi S\rangle$ and $\langle\phi\phi\rangle$ bootstrap results, while the second is obtained with the bulk bootstrap estimate $\lambda_{\phi\phi S}^{\mathrm{bulk}}$.}
\end{table}

For the $\langle\phi S\rangle$ bootstrap at $N=10,12$, we find solutions for the truncation family with $n_{\mathrm{diff}}=n_{\mathrm{bulk}}-n_{\mathrm{bdy}}=5$ only up to relatively low truncation orders in the main analysis. As a further check, we consider another truncation family with $n_{\mathrm{diff}}=6$, which reaches higher truncation orders.
The corresponding results are shown in Table~\ref{ta:21}. 
The uncertainty ranges obtained with the alternative truncation family lie within the final ranges quoted in Tables~\ref{ta:5}--\ref{ta:8}, indicating that the change of truncation family does not lead to sizable shifts in the results.

\begin{table}[tbp]
\small
\centering
\begin{tabular}{c|ccccc}
\noalign{\hrule height 1.0pt}
$N$ &$\Delta_{\phi'}$ & $\Delta_{\phi''}$ & $\Delta_{\phi'''}$ & $\hat{\Delta}_{\hat{S}}$ & $\hat{\Delta}_{\hat{S}'}$\\
\hline
10&$4.715(8)$&7.309(16)&10.357(18)&$5.390_{-0.017}^{+0.015}$& $7.942_{-0.019}^{+0.016}$ \\
\hline
12&$4.689(15)$&$7.268(29)$&$10.306_{-0.038}^{+0.034}$&$5.455_{-0.031}^{+0.029}$&$8.042_{-0.036}^{+0.032}$\\
\noalign{\hrule height 1.0pt}
\end{tabular}

\vspace{0.5em}
\begin{tabular}{c|ccccc}
\noalign{\hrule height 1.0pt}
$N$ & $\frac{a_S}{\lambda_{\phi\phi S}}$ & $\frac{b_{\phi D} b_{SD}}{\lambda_{\phi\phi S}a_{\phi}}$ & $\frac{b_{\phi \hat{S}} b_{S\hat{S}}}{\lambda_{\phi\phi S}a_{\phi}}$& $\frac{\lambda_{\phi S \phi'}a_{\phi'}}{\lambda_{\phi\phi S}a_{\phi}}$ & $\frac{\lambda_{\phi S \phi''}a_{\phi''}}{\lambda_{\phi\phi S}a_{\phi}}$ \\
\hline
10&$32.24_{-0.26}^{+0.27}$&0.2842(23)&$8.62_{-0.16}^{+0.18}\times 10^{-3}$&16.42(15)&5.068(52)  \\
\hline
12&$39.51_{-0.72}^{+0.76}$&$0.2986_{-0.0047}^{+0.0044}$&$8.62_{-0.30}^{+0.34}\times 10^{-3}$&$20.41_{-0.40}^{+0.42}$&6.46(13)\\
\noalign{\hrule height 1.0pt}
\end{tabular}
\caption{\label{ta:21}Results from the $\langle\phi S\rangle$ bootstrap for $N=10,12$ with truncation choices $(n_{\mathrm{bulk}},n_{\mathrm{bdy}})=\{(19,13),(20,14),(21,15),(22,16),(23,17)\}$.}
\end{table}
\FloatBarrier

\subsection{$\<SS\>$}
\label{app:c:3}
Table~\ref{ta:add:ss} includes data for higher-lying operators and additional combinations of OPE and BOE coefficients obtained directly from the $\langle SS\rangle$ bootstrap.

\begin{table}[htbp]
\small
\centering
\begin{tabular}{c|cccc}
\noalign{\hrule height 1.0pt}
$N$ &$\Delta_{S'''}$ & $b_{S\hat{S}'}$ & $\lambda_{SSS'}a_{S'}$ &$\lambda_{SSS''}a_{S''}$\\
\hline
1&10.080(17)&$0.055(5)^{*}$&65.226(33)&$28.553_{-0.010}^{+0.006}$\\
\hline
2&$9.44_{-0.10}^{+0.08}$&0.103(7)&43.52(25)&$24.69_{-0.14}^{+0.11}$\\
\hline
3&$9.25(8)$&0.115(6)&38.38(48)&$27.06_{-0.36}^{+0.33}$\\
\noalign{\hrule height 1.0pt}
\end{tabular}
\caption{\label{ta:add:ss}Selected direct results from the $\langle SS\rangle$ bootstrap. 
Data marked with an asterisk are obtained from linear fits, while the others are obtained from power-law fits.
}
\end{table}

For the $N=2,3$ $\langle SS\rangle$ bootstrap results presented in the main text, we use the values of $a_S$ given in Eq.~\eqref{eq:as:2}.
These $a_S$ values are obtained by combining $a_S/\lambda_{\phi\phi S}$ from the mixed-correlator bootstrap with the bulk-bootstrap estimates of $\lambda_{\phi\phi S}^{\text{bulk}}$. 
For comparison, we repeat the analysis using the two alternative estimates of $a_S$ in Eqs.~\eqref{eq:as:3} and \eqref{eq:as:1}.
The former is obtained by combining $\lambda_{\phi\phi S}a_S$ from $\langle\phi\phi\rangle$ with $\lambda_{\phi\phi S}^{\text{bulk}}$, while the latter is obtained by combining $\lambda_{\phi\phi S}a_S$ from $\langle\phi\phi\rangle$ with $a_S/\lambda_{\phi\phi S}$ from $\langle\phi S\rangle$.
The resulting $\langle SS\rangle$ bootstrap estimates are mutually consistent within their quoted uncertainties, with central values differing by no more than about $1\%$.
This comparison shows that the $\langle SS\rangle$ bootstrap results are weakly sensitive to the choice among the three estimates of $a_S$.

\begin{table}[tbp]
\small
\setlength{\tabcolsep}{3pt}
\centering
\begin{subtable}{\textwidth}
\centering
\begin{tabular}{c|cccc}
\noalign{\hrule height 1.0pt}
$a_S^{\mathrm{input}}$&$\Delta_{S''}$ &$\Delta_{S'''}$ &$\hat{\Delta}_{\hat{S}}$ & $\hat{\Delta}_{\hat{S}'}$\\
\hline
$5.586(4)$ from Eq.~\eqref{eq:as:2}&$6.470_{-0.031}^{+0.028}$&$9.44_{-0.10}^{+0.08}$&$5.14_{-0.07}^{+0.05}$&$7.38_{-0.14}^{+0.11}$\\
\hline
$5.571(12)$ from Eq.~\eqref{eq:as:3}&$6.462_{-0.036}^{+0.032}$&$9.42_{-0.11}^{+0.09}$&$5.15(7)$&$7.39(14)$\\
\hline
$5.578(7)$ from Eq.~\eqref{eq:as:1}&$6.466_{-0.033}^{+0.029}$&$9.42(10)$&$5.15_{-0.07}^{+0.05}$&$7.39_{-0.15}^{+0.11}$\\
\noalign{\hrule height 1.0pt}
\end{tabular}
\end{subtable}

\vspace{0.5em}
\begin{subtable}{\textwidth}
\centering
\begin{tabular}{c|cccccc}
\noalign{\hrule height 1.0pt}
$a_S^{\mathrm{input}}$&$b_{SD}$ & $b_{S\hat{S}}$ & $b_{S\hat{S}'}$ & $\lambda_{SSS'}a_{S'}$ &$\lambda_{SSS''}a_{S''}$&$\frac{\lambda_{SSS'}}{\lambda_{\phi\phi S'}}$\\
\hline
$5.586(4)$ from Eq.~\eqref{eq:as:2}&$1.4267_{-0.0049}^{+0.0043}$&0.359(6)&0.103(7)&43.52(25)&$24.69_{-0.14}^{+0.11}$&33.17(25)\\
\hline
$5.571(12)$ from Eq.~\eqref{eq:as:3}&$1.4281_{-0.0056}^{+0.0048}$&0.357(7)&$0.102_{-0.006}^{+0.008}$&43.24(41)&$24.53_{-0.24}^{+0.19}$&32.95(37)\\
\hline
$5.578(7)$ from Eq.~\eqref{eq:as:1}&$1.4274_{-0.0051}^{+0.0045}$&0.358(6)&$0.102_{-0.006}^{+0.008}$&43.37(31)&$24.60_{-0.17}^{+0.14}$&33.05(30)\\
\noalign{\hrule height 1.0pt}
\end{tabular}
\end{subtable}
\caption{\label{ta:22}The $N=2$ $\<S S\>$ bootstrap results obtained using three choices of the input $a_S$. The first row corresponds to the input choice used in the main text.}
\end{table}

\begin{table}[tbp]
\small
\setlength{\tabcolsep}{2.3pt}
\centering
\begin{subtable}{\textwidth}
\centering
\begin{tabular}{c|cccc}
\noalign{\hrule height 1.0pt}
$a_S^{\mathrm{input}}$&$\Delta_{S''}$ &$\Delta_{S'''}$ &$\hat{\Delta}_{\hat{S}}$ & $\hat{\Delta}_{\hat{S}'}$\\
\hline
$5.389(20)$ from Eq.~\eqref{eq:as:2}&$6.303_{-0.028}^{+0.024}$&9.25(8)&5.13(5)&7.49(12)\\
\hline
$5.369(12)$ from Eq.~\eqref{eq:as:3}&$6.297_{-0.026}^{+0.023}$&$9.24(8)$&$5.14(5)$&$7.50_{-0.12}^{+0.09}$\\
\hline
$5.379(12)$ from Eq.~\eqref{eq:as:1}&$6.300_{-0.027}^{+0.023}$&$9.25_{-0.08}^{+0.06}$&5.14(5)&$7.49_{-0.12}^{+0.10}$\\
\noalign{\hrule height 1.0pt}
\end{tabular}
\end{subtable}

\vspace{0.5em}
\begin{subtable}{\textwidth}
\centering
\begin{tabular}{c|cccccc}
\noalign{\hrule height 1.0pt}
$a_S^{\mathrm{input}}$&$b_{SD}$ & $b_{S\hat{S}}$ & $b_{S\hat{S}'}$ & $\lambda_{SSS'}a_{S'}$ &$\lambda_{SSS''}a_{S''}$&$\frac{\lambda_{SSS'}}{\lambda_{\phi\phi S'}}$\\
\hline
$5.389(20)$ from Eq.~\eqref{eq:as:2}&$1.3413_{-0.0070}^{+0.0061}$&0.422(6)&0.115(6)&38.38(48)&$27.06_{-0.36}^{+0.33}$&38.63(56)\\
\hline
$5.369(12)$ from Eq.~\eqref{eq:as:3}&$1.3426_{-0.0064}^{+0.0055}$&$0.420(6)$&0.114(6)&$38.05_{-0.35}^{+0.33}$&$26.85_{-0.27}^{+0.24}$&38.30(42)\\
\hline
$5.379(12)$ from Eq.~\eqref{eq:as:1}&$1.3419_{-0.0064}^{+0.0056}$&0.421(6)&$0.114_{-0.004}^{+0.007}$&38.21(34)&$26.95_{-0.27}^{+0.25}$&$38.46_{-0.41}^{+0.43}$\\
\noalign{\hrule height 1.0pt}
\end{tabular}
\end{subtable}
\caption{\label{ta:23}The $N=3$ $\<S S\>$ bootstrap results obtained using three choices of the input $a_S$. The first row corresponds to the input choice used in the main text.}
\end{table}


\begin{thebibliography}{999}
\bibitem{Andrei:2018die}
N. Andrei, A. O'Bannon, R. Parini, A. Bissi, M. Buican, J. Cardy, P. Dorey, N. Drukker, J. Erdmenger and D. Friedan, \textit{et al.}
\emph{Boundary and Defect CFT: Open Problems and Applications}, J. Phys. A \textbf{53}, no.45, 453002 (2020),
\href{https://doi.org/10.48550/arXiv.1810.05697}{arXiv:1810.05697}.

\bibitem{Casimir:1948dh}
H.~B.~G.~Casimir, \emph{On the attraction between two perfectly conducting plates},
Indag. Math. \textbf{10}, 261--263 (1948).

\bibitem{Bordag:2001qi}
M.~Bordag, U.~Mohideen and V.~M.~Mostepanenko,
\emph{New developments in the Casimir effect},
Phys. Rept. \textbf{353}, 1--205 (2001),
\href{https://arxiv.org/abs/quant-ph/0106045}{arXiv:quant-ph/0106045}.

\bibitem{Affleck:1995ge}
I.~Affleck, \emph{Conformal field theory approach to the Kondo effect},
Acta Phys. Polon. B \textbf{26}, 1869 (1995),
\href{https://arxiv.org/abs/cond-mat/9512099}{arXiv:cond-mat/9512099}.

\bibitem{Recknagel:2013uja}
A.~Recknagel and V.~Schomerus, \emph{Boundary Conformal Field Theory and the Worldsheet Approach to D-Branes}, Cambridge Monographs on Mathematical Physics, Cambridge University Press (2013),
\href{https://doi.org/10.1017/CBO9780511806476}{doi:10.1017/CBO9780511806476}.

\bibitem{Qi:2012ngv}
X.-L.~Qi, H.~Katsura and A.~W.~W.~Ludwig,
\emph{General relationship between the entanglement spectrum and the edge state spectrum of topological quantum states},
Phys. Rev. Lett. \textbf{108}, 196402 (2012),
\href{https://arxiv.org/abs/1103.5437}{arXiv:1103.5437}.

\bibitem{Wen:2016snr}
X.~Wen, S.~Matsuura and S.~Ryu,
\emph{Edge theory approach to topological entanglement entropy, mutual information, and entanglement negativity in Chern Simons theories},
Phys. Rev. B \textbf{93}, 245140 (2016),
\href{https://arxiv.org/abs/1603.08534}{arXiv:1603.08534}.

\bibitem{Cho:2016kcc}
G.~Y.~Cho, A.~W.~W.~Ludwig and S.~Ryu,
\emph{Universal entanglement spectra of gapped one-dimensional field theories},
Phys. Rev. B \textbf{95}, 115122 (2017),
\href{https://arxiv.org/abs/1603.04016}{arXiv:1603.04016}.

\bibitem{Cho:2016xjw}
G.~Y.~Cho, K.~Shiozaki, S.~Ryu and A.~W.~W.~Ludwig,
\emph{Relationship between symmetry protected topological phases and boundary conformal field theories via the entanglement spectrum},
J. Phys. A: Math. Theor. \textbf{50}, 304002 (2017),
\href{https://arxiv.org/abs/1606.06402}{arXiv:1606.06402}.

\bibitem{Han:2017hdv}
B.~Han, A.~Tiwari, C.-T.~Hsieh and S.~Ryu,
\emph{Boundary conformal field theory and symmetry-protected topological phases in $2+1$ dimensions},
Phys. Rev. B \textbf{96}, 125105 (2017),
\href{https://arxiv.org/abs/1704.01193}{arXiv:1704.01193}.

\bibitem{Fliss:2017wop}
J.~R.~Fliss, X.~Wen, O.~Parrikar, C.-T.~Hsieh, B.~Han, T.~L.~Hughes and R.~G.~Leigh,
\emph{Interface contributions to topological entanglement in Abelian Chern--Simons theory},
JHEP \textbf{09}, 056 (2017),
\href{https://arxiv.org/abs/1705.09611}{arXiv:1705.09611}.

\bibitem{Wong:2017pdm}
G.~Wong,
\emph{A note on entanglement edge modes in Chern Simons theory},
JHEP \textbf{08}, 020 (2018),
\href{https://arxiv.org/abs/1706.04666}{arXiv:1706.04666}.

\bibitem{Bray:1977fvl}
A.~J. Bray and M.~A. Moore, \emph{Critical behaviour of semi-infinite systems},
J. Phys. A \textbf{10}, no.~11, 1927 (1977),
\href{https://doi.org/10.1088/0305-4470/10/11/021}{doi:10.1088/0305-4470/10/11/021}.

\bibitem{Binder:1983}
K.~Binder, \emph{Critical behavior at surfaces}, in \emph{Phase Transitions and Critical Phenomena},
C.~Domb and J.~L.~Lebowitz, eds., vol.~8, pp.~1--144, Academic Press, London (1983).

\bibitem{Diehl:1986}
H.~W.~Diehl, \emph{Field-theoretical approach to critical behaviour at surfaces},
in \emph{Phase Transitions and Critical Phenomena},
C.~Domb and J.~L.~Lebowitz, eds., vol.~10, pp.~75--267, Academic Press, London (1986).

\bibitem{Burkhardt:1987}
T.~W. Burkhardt and J.~L. Cardy, \emph{Surface critical behaviour and local operators with boundary-induced critical profiles},
J. Phys. A \textbf{20}, no.~4, L233 (1987),
\href{https://doi.org/10.1088/0305-4470/20/4/010}{doi:10.1088/0305-4470/20/4/010}.

\bibitem{Diehl:1994a}
H.~W. Diehl, \emph{Universality classes for the dynamic surface critical behavior of systems with relaxational dynamics},
Phys. Rev. B \textbf{49}, no.~4, 2846 (1994),
\href{https://doi.org/10.1103/PhysRevB.49.2846}{doi:10.1103/PhysRevB.49.2846}.

\bibitem{Diehl:1994b}
H.~W. Diehl, \emph{Critical adsorption of fluids and the equivalence of extraordinary to normal surface transitions},
Ber. Bunsenges. Phys. Chem. \textbf{98}, no.~3, 466--471 (1994),
\href{https://doi.org/10.1002/bbpc.19940980344}{doi:10.1002/bbpc.19940980344}.

\bibitem{Burkhardt:1994}
T.~W. Burkhardt and H.~W. Diehl, \emph{Ordinary, extraordinary, and normal surface transitions: Extraordinary-normal equivalence and simple explanation of $|T-T_c|^{2-\alpha}$ singularities},
Phys. Rev. B \textbf{50}, no.~6, 3894 (1994),
\href{https://doi.org/10.1103/PhysRevB.50.3894}{doi:10.1103/PhysRevB.50.3894}.

\bibitem{Diehl:1996kd}
H.~W. Diehl, \emph{The theory of boundary critical phenomena},
Int. J. Mod. Phys. B \textbf{11}, 3503--3523 (1997),
\href{https://arxiv.org/abs/cond-mat/9610143}{arXiv:cond-mat/9610143}.

\bibitem{Metlitski:2020cqy}
M.~A. Metlitski, \emph{Boundary criticality of the $\mathrm{O}(N)$ model in $d=3$ critically revisited},
SciPost Phys. \textbf{12}, no.~4, 131 (2022),
\href{https://arxiv.org/abs/2009.05119}{arXiv:2009.05119}.

\bibitem{Toldin:2020wbn}
F.~P.~Toldin, \emph{Boundary critical behavior of the three-dimensional Heisenberg universality class},
Phys. Rev. Lett. \textbf{126}, 135701 (2021),
\href{https://arxiv.org/abs/2012.00039}{arXiv:2012.00039}.

\bibitem{Hu:2021xdy}
M.~Hu, Y.~Deng and J.~P.~Lv, \emph{Extraordinary-log surface phase transition in the three-dimensional XY model},
Phys. Rev. Lett. \textbf{127}, 120603 (2021),
\href{https://arxiv.org/abs/2104.05152}{arXiv:2104.05152}.

\bibitem{Toldin:2021kun}
F.~P.~Toldin and M.~A. Metlitski, \emph{Boundary criticality of the 3D $\mathrm{O}(N)$ model: From normal to extraordinary},
Phys. Rev. Lett. \textbf{128}, 215701 (2022),
\href{https://arxiv.org/abs/2111.03613}{arXiv:2111.03613}.

\bibitem{Padayasi:2021sik}
J.~Padayasi, A.~Krishnan, M.~A. Metlitski, I.~A. Gruzberg and M.~Meineri, \emph{The extraordinary boundary transition in the 3D $\mathrm{O}(N)$ model via conformal bootstrap},
SciPost Phys. \textbf{12}, no.~6, 190 (2022),
\href{https://arxiv.org/abs/2111.03071}{arXiv:2111.03071}.

\bibitem{Zhang:2022hpz}
L.~R.~Zhang, C.~Ding, Y.~Deng and L.~Zhang, \emph{Surface criticality of the antiferromagnetic Potts model},
Phys. Rev. B \textbf{105}, 224415 (2022),
\href{https://arxiv.org/abs/2204.11692}{arXiv:2204.11692}.

\bibitem{Zou:2022mhr}
X.~Zou, S.~Liu and W.~Guo, \emph{Surface critical properties of the three-dimensional clock model},
Phys. Rev. B \textbf{106}, 064420 (2022),
\href{https://arxiv.org/abs/2204.13612}{arXiv:2204.13612}.

\bibitem{Krishnan:2023cff}
A.~Krishnan and M.~A.~Metlitski, \emph{A plane defect in the 3d $\mathrm{O}(N)$ model},
SciPost Phys. \textbf{15}, 090 (2023),
\href{https://arxiv.org/abs/2301.05728}{arXiv:2301.05728}.

\bibitem{Sun:2023vwy}
Y.~Sun, M.~Hu, Y.~Deng and J.~P.~Lv, \emph{Extraordinary-log universality of critical phenomena in plane defects},
Phys. Rev. Lett. \textbf{131}, 207101 (2023),
\href{https://arxiv.org/abs/2301.11720}{arXiv:2301.11720}.

\bibitem{Cuomo:2023qvp}
G.~Cuomo and S.~Zhang, \emph{Spontaneous symmetry breaking on surface defects},
JHEP \textbf{03}, 022 (2024),
\href{https://arxiv.org/abs/2306.00085}{arXiv:2306.00085}.

\bibitem{Toldin:2024pqi}
F.~P.~Toldin, A.~Krishnan and M.~A.~Metlitski, \emph{Universal finite-size scaling in the extraordinary-log boundary phase of three-dimensional $\mathrm{O}(N)$ model},
Phys. Rev. Res. \textbf{7}, 023052 (2025),
\href{https://arxiv.org/abs/2411.05089}{arXiv:2411.05089}.

\bibitem{Diehl:1986mu}
H.~W.~Diehl and A.~N\"usser, \emph{Critical behavior of the nonlinear $\sigma$ model with a free surface: The ``ordinary'' transition in $(2+\epsilon)$ dimensions},
Phys. Rev. Lett. \textbf{56}, 2834--2837 (1986),
\href{https://doi.org/10.1103/PhysRevLett.56.2834}{doi:10.1103/PhysRevLett.56.2834}.

\bibitem{Dey:2020lwp}
P.~Dey, T.~Hansen and M.~Shpot, \emph{Operator expansions, layer susceptibility and two-point functions in BCFT},
JHEP \textbf{12}, 051 (2020),
\href{https://arxiv.org/abs/2006.11253}{arXiv:2006.11253}.

\bibitem{Giombi:2025pxx}
S.~Giombi and Z.~Sun,
\emph{Higher loops in AdS: applications to boundary CFT},
JHEP \textbf{12}, 011 (2025),
\href{https://arxiv.org/abs/2506.14699}{arXiv:2506.14699}.

\bibitem{Sun:2026mib}
X.~Sun, S.-K.~Jian and H.~Yao,
\emph{Analytic bootstrap for $\mathrm{O}(N)$ boundary conformal field theories with interacting boundaries},
\href{https://arxiv.org/abs/2605.28933}{arXiv:2605.28933}.

\bibitem{Ohno:1984}
K.~Ohno and Y.~Okabe, \emph{The $1/n$ expansion for the extraordinary transition of semi-infinite system},
Prog. Theor. Phys. \textbf{72}, 736--745 (1984),
\href{https://doi.org/10.1143/PTP.72.736}{doi:10.1143/PTP.72.736}.

\bibitem{Giombi:2020rmc}
S.~Giombi and H.~Khanchandani,
\emph{CFT in AdS and boundary RG flows},
JHEP \textbf{11}, 118 (2020),
\href{https://arxiv.org/abs/2007.04955}{arXiv:2007.04955}.

\bibitem{Ohno:2025qfv}
K.~Ohno and Y.~Okabe,
\emph{Comment on CFT in AdS and boundary RG flows: $O(1/N)$ result},
\href{https://arxiv.org/abs/2511.06577}{arXiv:2511.06577}.

\bibitem{Csipes:2026nyo}
J.~Csipes and P.~Va\v{s}ko,
\emph{$\mathrm{O}(N)$ BCFT: New data from conformal partial wave expansions},
JHEP \textbf{11}, 086 (2026),
\href{https://arxiv.org/abs/2606.14733}{arXiv:2606.14733}.

\bibitem{Deng:2005dh}
Y.~J.~Deng, H.~W.~J.~Bl\"ote and M.~P.~Nightingale, \emph{Surface and bulk transitions in three-dimensional $\mathrm{O}(n)$ models},
Phys. Rev. E \textbf{72}, 016128 (2005),
\href{https://arxiv.org/abs/cond-mat/0504173}{arXiv:cond-mat/0504173}.

\bibitem{Hasenbusch:2010}
M.~Hasenbusch, \emph{Thermodynamic Casimir effect for films in the three-dimensional Ising universality class: Symmetry-breaking boundary conditions},
Phys. Rev. B \textbf{82}, 104425 (2010),
\href{https://arxiv.org/abs/1005.4749}{arXiv:1005.4749}.

\bibitem{ParisenToldin:2010}
F.~P.~Toldin and S.~Dietrich, \emph{Critical Casimir forces and adsorption profiles in the presence of a chemically structured substrate},
J. Stat. Mech. \textbf{2010}, P11003 (2010),
\href{https://arxiv.org/abs/1007.3913}{arXiv:1007.3913}.

\bibitem{Toldin:2023fny}
F.~Parisen Toldin, \emph{The ordinary surface universality class of the three-dimensional $\mathrm{O}(N)$ model},
Phys. Rev. B \textbf{108}, L020404 (2023),
\href{https://arxiv.org/abs/2303.16683}{arXiv:2303.16683}.

\bibitem{Przetakiewicz:2025gzi}
D.~Przetakiewicz, S.~Wessel and F.~P.~Toldin, \emph{Boundary operator product expansion coefficients of the three-dimensional Ising universality class},
Phys. Rev. Res. \textbf{7}, L032051 (2025),
\href{https://arxiv.org/abs/2502.14965}{arXiv:2502.14965}.

\bibitem{Liendo:2012hy}
P.~Liendo, L.~Rastelli and B.~C. van Rees, \emph{The bootstrap program for boundary $\mathrm{CFT}_d$},
JHEP \textbf{07}, 113 (2013),
\href{https://arxiv.org/abs/1210.4258}{arXiv:1210.4258}.

\bibitem{Behan:2020nsf}
C.~Behan, L.~Di Pietro, E.~Lauria and B.~C. van Rees,
\emph{Bootstrapping boundary-localized interactions},
JHEP \textbf{12}, 182 (2020),
\href{https://arxiv.org/abs/2009.03336}{arXiv:2009.03336}.

\bibitem{Behan:2021tcn}
C.~Behan, L.~Di Pietro, E.~Lauria and B.~C. van Rees,
\emph{Bootstrapping boundary-localized interactions II: Minimal models at the boundary},
JHEP \textbf{03}, 146 (2022),
\href{https://arxiv.org/abs/2111.04747}{arXiv:2111.04747}.

\bibitem{Collier:2021ngi}
S.~Collier, D.~Maz\'a\v{c} and Y.~Wang,
\emph{Bootstrapping boundaries and branes},
JHEP \textbf{02}, 019 (2023),
\href{https://arxiv.org/abs/2112.00750}{arXiv:2112.00750}.

\bibitem{Meineri:2025qwz}
M.~Meineri and B.~Radhakrishnan,
\emph{The bootstrap of points and lines},
JHEP \textbf{04}, 040 (2026),
\href{https://arxiv.org/abs/2506.17382}{arXiv:2506.17382}.

\bibitem{Gliozzi:2015qsa}
F.~Gliozzi, P.~Liendo, M.~Meineri and A.~Rago, \emph{Boundary and interface CFTs from the conformal bootstrap},
JHEP \textbf{05}, 036 (2015),
\href{https://arxiv.org/abs/1502.07217}{arXiv:1502.07217}.

\bibitem{Gliozzi:2016cmg}
F.~Gliozzi, \emph{Truncatable bootstrap equations in algebraic form and critical surface exponents},
JHEP \textbf{10}, 037 (2016),
\href{https://arxiv.org/abs/1605.04175}{arXiv:1605.04175}.

\bibitem{Hu:2025yrs}
R.~Hu and W.~Li, \emph{Accurate boundary bootstrap for the three-dimensional $\mathrm{O}(N)$ normal universality class},
Phys. Rev. Lett. \textbf{136}, 221601 (2026),
\href{https://arxiv.org/abs/2508.20854}{arXiv:2508.20854}.

\bibitem{Dedushenko:2024nwi}
M.~Dedushenko, \emph{Ising BCFT from fuzzy hemisphere},
\href{https://arxiv.org/abs/2407.15948}{arXiv:2407.15948}.

\bibitem{Zhou:2024dbt}
Z.~Zhou and Y.~Zou, \emph{Studying the 3d Ising surface CFTs on the fuzzy sphere},
SciPost Phys. \textbf{18}, 031 (2025),
\href{https://arxiv.org/abs/2407.15914}{arXiv:2407.15914}.

\bibitem{Feng:2026iii}
J.~Feng and T.~Wang,
\emph{Studying 3D $\mathrm{O}(N)$ surface CFT on the fuzzy sphere},
\href{https://arxiv.org/abs/2604.21091}{arXiv:2604.21091}.

\bibitem{Rattazzi:2008pe}
R.~Rattazzi, V.~S. Rychkov, E.~Tonni and A.~Vichi, \emph{Bounding scalar operator dimensions in 4D CFT},
JHEP \textbf{12}, 031 (2008),
\href{https://arxiv.org/abs/0807.0004}{arXiv:0807.0004}.

\bibitem{Poland:2018epd}
D.~Poland, S.~Rychkov and A.~Vichi, \emph{The conformal bootstrap: Theory, numerical techniques, and applications},
Rev. Mod. Phys. \textbf{91}, 015002 (2019),
\href{https://arxiv.org/abs/1805.04405}{arXiv:1805.04405}.

\bibitem{Rychkov:2023wsd}
S.~Rychkov and N.~Su, \emph{New developments in the numerical conformal bootstrap},
Rev. Mod. Phys. \textbf{96}, 045004 (2024),
\href{https://arxiv.org/abs/2311.15844}{arXiv:2311.15844}.

\bibitem{El-Showk:2012cjh}
S.~El-Showk, M.~F. Paulos, D.~Poland, S.~Rychkov, D.~Simmons-Duffin and A.~Vichi, \emph{Solving the 3D Ising model with the conformal bootstrap},
Phys. Rev. D \textbf{86}, 025022 (2012),
\href{https://arxiv.org/abs/1203.6064}{arXiv:1203.6064}.

\bibitem{Kos:2013tga}
F.~Kos, D.~Poland and D.~Simmons-Duffin, \emph{Bootstrapping the $\mathrm{O}(N)$ vector models},
JHEP \textbf{06}, 091 (2014),
\href{https://arxiv.org/abs/1307.6856}{arXiv:1307.6856}.

\bibitem{El-Showk:2014dwa}
S.~El-Showk, M.~F. Paulos, D.~Poland, S.~Rychkov, D.~Simmons-Duffin and A.~Vichi, \emph{Solving the 3D Ising model with the conformal bootstrap II: $c$-minimization and precise critical exponents},
J. Stat. Phys. \textbf{157}, 869--914 (2014),
\href{https://arxiv.org/abs/1403.4545}{arXiv:1403.4545}.

\bibitem{Kos:2014bka}
F.~Kos, D.~Poland and D.~Simmons-Duffin, \emph{Bootstrapping mixed correlators in the 3D Ising model},
JHEP \textbf{11}, 109 (2014),
\href{https://arxiv.org/abs/1406.4858}{arXiv:1406.4858}.

\bibitem{Simmons-Duffin:2015qma}
D.~Simmons-Duffin, \emph{A semidefinite program solver for the conformal bootstrap},
JHEP \textbf{06}, 174 (2015),
\href{https://arxiv.org/abs/1502.02033}{arXiv:1502.02033}.

\bibitem{Kos:2015mba}
F.~Kos, D.~Poland, D.~Simmons-Duffin and A.~Vichi, \emph{Bootstrapping the $\mathrm{O}(N)$ archipelago},
JHEP \textbf{11}, 106 (2015),
\href{https://arxiv.org/abs/1504.07997}{arXiv:1504.07997}.

\bibitem{Kos:2016ysd}
F.~Kos, D.~Poland, D.~Simmons-Duffin and A.~Vichi, \emph{Precision islands in the Ising and $\mathrm{O}(N)$ models},
JHEP \textbf{08}, 036 (2016),
\href{https://arxiv.org/abs/1603.04436}{arXiv:1603.04436}.

\bibitem{Simmons-Duffin:2016wlq}
D.~Simmons-Duffin, \emph{The lightcone bootstrap and the spectrum of the 3D Ising CFT},
JHEP \textbf{03}, 086 (2017),
\href{https://arxiv.org/abs/1612.08471}{arXiv:1612.08471}.

\bibitem{Chester:2019ifh}
S.~M. Chester, W.~Landry, J.~Liu, D.~Poland, D.~Simmons-Duffin, N.~Su and A.~Vichi, \emph{Carving out OPE space and precise $\mathrm{O}(2)$ model critical exponents},
JHEP \textbf{06}, 142 (2020),
\href{https://arxiv.org/abs/1912.03324}{arXiv:1912.03324}.

\bibitem{Chester:2020iyt}
S.~M. Chester, W.~Landry, J.~Liu, D.~Poland, D.~Simmons-Duffin, N.~Su and A.~Vichi, \emph{Bootstrapping Heisenberg magnets and their cubic instability},
Phys. Rev. D \textbf{104}, 105013 (2021),
\href{https://arxiv.org/abs/2011.14647}{arXiv:2011.14647}.

\bibitem{Reehorst:2021hmp}
M.~Reehorst, \emph{Rigorous bounds on irrelevant operators in the 3D Ising model CFT},
JHEP \textbf{09}, 177 (2022),
\href{https://arxiv.org/abs/2111.12093}{arXiv:2111.12093}.

\bibitem{Chang:2024whx}
C.~H. Chang, V.~Dommes, R.~S. Erramilli, A.~Homrich, P.~Kravchuk, A.~Liu, M.~S. Mitchell, D.~Poland and D.~Simmons-Duffin, \emph{Bootstrapping the 3D Ising stress tensor},
JHEP \textbf{03}, 136 (2025),
\href{https://arxiv.org/abs/2411.15300}{arXiv:2411.15300}.

\bibitem{Gliozzi:2013ysa}
F.~Gliozzi, \emph{More constraining conformal bootstrap},
Phys. Rev. Lett. \textbf{111}, 161602 (2013),
\href{https://arxiv.org/abs/1307.3111}{arXiv:1307.3111}.

\bibitem{Esterlis:2016psv}
I.~Esterlis, A.~L. Fitzpatrick and D.~M. Ramirez,
\emph{Closure of the operator product expansion in the non-unitary bootstrap},
JHEP \textbf{11}, 030 (2016),
\href{https://arxiv.org/abs/1606.07458}{arXiv:1606.07458}.

\bibitem{Li:2017ukc}
W. Li, \emph{New method for the conformal bootstrap with OPE truncations},  \href{https://arxiv.org/abs/1711.09075}{arXiv:1711.09075}.

\bibitem{Poland:2023vpn}
D.~Poland, V.~Prilepina and P.~Tadi\'c,
\emph{The five-point bootstrap},
JHEP \textbf{10}, 153 (2023),
\href{https://arxiv.org/abs/2305.08914}{arXiv:2305.08914}.

\bibitem{Li:2023tic}
W.~Li,
\emph{Easy bootstrap for the 3D Ising model: a hybrid approach of the lightcone bootstrap and error minimization methods},
JHEP \textbf{07}, 047 (2024),
\href{https://arxiv.org/abs/2312.07866}{arXiv:2312.07866}.

\bibitem{Parga:1981}
N.~Parga and J.~E. van Himbergen, \emph{Renormalization group studies of phase diagrams of Ising and XY-model films of variable thickness},
Ann. Phys. \textbf{134}, 286 (1981),
\href{https://doi.org/10.1016/0003-4916(81)90212-8}{doi:10.1016/0003-4916(81)90212-8}.

\bibitem{Landau:1989}
D.~P. Landau, R.~Pandey and K.~Binder, \emph{Monte Carlo study of surface critical behavior in the XY model},
Phys. Rev. B \textbf{39}, no.~16, 12302 (1989),
\href{https://doi.org/10.1103/PhysRevB.39.12302}{doi:10.1103/PhysRevB.39.12302}.

\bibitem{Peczak:1991}
P.~Peczak and D.~P. Landau, \emph{Monte Carlo study of the surface special transition in the XY model in three dimensions},
Phys. Rev. B \textbf{43}, 1048 (1991),
\href{https://doi.org/10.1103/PhysRevB.43.1048}{doi:10.1103/PhysRevB.43.1048}.

\bibitem{McAvity:1993ue}
D.~M. McAvity and H.~Osborn, \emph{Energy-momentum tensor in conformal field theories near a boundary},
Nucl. Phys. B \textbf{406}, 655--680 (1993),
\href{https://arxiv.org/abs/hep-th/9302068}{arXiv:hep-th/9302068}.

\bibitem{McAvity:1995zd}
D.~M. McAvity and H.~Osborn, \emph{Conformal field theories near a boundary in general dimensions},
Nucl. Phys. B \textbf{455}, 522--576 (1995),
\href{https://arxiv.org/abs/cond-mat/9505127}{arXiv:cond-mat/9505127}.

\bibitem{Billo:2016cpy}
M.~Billò, V.~Goncalves, E.~Lauria and M.~Meineri, \emph{Defects in conformal field theory},
JHEP \textbf{04}, 091 (2016),
\href{https://arxiv.org/abs/1601.02883}{arXiv:1601.02883}.

\bibitem{Lauria:2017wav}
E.~Lauria, M.~Meineri and E.~Trevisani, \emph{Radial coordinates for defect CFTs},
JHEP \textbf{11}, 148 (2018),
\href{https://arxiv.org/abs/1712.07668}{arXiv:1712.07668}.

\bibitem{Ferrara:1973yt}
S.~Ferrara, A.~F.~Grillo and R.~Gatto, \emph{Tensor representations of conformal algebra and conformally covariant operator product expansion},
Ann. Phys. \textbf{76}, 161--188 (1973),
\href{https://doi.org/10.1016/0003-4916(73)90446-6}{doi:10.1016/0003-4916(73)90446-6}.

\bibitem{Grans-Samuelsson:2021uor}
L.~Grans-Samuelsson, R.~Nivesvivat, J.~L. Jacobsen, S.~Ribault and H.~Saleur, \emph{Global symmetry and conformal bootstrap in the two-dimensional $\mathrm{O}(n)$ model},
SciPost Phys. \textbf{12}, 147 (2022),
\href{https://arxiv.org/abs/2111.01106}{arXiv:2111.01106}.

\bibitem{Rychkov:2015naa}
S.~Rychkov and Z.~M.~Tan, \emph{The $\epsilon$-expansion from conformal field theory},
J. Phys. A \textbf{48}, 29FT01 (2015),
\href{https://arxiv.org/abs/1505.00963}{arXiv:1505.00963}.

\bibitem{Cardy:1990xm}
J.~L. Cardy, \emph{Universal critical-point amplitudes in parallel-plate geometries},
Phys. Rev. Lett. \textbf{65}, 1443 (1990),
\href{https://doi.org/10.1103/PhysRevLett.65.1443}{doi:10.1103/PhysRevLett.65.1443}.

\bibitem{Penedones:2015aga}
J.~Penedones, E.~Trevisani and M.~Yamazaki, \emph{Recursion relations for conformal blocks},
JHEP \textbf{09}, 070 (2016),
\href{https://arxiv.org/abs/1509.00428}{arXiv:1509.00428}.

\bibitem{Henriksson:2022rnm}
J.~Henriksson, \emph{The critical $\mathrm{O}(N)$ CFT: Methods and conformal data},
Phys. Rept. \textbf{1002}, 1--72 (2023),
\href{https://arxiv.org/abs/2201.09520}{arXiv:2201.09520}.

\bibitem{Henriksson:2025hwi}
J.~Henriksson, F.~Herzog, S.~R.~Kousvos and J.~Roosmale Nepveu,
\emph{Multiloop spectra in general scalar EFTs and CFTs},
Phys. Lett. B \textbf{874}, 140235 (2026),
\href{https://arxiv.org/abs/2507.12518}{arXiv:2507.12518}.

\bibitem{Dey:2026cso}
A.~Dey, L.~Herviou, C.~Mudry, S.~Rychkov and A.~M.~L\"auchli,
\emph{Conformal data for the $\mathrm{O}(2)$ Wilson--Fisher CFT in $(2+1)$-dimensional spacetime from exact diagonalization and matrix product states on the fuzzy sphere},
\href{https://arxiv.org/abs/2604.18705}{arXiv:2604.18705}.

\bibitem{Podo:2026hfh}
A.~Podo and S.~Rychkov, \emph{Direct experimental test of conformal invariance via grazing scattering: A proposal for X-ray and neutron experiments},
\href{https://arxiv.org/abs/2605.06773}{arXiv:2605.06773}.

\bibitem{Kangsabanik:2026tdm}
K.~Kangsabanik, A.~Kaviraj and A.~Tiwari,
\emph{Truncated Polyakov bootstrap for BCFTs: Neumann--Dirichlet flow and Ising special transition},
\href{https://arxiv.org/abs/2609.07841}{arXiv:2609.07841}.

\bibitem{Diehl:1984}
H.~W. Diehl and E.~Eisenriegler, \emph{Effects of surface exchange anisotropies on magnetic critical and multicritical behavior at surfaces},
Phys. Rev. B \textbf{30}, 300 (1984),
\href{https://doi.org/10.1103/PhysRevB.30.300}{doi:10.1103/PhysRevB.30.300}.

\bibitem{Giombi:2021cnr}
S.~Giombi, E.~Helfenberger and H.~Khanchandani, \emph{Fermions in AdS and Gross-Neveu BCFT},
JHEP \textbf{07}, 018 (2022),
\href{https://arxiv.org/abs/2110.04268}{arXiv:2110.04268}.

\bibitem{Herzog:2022jlx}
C.~P.~Herzog and V.~Schaub,
\emph{Fermions in boundary conformal field theory: crossing symmetry and $\epsilon$-expansion},
JHEP \textbf{02}, 129 (2023), 
\href{https://arxiv.org/abs/2209.05511}{arXiv:2209.05511}.

\bibitem{Shen:2024pve}
X.~Shen, Z.~Wu and S.-K.~Jian,
\emph{Boundary and defect criticality in topological insulators and superconductors},
Phys. Rev. B \textbf{112}, L041118 (2025),
\href{https://arxiv.org/abs/2407.15916}{arXiv:2407.15916}.

\bibitem{Jiang:2025sfb}
H.~Jiang, Y.~Ge and S.-K.~Jian,
\emph{Boundary criticality for the Gross--Neveu--Yukawa models},
Phys. Rev. Lett. \textbf{135}, 141602 (2025),
\href{https://arxiv.org/abs/2503.13247}{arXiv:2503.13247}.

\bibitem{Ge:2025cmk}
Y.~Ge, H.~Jiang, H.~Yao and S.-K.~Jian,
\emph{Boundary criticality in two-dimensional correlated topological superconductors},
\href{https://arxiv.org/abs/2510.05230}{arXiv:2510.05230}.

\bibitem{Diatlyk:2026gab}
O.~Diatlyk, Z.~Sun and Y.~Wang,
\emph{Extraordinary surface criticalities for interacting fermions},
\href{https://arxiv.org/abs/2604.15187}{arXiv:2604.15187}.

\bibitem{Diatlyk:2026eta}
O.~Diatlyk, S.~Giombi and Z.~Sun,
\emph{Boundary criticality in the Gross--Neveu--Yukawa model at higher orders},
\href{https://arxiv.org/abs/2606.07510}{arXiv:2606.07510}.

\bibitem{Atanasov:2022bpi}
A.~Atanasov, A.~Hillman, D.~Poland, J.~Rong and N.~Su,
\emph{Precision bootstrap for the $\mathcal{N}=1$ super-Ising model},
JHEP \textbf{08}, 136 (2022),
\href{https://arxiv.org/abs/2201.02206}{arXiv:2201.02206}.

\bibitem{Erramilli:2022kgp}
R.~S. Erramilli, L.~V. Iliesiu, P.~Kravchuk, A.~Liu, D.~Poland and D.~Simmons-Duffin, \emph{The Gross-Neveu-Yukawa archipelago},
JHEP \textbf{02}, 036 (2023),
\href{https://arxiv.org/abs/2210.02492}{arXiv:2210.02492}.

\bibitem{Mitchell:2024hix}
M.~S.~Mitchell and D.~Poland,
\emph{Bounding irrelevant operators in the 3d Gross--Neveu--Yukawa CFTs},
JHEP \textbf{09}, 134 (2024),
\href{https://arxiv.org/abs/2406.12974}{arXiv:2406.12974}.


\bibitem{Kantor:2021kbx}
G.~Kántor, V.~Niarchos and C.~Papageorgakis,
\emph{Solving conformal field theories with artificial intelligence},
Phys. Rev. Lett. \textbf{128}, no.~4, 041601 (2022),
\href{https://arxiv.org/abs/2108.08859}{arXiv:2108.08859}.

\bibitem{Kantor:2021jpz}
G.~Kántor, V.~Niarchos and C.~Papageorgakis,
\emph{Conformal bootstrap with reinforcement learning},
Phys. Rev. D \textbf{105}, no.~2, 025018 (2022),
\href{https://arxiv.org/abs/2108.09330}{arXiv:2108.09330}.

\bibitem{Kantor:2022epi}
G.~Kántor, V.~Niarchos, C.~Papageorgakis and P.~Richmond,
\emph{$6D$ $(2,0)$ bootstrap with the soft-actor-critic algorithm},
Phys. Rev. D \textbf{107}, no.~2, 025005 (2023),
\href{https://arxiv.org/abs/2209.02801}{arXiv:2209.02801}.

\bibitem{Niarchos:2023lot}
V.~Niarchos, C.~Papageorgakis, P.~Richmond, A.~G.~Stapleton and M.~Woolley,
\emph{Bootstrability in line-defect CFTs with improved truncation methods},
Phys. Rev. D \textbf{108}, no.~10, 105027 (2023),
\href{https://arxiv.org/abs/2306.15730}{arXiv:2306.15730}.

\bibitem{Niarchos:2025cdg}
V.~Niarchos, C.~Papageorgakis, A.~Stratoudakis and M.~Woolley,
\emph{Deep finite temperature bootstrap},
Phys. Rev. D \textbf{112}, no.~12, 126012 (2025),
\href{https://arxiv.org/abs/2508.08560}{arXiv:2508.08560}.

\bibitem{Niarchos:2026vow}
V.~Niarchos and C.~Papageorgakis,
\emph{Neural networks, dispersion relations and the thermal bootstrap},
\href{https://arxiv.org/abs/2605.13183}{arXiv:2605.13183}.

\bibitem{Poland:2023bny}
D.~Poland, V.~Prilepina and P.~Tadi\'c,
\emph{Improving the five-point bootstrap},
JHEP \textbf{05}, 299 (2024),
\href{https://arxiv.org/abs/2312.13344}{arXiv:2312.13344}.

\bibitem{Barrat:2025wbi}
J.~Barrat, E.~Marchetto, A.~Miscioscia and E.~Pomoni,
\emph{Thermal bootstrap for the critical $\mathrm{O}(N)$ model},
Phys. Rev. Lett. \textbf{134}, no.~21, 211604 (2025),
\href{https://arxiv.org/abs/2411.00978}{arXiv:2411.00978}.

\bibitem{Poland:2025ide}
D.~Poland, V.~Prilepina and P.~Tadi\'c,
\emph{Mixed five-point correlators in the 3d Ising model},
JHEP \textbf{10}, 237 (2025),
\href{https://arxiv.org/abs/2507.01223}{arXiv:2507.01223}.

\bibitem{Li:2021uki}
W.~Li,
\emph{Ising model close to $d=2$},
Phys. Rev. D \textbf{105}, no.~9, L091902 (2022),
\href{https://arxiv.org/abs/2107.13679}{arXiv:2107.13679}.

\bibitem{Laio:2022ayq}
A.~Laio, U.~Luviano Valenzuela and M.~Serone,
\emph{Monte Carlo approach to the conformal bootstrap},
Phys. Rev. D \textbf{106}, no.~2, 025019 (2022),
\href{https://arxiv.org/abs/2206.05193}{arXiv:2206.05193}.

\bibitem{Benjamin:2026lbj}
N.~Benjamin, A.~L.~Fitzpatrick, W.~Li and J.~Thaler,
\emph{Descending into the modular bootstrap},
JHEP \textbf{09}, 029 (2026),
\href{https://arxiv.org/abs/2604.01275}{arXiv:2604.01275}.

\bibitem{Li:2024ggr}
W.~Li, \emph{Analytic trajectory bootstrap for matrix models},
JHEP \textbf{02}, 098 (2025),
\href{https://arxiv.org/abs/2407.08593}{arXiv:2407.08593}.

\bibitem{Kangshabanik:2026onk}
K.~Kangshabanik, A.~Kaviraj and S.~Lal,
\emph{Truncated Polyakov bootstrap},
\href{https://arxiv.org/abs/2609.01732}{arXiv:2609.01732}.

\bibitem{Gliozzi:2014jsa}
F.~Gliozzi and A.~Rago,
\emph{Critical exponents of the 3d Ising and related models from conformal bootstrap},
JHEP \textbf{10}, 042 (2014),
\href{https://arxiv.org/abs/1403.6003}{arXiv:1403.6003}.

\bibitem{Shimada:2015gda}
H.~Shimada and S.~Hikami,
\emph{Fractal dimensions of self-avoiding walks and Ising high-temperature graphs in 3D conformal bootstrap},
J. Stat. Phys. \textbf{165}, 1006 (2016),
\href{https://arxiv.org/abs/1509.04039}{arXiv:1509.04039}.

\bibitem{Hogervorst:2016itc}
M.~Hogervorst, M.~Paulos and A.~Vichi,
\emph{The ABC (in any D) of logarithmic CFT},
JHEP \textbf{10}, 201 (2017),
\href{https://arxiv.org/abs/1605.03959}{arXiv:1605.03959}.

\bibitem{Hikami:2017hwv}
S.~Hikami,
\emph{Conformal bootstrap analysis for the Yang--Lee edge singularity},
PTEP \textbf{2018}, no.~5, 053I01 (2018),
\href{https://arxiv.org/abs/1707.04813}{arXiv:1707.04813}.

\bibitem{Hikami:2018qpz}
S.~Hikami,
\emph{Conformal bootstrap analysis for localization: symplectic case},
\href{https://arxiv.org/abs/1811.05918}{arXiv:1811.05918}.

\bibitem{Leclair:2018trn}
A.~Leclair and J.~Squires,
\emph{Conformal bootstrap for percolation and polymers},
J. Stat. Mech. \textbf{1812}, 123105 (2018),
\href{https://arxiv.org/abs/1802.08911}{arXiv:1802.08911}.

\bibitem{Hikami:2017sbg}
S.~Hikami,
\emph{Conformal bootstrap analysis for single and branched polymers},
PTEP \textbf{2018}, no.~12, 123I01 (2018),
\href{https://arxiv.org/abs/1708.03072}{arXiv:1708.03072}.

\bibitem{Padayasi:2023hpd}
J.~Padayasi and I.~A.~Gruzberg,
\emph{Conformal invariance and multifractality at Anderson transitions in arbitrary dimensions},
Phys. Rev. Lett. \textbf{131}, no.~26, 266401 (2023),
\href{https://arxiv.org/abs/2306.07340}{arXiv:2306.07340}.

\bibitem{Shimamori:2025afk}
S.~Shimamori and Y.~Wang,
\emph{Subdimensional disorder and logarithmic defect},
JHEP \textbf{07}, 129 (2026),
\href{https://arxiv.org/abs/2510.13964}{arXiv:2510.13964}.

\bibitem{Hasenbusch:2025yrl}
M.~Hasenbusch, \emph{Eliminating leading and subleading corrections to scaling in the three-dimensional XY universality class},
Phys. Rev. B \textbf{112}, no.~18, 184512 (2025),
\href{https://arxiv.org/abs/2507.19265}{arXiv:2507.19265}.

\bibitem{Hasenbusch:2019jkj}
M. Hasenbusch, \emph{Monte Carlo study of an improved clock model in three dimensions},
Phys. Rev. B \textbf{100}, no.~22, 224517 (2019),
\href{https://arxiv.org/abs/1910.05916}{arXiv:1910.05916}.

\bibitem{Hasenbusch:2020pwj}
M.~Hasenbusch, \emph{Monte Carlo study of a generalized icosahedral model on the simple cubic lattice},
Phys. Rev. B \textbf{102}, no.~2, 024406 (2020),
\href{https://arxiv.org/abs/2005.04448}{arXiv:2005.04448}.

\bibitem{Hasenbusch:2021rse}
M.~Hasenbusch, \emph{Three-dimensional $\mathrm{O}(N)$-invariant $\phi^4$ models at criticality for $N\geq 4$},
Phys. Rev. B \textbf{105}, no.~5, 054428 (2022),
\href{https://arxiv.org/abs/2112.03783}{arXiv:2112.03783}.

\end{thebibliography}
\end{document}